\documentclass[fleqn,usenatbib]{mnras}

\usepackage{newtxtext,newtxmath}
\usepackage{physics}

\usepackage[T1]{fontenc}

\DeclareRobustCommand{\VAN}[3]{#2}
\let\VANthebibliography\thebibliography
\def\thebibliography{\DeclareRobustCommand{\VAN}[3]{##3}\VANthebibliography}
\def\beq{\begin{equation}}
\def\eeq{{\end{equation}}}

\def\softwarenamestyle[#1]{\textsc{#1}}

\usepackage{graphicx}	
\usepackage{amsmath}	
\usepackage[dvipsnames]{xcolor}
\usepackage{enumitem}
\usepackage{url}
\usepackage{braket}
\usepackage{threeparttable}
\usepackage{hyperref}
\usepackage{multirow, tabularx}
\usepackage{siunitx}

\def\mpc{\, h^{-1}{\rm {Mpc}}}

\def\Kpc{\, {\rm {kpc}}}
\def\kpc{\, h^{-1}{\rm {kpc}}}

\def\Myr{\,{\rm Myr}}

\def\gyr{\, h^{-1}{\rm Gyr}}

\def\gyri{\, h\,{\rm Gyr}^{-1}}
\def\kms{\,{\rm {km\, s^{-1}}}}
\def\Msun{{\rm M_\odot}}
\def\msun{\, h^{-1}{\rm M_\odot}}
\def\msunperyr{\, {\rm M_\odot}/{\rm yr}}

\title[A two-phase model of galaxy formation]{A two-phase model of galaxy formation: 
I. The growth of galaxies and supermassive black holes}

\author[Mo, Chen, Wang]{
Houjun Mo,$^{1,2}$\thanks{E-mail: hjmo@umass.edu}
Yangyao Chen,$^{3,4}$
and Huiyuan Wang$^{3,4}$ 
\\
$^{1}$Department of Astronomy, University of Massachusetts, Amherst MA01003, USA\\
$^{2}$Tsung-Dao Lee Institute, Shanghai Jiao Tong University, Shanghai 200240, China\\
$^{3}$School of Astronomy and Space Science, University of Science and Technology of China, Hefei, Anhui 230026, China\\
$^{4}$Key Laboratory for Research in Galaxies and Cosmology, Department of Astronomy, University of Science and Technology of China, Hefei, Anhui 230026, China \\
}

\date{Accepted XXX. Received YYY; in original form ZZZ}

\pubyear{2023}

\begin{document}
\label{firstpage}
\pagerange{\pageref{firstpage}--\pageref{lastpage}}
\maketitle

\begin{abstract}
We develop a model for galaxy formation and 
the growth of supermassive black holes (SMBHs), based on the fact that cold dark 
matter (CDM) halos form their gravitational potential wells 
through a fast phase with rapid change in the potential, and that 
the high universal baryon fraction makes cooled gas in halos self-gravitating 
and turbulent before it can form rotation-supported disks. Gas fragmentation 
produces sub-clouds so dense that cloud-cloud collision and drag 
on clouds are not significant, producing a dynamically hot system of 
sub-clouds that form stars and move ballistically to feed the central 
SMBH. Active galactic nucleus (AGN) and supernova (SN) feedback is effective 
only in the fast phase, and the cumulative effects are to regulate star formation and 
SMBH growth, as well as to reduce the amount of cold gas in halos to allow the 
formation of globally stable disks. Using a set of 
halo assembly histories, we demonstrate that the model can reproduce a 
number of observations, including correlations among SMBH mass, stellar mass 
of galaxies and halo mass, the number densities of galaxies and SMBH, 
as well as their evolution over the cosmic time. 
\end{abstract}

\begin{keywords}
galaxies: haloes -- galaxies: formation -- quasars: supermassive black holes
\end{keywords}



\section{Introduction}

In the current $\Lambda$CDM scenario of structure formation, galaxies
and supermassive black holes (SMBHs) hosted by them 
form in the gravitational potential wells of dark matter halos in the cosmic density field
\citep[e.g.][]{moGalaxyFormationEvolution2010}. It is thus crucial to understand  
how the formation and evolution of galaxies and SMBHs are determined by the growth 
and structure of dark matter halos. Despite the progress made so far  
using methods ranging from full hydro simulations 
\citep[e.g.][]{springelCosmologicalSmoothedParticle2003, springelPurSiMuove2010,
genelIntroducingIllustrisProject2014,
vogelsbergerPropertiesGalaxiesReproduced2014,
schayeEAGLEProjectSimulating2015,
pillepichFirstResultsIllustrisTNG2018,springelFirstResultsIllustrisTNG2018,
daveSimbaCosmologicalSimulations2019}
to simple analytical modeling \citep[e.g.][]{moFormationGalacticDiscs1998,
dekelToyModelsGalaxy2013}, 
the details remain poorly understood. For example, we still do not  
understand fully what processes separate the two basic types of galaxies, 
elliptical and disk, and how SMBHs fit into and affect the formation processes.   

The difficulty arises from the complexity of the problem.  Numerical simulations 
show that the halo population is very diverse not only in their mass, but 
also in other important properties relevant to galaxy formation, such as the spin, shape, 
and formation history \citep[e.g.][]{maccioConcentrationSpinShape2007,
liHaloFormationTimes2008,wongWhatDarkMatter2012,maoAssemblyBiasExploring2018,
chenRelatingStructureDark2020,chenMassiveDarkMatter2023,
wangCharacterizingAssemblyDark2024}. 
Even more complex is the formation and evolution of the baryonic 
components, such as gas, stars and SMBHs, as they are affected not only by gravity but also by 
many gas-dynamical and radiative processes. These processes are far from linear; they are 
strongly coupled with each other, and could even be self-interacting through feedback loops 
\citep[e.g.][]{crainEAGLESimulationsGalaxy2015,weinbergerSimulatingGalaxyFormation2017,
pillepichSimulatingGalaxyFormation2018,daveSimbaCosmologicalSimulations2019}.  

As in dealing with any complex system, the best approach is to move step by step, 
starting from the simplest model that involves the most important properties
of the driver population of the evolution, in our case dark matter halos. 
One then moves forward by adding more complexities when needed. The addition 
of complexity must be done intelligently, so that our goal of understanding the 
underlying physical processes is not lost in our endeavor.  

One of the key properties of the halo population is the regularity 
in their formation histories. As shown by numerical simulations 
\citep{wechslerConcentrationsDarkHalos2002, zhaoGrowthStructureDark2003, 
zhaoAccurateUniversalModels2009}, the assembly of dark halos in a cold dark matter (CDM) 
cosmology typically consists of two distinct phases: a fast assembly phase in which 
the gravitational potential well of the halo is established, and a slow assembly 
phase during which mass is added into the halo gently without changing  the potential 
well significantly. The early fast assembly phase is also found to be 
associated with violent changes in the gravitational potential, which can cause 
energy exchanges between mass particles and make the velocity dispersion of these 
particles more isotropic \citep[e.g.][]{zhaoGrowthStructureDark2003,luOriginColdDark2006}. 

Another key condition for galaxy formation is set by current cosmology. 
Although the mass in the universe is dominated by cold dark matter (CDM),  
the mass fraction contained in baryons, $f_{\rm B} \approx 0.15$, 
is significant gravitationally. Furthermore, the gas component is 
dissipative, so that it can cool and collapse further in a dark matter 
halo, making it gravitationally dominating on small scales where such dissipation 
is important. The typical specific angular momentum of dark matter halos, as described 
by the spin parameter $\lambda$, is in general too low to significantly impede the dissipation 
and collapse of the gas before it becomes self-gravitating. This, combined with the 
rapid change of the gravitational potential during the fast assembly phase, 
is expected to produce self-gravitating, dynamically hot (turbulent) clouds. 
These clouds can then fragment to form sub-clouds within which stars form 
and SMBHs grow. This early formation may also be accompanied by strong feedback that can 
reduce the cold gas fraction to a level $\sim \lambda$ so that globally stable disks 
with realistic rotation curves can form \citep[e.g.][]{moFormationGalacticDiscs1998}. 
In this scenario, the feedback is `ejective' in the fast assembly phase, but becomes 
`preventive' in the subsequent slow assembly phase.  Such a two-phase scenario of galaxy 
formation has been put forward by \citet{moGalaxyFormationPreheated2002} 
and \citet{moGalaxyFormationPreprocessed2004}, and worked out in some detail by 
\citet{cookTwophaseGalaxyFormation2009} and \citet{luFormationDiscGalaxies2015} 
using an analytical approach, and 
by \citet{bocoTOPSEMTwOParameters2023} using an abundance-matching scheme.  
The turbulence and fragmentation of the gas associated with the fast assembly phase 
may also create conditions for seeding and growing SMBHs 
\citep[e.g.][]{hobbsFeedingSupermassiveBlack2011,latifTurbulentColdFlows2022}, 
which allows us to treat the growth of the SMBHs and active galactic nuclei (AGN) 
feedback in the same framework as galaxy formation.

In this paper, we expand the two-phase formation scenario by developing a  
framework to model galaxy formation and the growth of SMBHs in 
dark matter halos predicted by current cosmology. The presentation of our framework 
will be based on a set of simple assumptions that make our model transparent
to key processes driving the evolution. Our model is also quantitative so that 
it can be applied to make model predictions that can be tested with observations. 
The structure of the paper is as follows. In Section~\ref{sec:halos} we describe basic halo 
properties used in our model, including a scheme to separate the fast 
and slow assembly phases. In Section~\ref{sec:gas-collape} we analyze the cooling and collapse of the gas component
in dark matter halos, and present the case for the formation of self-gravitating, 
turbulent gas clouds (SGC) and their fragmentation to form gas sub-clouds.   
Section~\ref{sec:bh_growth} describes how stars form and the SMBH grows in an SGC. 
The application of our model to simulated halos is given in Section~\ref{sec_applications},
together with model predictions and comparisons with observational data.
Finally, in Section~\ref{sec:summary}, we summarize the main components and 
predictions of our model, and make further discussions. 
A code library that implements our model to populate halos with galaxies and SMBHs, 
along with the data used to make all the figures in the paper, is described in the 
Data Availability section.

\section{Dark matter halos}
\label{sec:halos}

\begin{figure*} \centering
    \includegraphics[width=\textwidth]{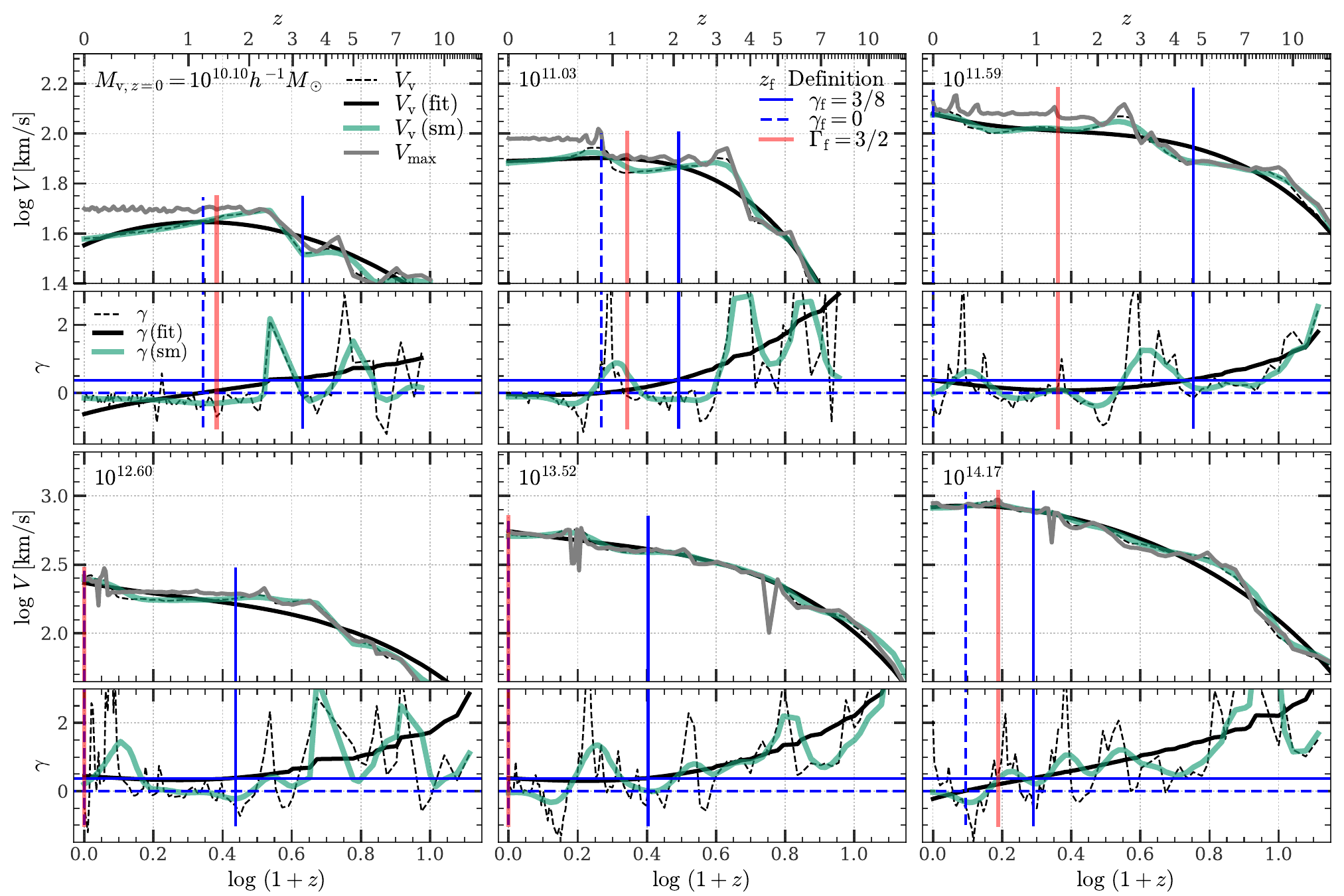}
    \caption{
        Assembly histories of individual halos (each pair of panels) 
        described by circular velocity as a function of redshift. 
        For each halo, the virial velocity, $V_{\rm v}(z)$, is shown by 
        the {\bf black dashed curve}, obtained directly from the simulation.
        The {\bf black solid curve} shows the virial velocity obtained from 
        a parametric fitting of halo mass assembly history (Eq.~\ref{eq:mah-fitting}),
        while the {\bf green curve} shows that obtained from a smoothing 
        of halo mass assembly history by a Gaussian kernel with $\sigma = \tau_{\rm dyn}(z)$, 
        the dynamical time scale of the halo.
        The {\bf gray curve} shows the maximal circular velocity, 
        $V_{\rm max}(z)$, obtained from the simulation.
        The virial mass, $M_{\rm v} \equiv M_{\rm 200c}$, at $z=0$ is 
        indicated in the top-left corner of each top panel. 
        The velocity is shown in each {\bf top panel}, while the corresponding 
        {\bf bottom panel} displays $\gamma(z)$, which represents the 
        specific growth rate of $V_{\rm v}(z)$. 
        Transition redshifts, $z_{\rm f}$, obtained using three definitions, 
        as indicated in the legend of the top-center panel, are 
        shown by three {\bf colored vertical lines}. In each bottom panel, 
        the {\bf blue solid and dashed horizontal lines} denote the 
        threshold $\gamma_{\rm f}=3/8$ and $\gamma_{\rm f}=0$ used to 
        define $z_{\rm f}$ in the High-$z_{\rm f}$ and Low-$z_{\rm f}$ variants 
        of our model, respectively. All the halos presented here are selected 
        from the small sample, $S_{\rm h,small}$ (see \S\ref{ssec:halo-sample}). 
        For detailed information about halo assembly history and transition 
        redshift, see \S\ref{sec:halos}. This figure clearly demonstrates 
        the two-phase nature of CDM halo assembly, which forms the foundation 
        for our two-phase model of the growth of galaxies and SMBHs.
    }
    \label{fig:transition_points}
\end{figure*}

We consider galaxy formation in cold dark matter (CDM) halos. 
We model the halo mass density distribution using the NFW profile 
\begin{equation}
\rho(r)=\rho_{\rm crit} {\delta_{\rm v}\over (r/r_{\rm s})(1+r/r_{\rm s})^2}\,,
\end{equation}
where $\rho_{\rm crit} \equiv 3H^2(z)/(8 \pi G)$ is the critical density of the universe 
at the redshift $z$ in question, $r_{\rm s}$ is a scale radius, 
and $\delta_{\rm v}$ is a characteristic 
over-density \citep[e.g.][]{navarroUniversalDensityProfile1997}.
Based on the spherical collapse model and virial theorem, we define $\delta_{\rm v}$ 
so that the mean density within the halo radius, $r_{\rm v}$, is ${\overline\rho}_{\rm v} = 200 \rho_{\rm crit}$.
In this case,  
\begin{equation}
\delta_{\rm v} ={200\over 3} {c^3\over \mu(c) }\,,    
\label{eq:nfw_amplitude}
\end{equation}
where 
\begin{equation}
c\equiv {r_{\rm v}\over r_{\rm s}}    
\end{equation}
is the halo concentration, and 
\begin{equation} \label{eq:mu_nfw}
\mu(y) \equiv \ln (1+y) -y/(1+y)\,.
\end{equation}
The halo mass within a radius $r$ is
\begin{equation}
M(<r) =4\pi\rho_{\rm crit} r_{\rm s}^3\mu(cx)\,,
\label{eq:nfw_m_enclosed}
\end{equation}
where $x=r/r_{\rm v}$. We also define a circular velocity at radius $r$, 
\begin{equation}
V_{\rm c}(r)= \sqrt{GM (<r)\over r}\,.     
\end{equation}
The halo (virial) mass, $M_{\rm v}$, and circular velocity, $V_{\rm v}$, are defined as
$M(<r)$ and $V_{\rm c}(r)$ evaluated at $r=r_{\rm v}$, respectively:
\begin{equation}
M_{\rm v} \equiv M(<r_{\rm v})\,;~~~~
V_{\rm v} \equiv \sqrt{GM_{\rm v} \over r_{\rm v}}\,.
\end{equation}

The formation history of a dark matter halo is represented by its 
merger tree. The mass growth of a halo is described by its mass assembly 
history, $M_{\rm v} (z)$, which is the virial mass of the progenitor halo 
at redshift $z$ along the main branch of the halo. 
The mass growth of the halo at $z$, over an interval $\Delta z$, can then be defined as 
\begin{equation}
\Delta M_{\rm v} (z) = {{\rm d} M_{\rm v}(z)\over {\rm d} z} \Delta z\,.
\end{equation}
As shown in \citet{zhaoGrowthStructureDark2003}, the formation of CDM halos 
in general consists of a fast phase, where the mass increases with time rapidly,  
followed by a slow phase, where the mass assembly rate is slower. 
The distinction between these two phases is more clearly seen in the 
redshift evolution of $V_{\rm v} (z)$. In the fast assembly regime, 
$V_{\rm v} (z)$ increases rapidly with time (i.e. with decreasing $z$), 
while $V_{\rm v}(z)$ remains almost constant or even declines with time in the slow 
phase. These behaviors can be seen in the examples shown in the top panels of
Fig.~\ref{fig:transition_points} 
(see \S\ref{ssec:halo-sample} and Appendix~\ref{app:method-halo-sampling} for 
halo samples used in this paper).

For convenience, we denote the transition redshift of the two phases as $z_{\rm f}$
(`f' means fast). To obtain $z_{\rm f}$ of a halo, we trace its main branch assembly history and use  
\begin{equation}\label{eq:def_gamma}
\frac{\dot{V}_{\rm v}(z)}{V_{\rm v}(z)} = \gamma (z) H(z)
\end{equation}
to obtain $\gamma (z)$, the specific growth rate of the virial velocity. 
Following the arguments given in \S4 of 
\citet{zhaoGrowthStructureDark2003}, we use the redshift when 
$\gamma (z)$ first reaches the threshold $\gamma_{\rm f} = 3/8$ to define $z_{\rm f}$:
\begin{equation}\label{eq:def-high-zf}
\gamma (z_{\rm f}) = \gamma_{\rm f} =3/8 
~~~({\rm High-}z_{\rm f})\,.
\end{equation}
Models using this definition of $z_{\rm f}$ will be referred to as the High-$z_{\rm f}$ variant, since 
$z_{\rm f}$ so obtained is in general higher than those obtained by other definitions (see below). 
\citet{moreSplashbackRadiusPhysical2015} and \citet{bocoTOPSEMTwOParameters2023}
both adopted $\Gamma_{\rm f} \equiv \Gamma(z_{\rm f}) = 3/2$ in their modeling based on the mass history, 
${\dot M}_{\rm v}(z)/ M_{\rm v}(z)= \Gamma (z) H(z)$. This corresponds to 
$\gamma_{\rm f} = 0$ at high $z$. We thus also consider another case, 
\begin{equation}\label{eq:def-low-zf}
\gamma (z_{\rm f}) = \gamma_{\rm f} =0
~~~({\rm Low-}z_{\rm f})\,,
\end{equation}
where Low-$z_{\rm f}$ indicates that the value of $z_{\rm f}$ defined 
with $\gamma_{\rm f} =0$ is in general lower than that defined with 
$\gamma_{\rm f} =3/8$ (see Appendix~\ref{app:transition-redshift} for a comparison 
of different definitions).

Halo assembly histories in simulations are in general quite noisy, as can be seen from the 
simulated $V_{\rm v}$ and $V_{\rm max}$ curves shown in the top panels of Fig.~\ref{fig:transition_points}. 
The time differentiation, $\gamma(z)$, is even noisier, as seen from the bottom 
panels of the same figure. To identify the transition redshift reliably, we fit 
each halo assembly history, $M_{\rm v}(z)$, by a smooth function
\footnote{To avoid confusion, we use `$\log$' to denote 
10-based logarithm, and `$\ln$' to denote $e$-based logarithm.},
\begin{equation}\label{eq:mah-fitting}
    \ln\,M_{\rm v}(z) = c_0 + c_1 {z\over 1+z} + c_2 \ln(1+z) + c_3 z\,,
\end{equation}
where $(c_0,\,c_1,\,c_2,\,c_3)$ are four free parameters. 
This functional form is similar to that used by \citet{behrooziUniverseMachineCorrelationGalaxy2019}, 
and it can be viewed as an extension of the fitting methods used  
by, e.g. \citet{wechslerConcentrationsDarkHalos2002},
\citet{vandenboschUniversalMassAccretion2002}, \citet{mcbrideMassAccretionRates2009} and
\citet{correaAccretionHistoryDark2015}. Our tests show that the four-parameter fitting
gives more stable and accurate results for the problem interested here.
The evolution of $V_{\rm v}(z)$ is then obtained by
\begin{equation}\label{eq:v-vir-fitting}
    V_{\rm v}(z) = \left[ \frac{\Delta_{\rm v}(z)}{2} \right]^{1/6} 
    \left[ G M_{\rm v}(z)H(z) \right]^{1/3}\,,
\end{equation}
where $\Delta_{\rm v}(z) \equiv 200$. The dependence of $\gamma$ and $\Gamma$ on $z$ can 
be obtained from the fittings to $V_{\rm v}(z)$ and $M_{\rm v}$, respectively. 
The fitting results of $V_{\rm v}(z)$ and $\gamma(z)$ are shown by the black solid curves 
in Fig.~\ref{fig:transition_points}. Finally, to reduce the ambiguity due to multiple 
solutions, we search for $z_{\rm f}$ by maximizing a `loss' function defined as
\begin{equation}\label{eq:loss_fn_of_gamma}
    l_\gamma(z \vert \gamma_{\rm f}) = \ln \frac{V_{\rm v}(z)}{V_{\rm v}(z_{\rm f})} + \gamma_{\rm f} \ln (1+z) \,.
\end{equation}
It is easy to show that the extreme location, ${\rm argmax}_{z}\left[ l_\gamma(z \vert \gamma_{\rm f}) \right]$,
is equivalent to the solution of $\gamma(z)=\gamma_{\rm f}$.

The vertical lines in Fig.\,\ref{fig:transition_points}
mark the values of $z_{\rm f}$ obtained using different definitions. 
For most halos, the value represented by the blue dashed line ($\gamma_{\rm f} =0$) is significantly
larger than that by the blue solid line ($\gamma_{\rm f} =3/8$), and the red line 
($\Gamma_{\rm f} = 3/2$) lies in between. In Appendix~\ref{app:transition-redshift},
we show that the transition redshift obtained by all these definitions 
is a monotonically decreasing function of the halo mass. We also compare the 
transition redshift with the commonly used half-mass formation redshift 
to obtain some idea about the different definitions of the transition redshift.

For halos that are still in the fast assembly regime, \citet{zhaoGrowthStructureDark2003} 
found that the halo concentration is about $c=c_{\rm f} \approx 4$, 
with the exact value of $c_{\rm f}$ depending on the definition of the halo 
radius $r_{\rm v}$. \citet{zhaoGrowthStructureDark2003} also found that 
particles accreted at $z<z_{\rm f}$ are distributed in the 
outer part of the halo, at $r>c_{\rm f} r_{\rm s}$, while those accreted in the 
fast assembly regime stay in the inner part, at $r\le c_{\rm f} r_{\rm s}$. 
Note that $r=c_{\rm f} r_{\rm s}$ is close to the radius where 
the halo circular velocity reaches its peak and the density profile is 
roughly the isothermal profile, $1/r^2$. Thus, for a given halo, we may define a radius
\begin{equation}
r_{\rm f} \approx c_{\rm f} r_{\rm s}= (c_{\rm f}/c) r_{\rm v}\,,
\end{equation}
which separates an inner part, where mass is assembled 
into the halo by fast accretion and an outer part, 
where the mass is assembled mainly through slow accretion. 
So defined, $r_{\rm f}=r_{\rm v}$ for halos that are still in their fast 
assembly regime. The mass within $r_{\rm f}$ is 
\begin{equation}
M_{\rm f} = M(<r_{\rm f}) 
=M_{\rm v} {\mu(c_{\rm f}) \over \mu(c)}\,.
\label{eq:def_m_f}
\end{equation}
Since $M_{\rm f}$ is the halo mass assembled in the fast regime, 
the mass assembled in the slow accretion regime is, by definition, 
$M_{\rm v} - M_{\rm f}$.    

The average density within $r_{\rm f}$ is
\begin{equation}
{\overline\rho}_{\rm f}
= {3 M_{\rm f}\over 4\pi r_{\rm f}^3} =
{\mu(c_{\rm f}) \over \mu(c)} \left(c\over c_{\rm f}\right)^3 
\times {\overline\rho}_{\rm v}\,.
\label{eq_rhof}
\end{equation}
Given that the mean density is proportional to the critical density of the universe, 
the time, $t_{\rm f}$, when the mean density of a halo reaches ${\overline\rho}_{\rm f}$
is related to the current time, $t$, by
\begin{equation}
H^2(t_{\rm f}) =
{\mu(c_{\rm f})\over \mu(c)}\left({c\over c_{\rm f}}\right)^3 \times H^2 (t)  \,.
\label{eq:h_sq_to_c}
\end{equation}
For an Einstein-de Sitter universe, or for $z\gg 1$,  
$H^2 (t_{\rm f})/H^2(t) = (1+z_{\rm f})^3/(1+z)^3$, 
and so $(1+z_{\rm f}) \sim (c/c_{\rm f}) (1+z)$.

The dynamical time scale of the gravitational collapse of a halo is about 
$\tau_{\rm dyn} \equiv R_{\rm v}/V_{\rm v}$. Thus the typical time 
for the halo structure to adjust to a new state corresponding to 
the change in the mass accretion is $\tau_{\rm dyn}$.     
For reference, we smooth $M_{\rm v}(z)$ 
using a Gaussian kernel with standard deviation  
$\sigma (z) = \tau_{\rm dyn}(z) = R_{\rm v}(z)/V_{\rm v}(z)$, where 
$z$ is the redshift of the snapshot in question,
and we obtain other virial quantities, such as $V_{\rm v}(z)$ and $\gamma(z)$,
from the smoothed $M_{\rm v}(z)$.
The results are shown by the green curves in Fig.~\ref{fig:transition_points}. 
The smoothed $\gamma(z)$ follows the general trend of that obtained by 
the parametric fitting, but contains significant fluctuations even within a 
given phase of mass assembly. This suggests that a halo in the fast 
assembly phase can sometimes make an excursion to the slow phase and 
vice versa. We will discuss in \S\ref{ssec:four-quadrant-gas-evolution}
the implications of such excursions for the formation of galaxies in halos.

In the following, we will use the features of dark matter halos 
described above to develop a model for the formation and evolution of galaxies 
and SMBHs. Our modeling adopts a `step-wise' strategy 
\citep[e.g.][]{luEmpiricalModelStar2014}, deliberately designed to 
avoid over-fitting. Briefly, for each relevant process, we start with a physically 
motivated, heuristic functional form with all parameters held constant. 
We then introduce secondary dependencies, such as redshift, only when they are 
implied by observational evidence. An assumption made in this paper is the 
instantaneous recycling of baryons, where gas components are quantified by their 
`effective' amounts at a given time that describe the net results of multiple cycles of 
inflow and outflow. Thus, the modeled quantities should be interpreted as 
averages over a period defined by the recycling timescale, which is typically 
the dynamical time or the cooling time, depending on the specific properties and scales 
in question. The details of baryon cycles remain poorly constrained by observations, 
particularly at high $z$  \citep[see, e.g.][]{zhangInspiralingStreamsEnriched2023,linMetalenrichedNeutralGas2023}, 
and are not included in our model. To account for potential short-term fluctuations 
that are omitted in our model, we incorporate some random noise based on our best 
understanding of the uncertainties when conducting comparisons between model predictions 
and observational data.

\section{The collapse of the gas component}
\label{sec:gas-collape}

\begin{figure} \centering
    \includegraphics[width=0.98\columnwidth]{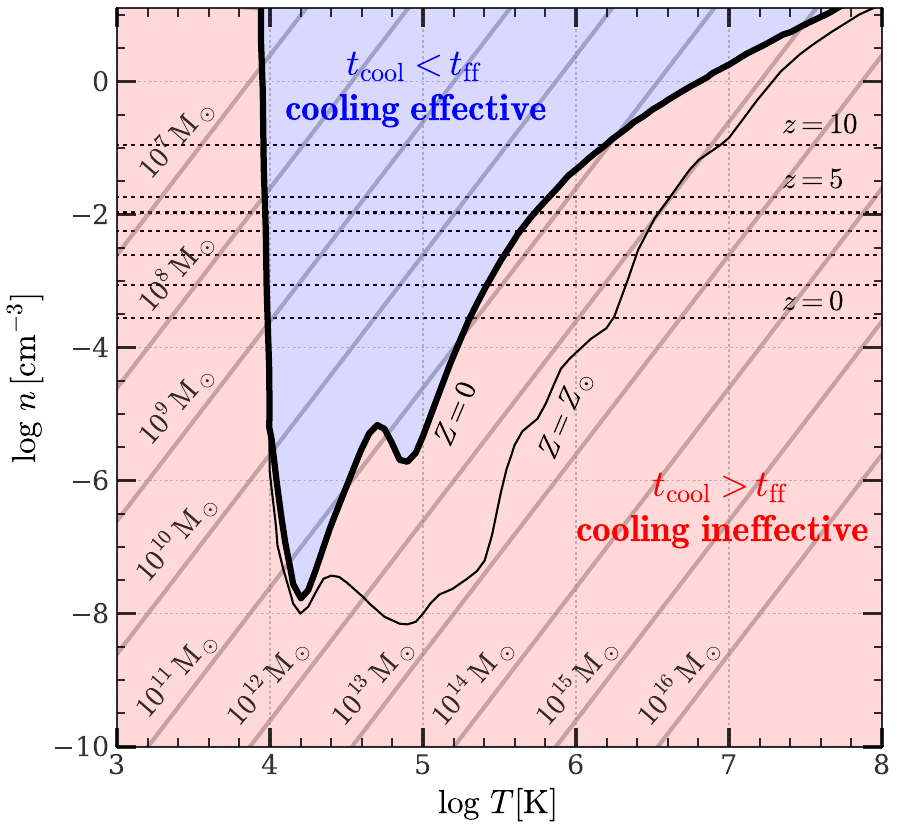}
    \caption{
        The cooling diagram. The {\bf thick and thin black curves} show the locus 
        of $t_{\rm cool}=t_{\rm ff}$ in the plane of gas particle number density 
        ($n$) and gas temperature ($T$) for halo gas with zero and solar metallicity, 
        respectively. The {\bf gray tilted solid lines} are lines of 
        constant halo mass ($M_{\rm v} \equiv M_{\rm 200c}$). 
        {\bf Horizontal dashed lines}, from lower to upper, indicate the density of virialized 
        halos at redshifts $z=0$, $1$, $2$, $3$, $4$, $5$, and $10$. The blue and red 
        shaded regions indicate the regimes of effective and ineffective cooling, 
        respectively. These calculations are based on 
        a $\Lambda$CDM cosmology with $f_{\rm gas}=0.15$, $h=0.7$, and 
        $\Omega_{\rm M,0}=0.3$. For a more detailed description, 
        see \S\ref{ssec:role-of-cooling}.
    }
    \label{fig:cooling_diagram}
\end{figure}

\subsection{The role of radiative cooling}
\label{ssec:role-of-cooling}

Cosmic gas is originally mixed uniformly with dark matter and collapses together  
with the dark matter as a halo forms. If the gas is initially heated
by accretion shocks associated with the collapse, radiative processes can cool it down.
The effectiveness of radiative cooling is represented by a cooling 
time scale, $t_{\rm cool}$, under the assumption that the gas is at the virial 
temperature, $T=T_{\rm v}$:
\begin{equation}
t_{\rm cool}
\approx 3.3\times 10^9 {T_6 \over n_{-3} \Lambda_{-23} (T)} 
{\,\rm yrs}\,,
\end{equation}
where $n_{-3}$ is the gas number density in units of $10^{-3}{\rm cm}^{-3}$, 
$T_6 \equiv T_{\rm v}/(10^6{\rm K})$, and $\Lambda_{-23}$ is the cooling 
function in units of $10^{-23}\,{\rm erg\,s^{-1} cm^3}$.
This time scale is compared with the free-fall time scale of the gas, 
\begin{equation}
t_{\rm ff} 
=\sqrt{3\pi \over 32 G {\overline\rho}}
\approx 
2.1\times 10^9 
f_{\rm gas}^{1/2} n_{-3} ^{-1/2}\,{\rm yrs}\,,
\end{equation}
where ${\overline\rho}$ is the typical total mass density of the halo,  
and $f_{\rm gas}$ is the gas mass fraction. 
If we approximate the cooling function by a power law, 
$\Lambda (T) \propto T^{-\mu}$, then 
\begin{equation}
{t_{\rm cool}\over t_{\rm ff}}
\propto 
{1\over f_{\rm gas}} M_{\rm v}^{2(1+\mu)/3} {\overline\rho}^{(2\mu -1)/6}\,,
\end{equation}
where $\mu$ is the mean molecular mass of the gas particle in units of the proton mass.
For an enriched gas at temperature $T\sim 10^6\,{\rm K}$, 
$\mu\sim 0.5$, the following interesting relation holds roughly:
\begin{equation}
{t_{\rm cool}\over t_{\rm ff}}
\sim \left({f_{\rm gas} \over 0.15}\right)^{-1}
\left({M_{\rm v}\over 10^{12}{\rm M}_\odot}\right)\,.
\end{equation}
Thus, radiative cooling is expected to be effective, 
i.e. $t_{\rm cool}< t_{\rm ff}$, for all halos with masses below 
$M_{\rm cool}\sim 10^{12}M_\odot$, quite independent of redshift, as 
shown in Fig.\,\ref{fig:cooling_diagram}. This suggests that the gas 
that can cool is a fraction of the total, and this fraction can be modeled  as 
\begin{equation}
F_{\rm cool} (M_{\rm v}\vert M_{\rm cool},\beta_{\rm cool})=
{1 \over 1+(M_{\rm v}/M_{\rm cool})^{\beta_{\rm cool}}}\,,
\label{eq:def-f-cool}
\end{equation}
where $\beta_{\rm cool}\sim 1$ and $M_{\rm cool} \sim 10^{12}{\rm M}_\odot$ 
are two model parameters. 
So modeled, $F_{\rm cool} \to 1$ for $M_{\rm v}\ll M_{\rm cool}$ and 
$\to 0$ for $M_{\rm v}\gg M_{\rm cool}$.   
Note that the cooled amount of gas described by $F_{\rm cool}$ is an 
upper limit, assuming that the gas inflow is not coupled with, for example, 
hot outflows resulting from feedback processes. The actual amount of gas entering 
the galaxy and forming stars is modified by additional operators to be described 
later.

Note that 
$t_{\rm cool}/ t_{\rm ff}$ scales inversely with $f_{\rm gas}$,  
and that $t_{\rm ff}$ defined above is about $1/10$ times the Hubble time: 
$t_{\rm ff} \sim t_{\rm H}/10$. Thus, for example, the locus of 
$M_{\rm v}=10^{11}{\rm M}_\odot$ will be shifted to that labeled by
$M_{\rm v}=10^{12}{\rm M}_\odot$ in the figure, either 
if the Hubble time is used instead of the free-fall time, or if $f_{\rm gas}$ is 
ten times smaller than the cosmic baryon fraction. Similarly, 
such changes will cause a shift of the locus of 
$M_{\rm v}=10^{12}{\rm M}_\odot$ to that labeled 
by $M_{\rm v}=10^{13}{\rm M}_\odot$ in the figure.

Fig.\,\ref{fig:cooling_diagram} also shows that radiative cooling 
is very effective in the temperature range $10^4{\rm K}< T<10^5{\rm K}$. 
This range roughly corresponds to 
$10^{10}{\rm M}_\odot< M_{\rm v}<10^{11}{\rm M}_\odot$ at $z \approx 0$, and 
$10^{9}{\rm M}_\odot< M_{\rm v}<10^{10}{\rm M}_\odot$ at $z \approx 5$. 

\subsection{The role of angular momentum}

In the presence of angular momentum, gas can be supported by its angular 
momentum in the host halo to form a rotation-supported disk. 
According to the model described in 
\cite{moFormationGalacticDiscs1998}, halo gas typically needs to collapse by 
a factor of $\lambda$ (the effective spin parameter of the gas) to become supported 
by rotation. Thus, as long as the cooled gas fraction, $f_{\rm gas}$, 
is larger than $\lambda$, the collapsing gas will become self-gravitating before it 
can settle into a rotation-supported disk. In contrast, if the gas fraction 
is comparable to or smaller than the spin parameter, the collapse of the cooled gas 
will settle down to a rotation-supported disk before the gas can fragment and form stars.
Previous modeling of the observed size distribution of disk galaxies in the Universe  
indicates that the required $\lambda$-distribution is roughly log-normal,  
with a median $\sim 0.04$ and a dispersion $\sigma_{\ln\lambda}\sim 0.5$
\citep[e.g.][]{shenSizeDistributionGalaxies2003MassSizeRelation,
somervilleExplanationObservedWeak2008,
desmondTullyFisherMasssizeRelations2015,
burkertANGULARMOMENTUMDISTRIBUTION2016,
somervilleRelationshipGalaxyDark2018}. 
This distribution is similar to that of dark matter halos 
\citep[e.g.][]{bullockUniversalAngularMomentum2001,
bettSpinShapeDark2007,maccioConcentrationSpinShape2007}, 
but it does not necessarily mean that individual galaxies have 
similar spins to their host halos \citep[see, e.g.][]{danovichFourPhasesAngularmomentum2015,
jiangDarkmatterHaloSpin2019}.
Regardless of the origin of spins of galaxies, the typical value obtained from 
observations, $\lambda\sim 0.04$, is much smaller than the cosmic baryon fraction, $f_{\rm B}=0.15$.
Thus, as long as radiative cooling is effective and $f_{\rm gas}$ is not much smaller 
than the cosmic baryon fraction, the collapse of the halo gas will not be 
affected much by the angular momentum before it becomes self-gravitating. 
However, if $f_{\rm gas}$ is reduced to a value comparable or smaller than $\lambda$
due to, e.g. ineffective cooling and other (e.g. feedback) processes associated with galaxy 
formation,  then the gas will be able to form a rotation-supported disk before it 
collapses further to form stars in the gravitational potential of the dark matter halo.

\subsection{The four quadrants of gaseous structure formation} 
\label{ssec:four-quadrant-gas-evolution}

\begin{figure} \centering
    \includegraphics[width=0.99\columnwidth]{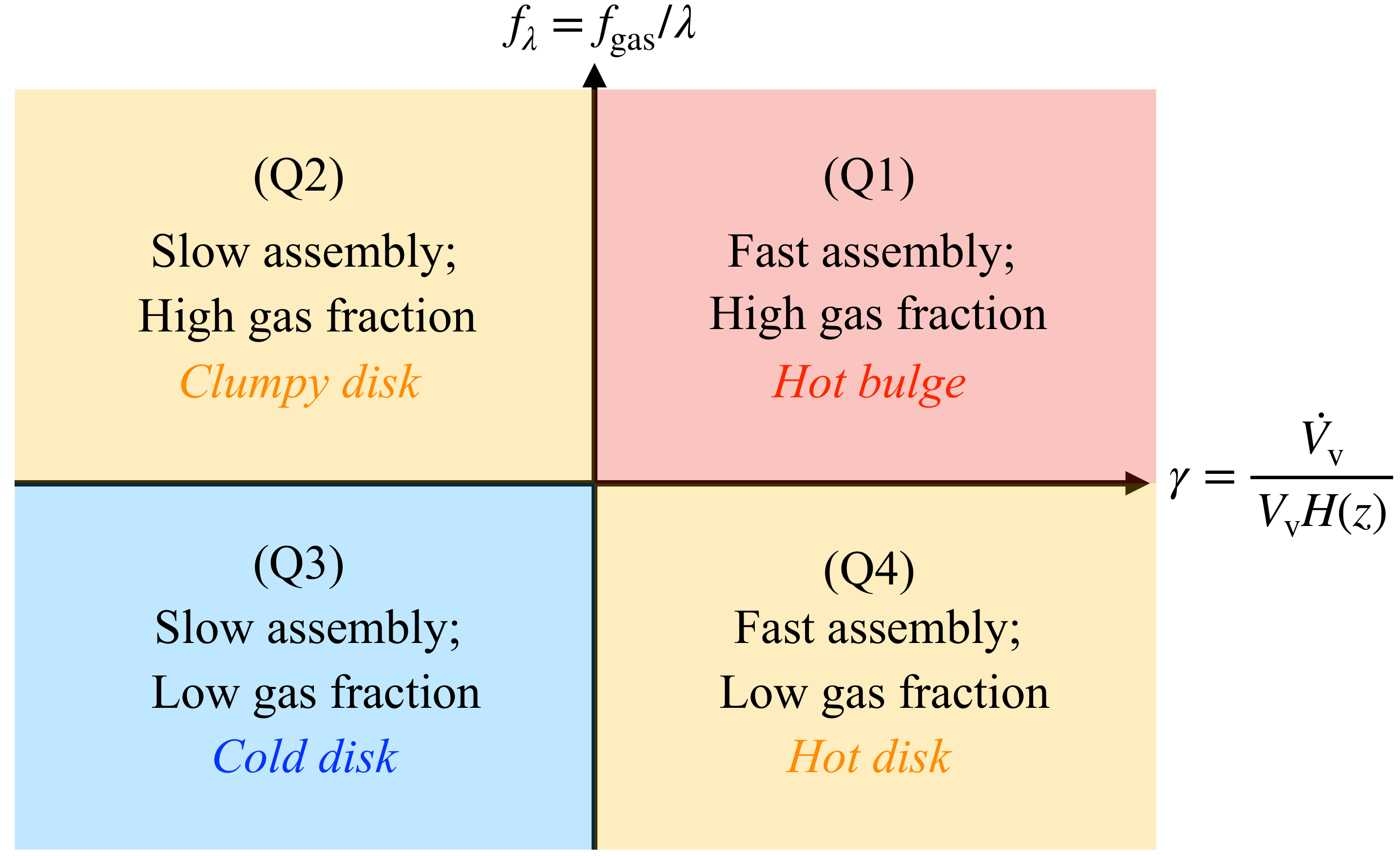}
    \caption{
    The quadrant diagram showing four combinations of 
    $\gamma \equiv \dot{V}_{\rm v}/\left[ V_{\rm v} H(z) \right] $, 
    which describes the rate of assembly, and 
    $f_{\lambda} \equiv f_{\rm gas}/\lambda$, which describes the importance of angular 
    momentum in supporting gas. The formation of a galaxy in 
    a halo can be represented by a trajectory in the
    $f_{\lambda}(t)$-$\gamma (t)$ space, and the formation of different 
    structural components depends on the location in the diagram. 
    See \S\ref{ssec:four-quadrant-gas-evolution} for a detailed discussion 
    for the quadrants.
    This figure motivates the use of $\gamma$ as a key parameter to separate
    the two phases of halo assembly and build the model for galaxy formation
    and SMBH growth based on the separation.
    }
    \label{fig:quadrant}
\end{figure}

The discussion above suggests that the collapse of cooled gas in dark matter halos 
may be characterized by the value of $\gamma$ that describes the
change of the gravitational potential and 
\begin{equation}
f_\lambda\equiv {f_{\rm gas}\over\lambda}\,,
\end{equation}
which describes the importance of the angular momentum. If we divide each of the two 
parameters into two portions, high and low, according to their critical values
at $(\gamma, f_\lambda) = (3/8, 1)$ based on the discussion above, we have four different 
combinations corresponding to the four quadrants in the $\gamma$-$f_\lambda$ plane 
shown in Fig.\,\ref{fig:quadrant}. In what follows, we discuss the structure
of the gas expected in each of the quadrants. The formation of a galaxy
can then be considered as a trajectory in this plane. The quadrant diagram may, 
therefore, be used to understand the evolution and structure of individual galaxies.    

\subsubsection*{Q1: Self-gravitating and turbulent gaseous spheroids}
\label{sssec:q1-phase}

Let us first look at the first quadrant where $f_\lambda>1$ and the collapse is in the 
fast regime, with $\gamma> \gamma_{\rm f}$. We note that the gravitational potential is 
dominated by dark matter as long as the gas and dark matter are well mixed. As the gas 
cools and collapses in its halo, the gravity of the gas component will become more important. 
The gas is expected to become self-gravitating when the circular velocity 
it generates is larger than that of the dark matter. Using the definition
of the circular velocity, we can write  
\begin{equation}
V_{\rm c}^2(r) = {GM(<r)\over r}
={GM_{\rm g} (<r) \over f_{\rm gas} r}   \,.
\end{equation}
Thus, for a roughly flat rotation curve, as is the case of CDM halos,   
the above relation suggests that the gas has to collapse by a factor of about 
$1/f_{\rm gas}$ relative to the dark matter to become self-gravitating. 
Since $f_\lambda>1$, rotation support is not important with such a collapse 
factor and the structure of the gas is roughly spheroidal. 
For convenience, we will refer to such a self-gravitating gas cloud as the  
SGC. The gas mass density will then be comparable to that of dark matter. 
For the gas assembled in the fast accretion phase, the radius at self-gravitating 
is thus 
\begin{equation} \label{eq:r_f_SGC}
r_{\rm f,SGC} = f_{\rm gas} r_{\rm f}\,,
\end{equation}
and the gas mass is
\begin{equation}
M_{\rm f,SGC}
= f_{\rm gas} M_{\rm f}\,.
\end{equation}
The mean gas density at this point is then 
\begin{equation}\label{eq_rhosgc}
{\overline\rho}_{\rm f,SGC} = {3 M_{\rm f,SGC}\over 4\pi r_{\rm f,SGC}^3}
={{\overline\rho}_{\rm f}\over f_{\rm gas}^2}.
\end{equation}
The free-fall time scale of the gas, which is determined by the gas density, 
is now about $f_{\rm gas}$ times that given by ${\overline\rho}_{\rm f}$.

We can also obtain the relation between the size and mass for 
the SGC. At a given redshift, the halo radius scales with the halo mass as 
$r_{\rm v}\propto M_{\rm v}^{1/3}$. Thus 
\begin{equation}
r_{\rm f} \propto \left({M_{\rm v} \over M_{\rm f}}\right)^{1/3} {r_{\rm f}\over r_{\rm v}} M_{\rm f}^{1/3}
\propto \left({{\overline\rho}_{\rm v}
\over{\overline\rho}_{\rm f}}\right)^{1/3} M_{\rm f}^{1/3}
\propto {c_{\rm f}\over c_{\rm v}} M_{\rm f}^{1/3}\,.
\end{equation}
We may write  
\begin{equation}
c_{\rm f}/c \propto M_{\rm v}^{\beta_c}~~~~ (\beta_c>0)\,,   
\end{equation}
so that 
\begin{equation}
r_{\rm f} \propto \left({M_{\rm v}\over M_{\rm f}}\right)^{\beta_c} M_{\rm f}^{\beta_c+{1/3}}  \,.
\end{equation}
Since $M_{\rm f} /M_{\rm v}$ depends on $M_{\rm v}$ only logarithmically, we have
\begin{equation}
r_{\rm f} \propto M_{\rm f} ^{\beta_c+1/3}\,.    
\label{eq:r_f_to_M_f_slope}
\end{equation}
This will lead to a relation with a logarithmic slope steeper than $1/3$. 
\cite{shenSizeDistributionGalaxies2003MassSizeRelation} found that
the size of an early-type galaxy scales with its stellar mass roughly as  
$R\propto M_*^{0.55}$. If these galaxies are formed from the SGC
with a constant star formation efficiency, we need $\beta_c=0.25$, 
which is consistent with the decrease of $c$ with the halo mass.  
In reality, the star formation efficiency is expected to depend on both 
redshift and halo mass, and thus different values of $\beta_{c}$ may be
possible to reproduce the observed size-mass relation. We will come 
back to this topic in the second paper of this series \citep{chenTwophaseModelGalaxy2023}.

The Jeans mass of a gas at temperature $T$ can be written as
\begin{equation}
M_{\rm J}
= 5\times 10^7 \left({c_s \over 10{\rm \,km\,s^{-1}}}\right)^3 
\left({n_{\rm gas}\over 1\,{\rm cm}^{-3}}\right)^{-1/2}
{\rm M}_\odot\,,
\end{equation}
where $c_s(T)$ and $n_{\rm gas}$ are the gas sound speed and density, respectively. 
As discussed above, effective atomic cooling can cool proto-galaxy gas to a 
temperature of about $10^4{\rm K}$. For gas assembled in the fast accretion region, 
the density is initially a factor of $(c/c_{\rm f})^3$ as large as the virial 
density. The gas density is further increased by a factor of $1/f_{\rm gas}^3$ 
as it becomes self-gravitation. Thus, 
\begin{equation}
n_{\rm gas} = {200 n_{\rm B,0} \over \Omega_{m,0}} {c^3\over c_{\rm f}^3} {1\over f_{\rm gas}^3}\,,        
\end{equation}
where $n_{\rm B,0}\approx 2.5\times 10^{-7}{\rm cm}^{-3}$
is the mean cosmic baryon number density of the present-day universe.
Assuming $c/c_{\rm f}= 4$ and $f_{\rm gas}=0.15$, we get 
$n_{\rm gas} \sim 1.0 {\rm cm}^{-3}$.
The corresponding Jeans mass is 
\begin{equation}
M_{\rm J}\sim 5\times 10^7 {\rm M}_\odot\,.
\end{equation}
This suggests that the original gas in an SGC can fragment to form gas 
sub-clouds with masses similar to $M_{\rm J}$. These clouds will collapse 
and cool further to form stars. Note that the mass is similar to that of 
a giant molecular cloud (GMC) complex observed in star-forming galaxies.
In what follows, we refer to such clouds as sub-clouds, or `sc' for short. 

When gas cooling is effective, the density of a sub-cloud generated by shock compression is typically 
\begin{equation}
\rho_{\rm sc} \sim {\cal M}^2 \rho_{\rm SGC} \sim \left({V_{\rm SGC}\over c_s}\right)^2
\rho_{\rm SGC}\,,
\end{equation}
where ${\cal M}$ is the Mach number of the shock, 
$c_s$ is the sound speed of the gas, and $V_{\rm SGC}$ is the virial velocity of 
the SGC. 
The number of collisions experienced by a sub-cloud as it moves over the size of an SGC
is expected to be
\begin{equation}
{\cal N}_c 
\sim {\rho_{\rm sc} r_{\rm sc}\over\rho_{\rm SGC} r_{\rm SGC}}
\sim \left(r_{\rm sc}\over r_{\rm SGC}\right)
\left({V_{\rm SGC}\over c_s}\right)^2
\sim  100 \left(r_{\rm sc}\over r_{\rm SGC}\right)\,.
\end{equation}
To avoid frequent collisions, sub-clouds must have sizes
\begin{equation} \label{eq:r_sc_to_r_SGC}
r_{\rm sc}< r_{\rm SGC}/100\,.
\end{equation}
The typical mass of such a sub-cloud is 
\begin{align}
M_{\rm sc}&\sim \rho_{\rm sc} r_{\rm sc}^3\\
&\sim 
\left({V_{\rm SGC}\over c_s}\right)^2 \left({r_{\rm sc}\over r_{\rm SGC}}\right)^3 M_{\rm SGC}\\
&\sim
\left({V_{\rm SGC}\over c_s}\right)^2 
\left({r_{\rm sc}\over r_{\rm SGC}}\right) 
M_{\rm SGC}
\left({r_{\rm sc}\over r_{\rm SGC}}\right)^2\\ 
&\sim 
M_{\rm SGC}
\left({r_{\rm sc}\over r_{\rm SGC}}\right)^2
\sim 10^{-4} M_{\rm SGC}\,.
\end{align}
This gives $M_{\rm sc}\sim 10^7{\rm M}_\odot$ for 
$M_{\rm SGC}\sim 10^{11}{\rm M}_\odot$, which is similar to the 
Jeans mass, $M_{\rm J}$, expected in an SGC. Thus, collisions of sub-clouds is 
expected to be infrequent. The same argument also suggests that the ram pressure on 
such sub-clouds is unimportant. These together indicate that the high-density sub-clouds, 
which inherit the large-scale turbulent motion generated by the gravitational collapse, 
can move roughly ballistically with velocities similar to that expected from the 
gravitational potential. In this case, the SGC as a whole is supported by 
random motion of sub-clouds, forming a dynamically hot system with a spheroidal 
morphology.

\subsubsection*{Q2: Self-gravitating unstable disks}
\label{sssec:q2-phase}

In the second quadrant where the collapse is in the slow accretion phase, the 
gravitational potential is roughly static, and the collapse is similar to that of secondary 
infall onto an existing static potential. Since all halos must start from the fast assembly 
phase, we expect that a galaxy in this quadrant has evolved from the first quadrant. 
The value of $f_\lambda$ remains high, indicating that $f_{\rm gas}$ has not been reduced 
significantly by feedback and ineffective cooling. The radiative cooling efficiency 
described in \S\ref{ssec:role-of-cooling} indicates that only low-mass halos with 
$M_{\rm v}<10^{13}{\rm M}_\odot$ can evolve to this quadrant. 

As the mass assembly is dominated by radial infall in this case, gas clouds 
can reach the central halo region to dissipate and become supported by angular momentum 
to form a disk-like structure. However, since $f_\lambda >1$, such a disk is dominated 
by self-gravity and is thus unstable both locally and globally. This is expected to lead to 
the formation of clumpy disks due to local instabilities of the self-gravitating disk, 
and/or bar-like structures due to the global instability of the disk. Both of these can cause 
migrations of cold gas towards the central region of a galaxy, a process referred to as 
the compactification \citep[e.g.][]{dekelFormationMassiveGalaxies2009,
dekelWetDiscContraction2014,
zolotovCompactionQuenchingHighz2015,
tacchellaConfinementStarformingGalaxies2016,
jiReconstructingAssemblyMassive2023}, leading to star formation and perhaps 
the growth of a bulge-like component near the center. The reduction of cold gas causes 
$f_\lambda$ to decrease. Since $\gamma$ changes only slowly in the slow assembly phase, 
a nearly vertical, downward evolution is expected in the quadrant diagram.  

\subsubsection*{Q3: Rotation-supported cold disks}

In the third quadrant, the gravitational potential is roughly static and $f_\lambda$ 
is about one or below. These are the conditions for the formation of a stable cold disk. 
In order for the disk to fragment and form stars, it needs to be locally unstable, 
i.e. $f_\lambda\sim 1$. If $\lambda$ is too large and/or $f_{\rm gas}$ too small
so that $f_\lambda\ll 1$, the gaseous disk will become too stable to form stars. 
This can lead to the formation of a low surface brightness disk if most of the disk 
material is assembled in such a way. 

A galaxy can evolve to this quadrant from Q2 as the amount of cold gas in the disk is 
reduced by the compactification. As to be described below, a galaxy may also 
evolve to this quadrant from Q4 when the amount of cold gas is reduced by processes 
associated with the fast assembly phase before the halo moves to the slow assembly 
phase. This evolutionary path is similar to the preventative scenario investigated in 
\citet{moGalaxyFormationPreprocessed2004} and \citet{luFormationDiscGalaxies2015}, where 
starburst and AGN feedback associated with the fast assembly phase can reduce the amount 
of gas to be accreted by the halo in the slow assembly phase. 

\subsubsection*{Q4: Disturbed and quenched galaxies}

In the fourth quadrant, the halo assembly is still in the fast phase so the 
gravitational potential can change significantly. However, since $f_\lambda < 1$, 
the collapse of the gas for it to become dense enough to form stars will 
be halted by angular momentum. A disk can then be produced in the interval where 
major mergers are not involved, but the disk cannot remain cold because of the change 
of the gravitational potential. This may produce a thick disk with significant 
random motion. In the extreme case where a significant merger is involved in the 
assembly, which causes a spike in $\gamma$, the disk that has formed may be disrupted, 
producing a merger remnant. The above is expected to be true in halos where 
the amount of cold gas is still significant. If, on the other hand,  radiative cooling 
is inefficient, such as in a massive halo, and/or feedback has eliminated the 
cold gas in the galaxy, the galaxy will be quenched. 
As mentioned above, if the halo makes a transition to the slow 
assembly regime, i.e. moves to Q3, a thin disk may grow around the 
structure that has formed in Q4.  

\bigskip

As one can see from Fig.\,\ref{fig:transition_points}, although the overall assembly history 
can be separated into a fast phase followed by a slow phase, individual histories in general contain 
fluctuations on small time scales. Thus, a halo in the fast assembly phase can   
make excursions to the slow assembly regime and vice versa. Since the typical time for a 
perturbed self-gravitating system to settle to a new quasi-static structure is its dynamical 
time scale, a bulge-like galaxy in Q1 may thus make an excursion to a disk-like morphology 
in Q2 if the time interval of the excursion in Q2 is longer than the dynamical time, 
and vise versa. 
In Appendix~\ref{app:fluctuation-mah}, we present various statistics for the 
fluctuations and the excursions based on halo assembly histories and the 
change of binding energy of dark matter particles.
Clearly, the binding energies of particles respond well to fluctuations in 
$\gamma$, even for particles in the inner part of the halo, indicating that 
the value of $\gamma$ indeed characterizes the dynamic state of the halo. 
The number of excursions to the slow regime ($\gamma < \gamma_{\rm f}$) 
during the fast phase ($z \geqslant z_{\rm f}$) is $\leq 2$ for about $80\%$
halos, and $\leq 3$ for all halos, regardless of the transition threshold 
$\gamma_{\rm f}$ adopted. The total time spent in the slow regime takes up to 
$30\%$ of the fast phase for about $80\%$ halos, 
and $40\%$ for all halos. This indicates that most halos have a chance 
of excursion to a disk-like morphology, but then return back to a bulge-like 
morphology after a short period.
In a forthcoming paper (Chen et al. in prep) we will 
examine in detail how galaxies evolve in the four-quadrant diagram using numerical simulations, 
and develop a detailed model for such evolution. 
In this paper, we only focus on some key points  
and demonstrate generic trends expected from the scenario.

\subsection{Further justifications of model assumptions}

One assumption in our modeling is that SGC is spherical.  This is clearly 
a simplification given that the assembly of a dark matter halo is typically 
non-spherical and clumpy. However, in the fast assembly phase, the gravitational 
potential well deepens rapidly with time. Since feedback from star formation and 
SMBH growth (see below) is expected to be more effective in heating and dispersing 
gas in lower-mass halos with shallower gravitational potential wells, most of the 
gas associated with the progenitors of a halo cannot form stars before the assembly 
of the final halo in the end of the fast assembly phase, where the gravitational 
potential well is able to retain more gas. It is thus likely that most of the gas 
in a halo is not bound to dark matter substructures, such as subhalos,  
as it cools and collapses to form the SGC \citep[see, e.g. figure 7 of][]{tillsonAngularMomentumTransfer2015}. The situation is similar to what 
\citet{latifTurbulentColdFlows2022} found in their simulations of the formation of gas 
collapse in galaxies, where ineffective cooling of metal-poor gas prevents star formation 
in substructures until the whole cloud becomes self-gravitating and collapses. 
Such a cloud is more or less spherical in the absence of angular momentum support.      
Consequently, the details of the halo assembly history may be irrelevant to the 
formation of the SGC. 

A similar argument may also be made for the enrichment 
history of an SGC. If most stars formed in the SGC, instead of in substructures 
before the assembly of the SGC, then the metallicity of the gas is expected 
to increase quickly to a roughly constant value as stars form and evolve in the 
SGC. This is actually consistent with the observed uniformity in the stellar population 
of early-type galaxies. However, if one is interested in the formation of different stellar 
populations in galaxies, e.g. globular clusters, the details of the chemical enrichment 
need to be followed, as described in \citet{chenTwophasePaper3-2024} .

Another simplification in \S\ref{ssec:four-quadrant-gas-evolution} for the formation
of SGC is the omission of the response of the dark matter to the collapse of gas. 
Following the model of \citet{moFormationGalacticDiscs1998}, 
we used adiabatic contraction to estimate this response, assuming  
spherical symmetry. We found that, for the typical concentration, 
$c_{\rm f} \approx 4$, the change in the radius of an SGC by adiabatic 
contraction is within 20 percent as $f_{\rm gas}$ changes from $0.025$
to $0.16$. However, during the fast assembly phase, the rapid change of the 
gravitational potential and the fast growth of the SGC may 
make the adiabatic assumption invalid. Another complication, found in 
numerical simulations \citep[e.g.][]{duttonNIHAOTooBig2016,el-badryBreathingFIREHow2016} 
arises from feedback-driven outflow, which may `heat' dark matter particles and 
reverse the contraction. Given the small effect of adiabatic contraction on the size 
of SGC and uncertainties in modeling it, we will ignore the effect in our model.

The detailed flow of the halo gas feeding the SGC is difficult to model 
analytically. However, high-resolution zoom-in simulations of individual galaxies 
published previously may provide some clues. 
\citet{ceverinoHighredshiftClumpyDiscs2010} performed simulations of three MW-size galaxies 
using a spatial resolution of $35$--$70\, {\rm pc}$. They found that, at $z > 2$ these 
galaxies are fed by steady cold streams from locations well outside the virial radius of the host 
halo to the inner core. The streams appear to converge at about $0.3 R_{\rm v}$, 
resulting in violent interactions and producing strong turbulence (see their figure~3). 
In all cases, the gas medium generated is clumpy and turbulent, consisting of 
dense clumps with active star formation. For the three cases simulated, 
the structure of the stellar component appears flattened (resembling a thick 
disk) and contains strong turbulent motion, with a radius about $0.05$--$0.1R_{\rm v}$.
This is consistent with the expectation for a system in Q2 or Q4 described above.
\citet{danovichFourPhasesAngularmomentum2015} expanded the investigation 
of \citet{ceverinoHighredshiftClumpyDiscs2010} by using 
29 galaxies covering a wider range of masses and simulated with comparable 
spatial resolutions. They found that these galaxies are roughly fed by three cold 
streams. The major stream usually has an impact parameter of $0.3 R_{\rm v}$, while 
the other two streams have smaller impact parameters. All the streams penetrate 
the inner core of the halo and are stopped at about $0.3 R_{\rm v}$ where 
they collide. They also found that about $30\%$ of the steam mass is 
counter-rotating to the net angular momentum of the gas
so that gas can be strongly compressed. This provides the condition for the 
formation of dense sub-clouds envisaged in \S\ref{ssec:four-quadrant-gas-evolution}. 
The gaseous structures in their simulated galaxies are in general very 
clumpy and turbulent, particularly at high redshift (see their figure~17), 
which reflects the fact that halo accretion is faster at higher $z$, 
as described in \S\ref{sec:halos}. 

As we assume in our model below, the star formation in the 
SGC is associated with dense sub-clouds. At high redshift, 
e.g. $z\sim 10$, the typical scale of an SGC is  $\lesssim 1\Kpc$
\citep[see also][]{dekelEfficientFormationMassive2023}. 
Using Eq.~\eqref{eq:r_sc_to_r_SGC} for $r_{\rm sc}/r_{\rm SGC}$, we expect that the size of a sub-cloud is typically 
$\lesssim 10\,{\rm pc}$, comparable to that of a young 
massive star cluster (YMSC) or a globular cluster (GC) observed in the local Universe \citep[see e.g.][for a review]{krumholzStarClustersCosmic2019}. 
The detailed star formation processes at this scale involve complex physics, 
such as the cooling by dust, metals and molecules, and the feedback from stars. 
These are difficult to resolve computationally, 
even in zoom-in simulations of individual galaxies discussed above. 
\citet{mandelkerColdFilamentaryAccretion2018} used an analytical 
approach to model the stream accretion and star formation in 
SGCs. Assuming cylindrical contraction and angular-momentum
conservation, they found that high-density sub-clouds
can form in a metal-poor ($Z\approx 0.02 Z_\odot$) SGC 
at $z \gtrsim 6$ in the core region ($R \lesssim 0.3 R_{\rm v}$) of a stream-fed halo, 
prior to the onset of gravitational instability. The criteria can be further relaxed in the 
presence of gravitational instability in sub-clouds, or if cold streams collide with each other 
to produce supersonic shocks that compress the gas.
\citet{maSimulatingGalaxiesReionization2018} used GIZMO to perform
a set of zoom-in simulations in volumes around 15 halos at $z \geqslant 5$ 
selected from the FIRE-2 project, achieving a sub-pc (in physical units) 
spatial resolution. All the central galaxies in their sample appear clumpy 
and irregular in the redshift range covered by their simulations 
(see the mock images in their figures~2 and 3 for examples at $z=5$ and $10$, 
respectively), and none of them resembles a disk or elliptical galaxy in the local 
Universe. The star formation histories of these galaxies all contain 
frequent bursts that typically last a period of $50$--$100\,{\rm Myr}$.
These results are consistent with the formation of dynamically hot systems in the Q1 
phase described in \S\ref{sssec:q1-phase}.
\citet{maSelfconsistentProtoglobularCluster2020} re-ran simulations for 
a number of the 15 halos using a mass resolution eight times better than that 
in the original simulations, focusing on narrow time windows of $100\,{\rm Myr}$ 
containing starbursts. They found that the gas-rich, turbulent environment
originated from fast halo accretion and/or feedback-driven winds naturally
give birth to high-density sub-clouds from which bound star clusters form.
This supports the scenario of the formation of dense sub-clouds  
described in \S\ref{ssec:four-quadrant-gas-evolution}. They also confirmed
that bound cluster formation ceases at lower redshift, due to the less 
dense and less turbulent galactic environment (see their figure~17, at $z=1$).
However, the sizes of their simulated bound star clusters (see their figure~13) are much 
larger than the observed size of YMSCs and GCs, and the number of bound clusters 
does not appear to reach convergence even at the highest resolution they used
(see their figure~16), indicating that the conditions for the formation of sub-clouds 
and star clusters are not reproduced in the simulation. Instead of simulating the 
entire galaxies, \citet{grudicModelFormationStellar2021} focused on individual 
GMCs with turbulence field injected artificially. The sizes of simulated bound 
clusters now appear to converge to the observed range (see their figure~7).
A related investigation was carried out by \citet{latifTurbulentColdFlows2022},
who focused on extremely metal-poor gas (with pristine composition) 
at $z \gtrsim 20$ using the Enzo code with a spatial resolution that reaches 
$\sim 20\,{\rm AU}$. They found that turbulence compression due to global collapse 
and cloud mergers, and the delayed star formation due to inefficient cooling, 
give birth to massive and dense gas clumps that can host massive stars
of $\gtrsim 10^4 M_\odot$ and probably provide a channel to seed SMBHs. 


The above presentation shows that the scenario of SGC formation and fragmentation 
to produce sub-clouds envisaged in our model has support from high-resolution 
hydro simulations. It is also clear that there are still substantial amounts of uncertainties   
in current simulations, related particularly to resolving sub-clouds 
and sub-grid physics in modeling star formation. These not only affect the 
predicted properties of sub-clouds and star clusters, but also have impacts on 
the structure and dynamics of galaxies on large scales. For instance, if   
the fragmentation of an SGC to form dense sub-clouds is not properly resolved, 
the dissipation of turbulent motion by cloud-cloud collision may be overestimated.   
This, in turn, may lead to an overestimate of dynamically colder galaxies.
Similarly, uncertainties in modeling the feedback from star formation in a sub-cloud 
can lead to incorrect predictions for the dissipation of the orbital energy of 
sub-clouds, again affecting the hotness of the predicted galaxies.   

Some justifications may also be obtained from observations. The newly-launched JWST has already 
been able to resolve, albeit only for a few cases in strongly lensed fields, the formation of 
dense star clusters at high redshift. Together with multi-wavelength observations from, e.g., 
HST, ALMA and VLT, the JWST data can be used to measure  a rich set of properties of high-$z$ 
galaxies. Current samples are still small. The limited analyses indicate that high-$z$
galaxies are in general very clumpy, containing clumps with sizes similar to those of YMSCs 
and GCs observed in the local universe 
\citep[e.g.][]{vanzellaEarlyResultsGLASSJWST2022,
linMetalenrichedNeutralGas2023,vanzellaJWSTNIRCamProbes2023,
fujimotoPrimordialRotatingDisk2024,
mowlaFireflySparkleEarliest2024,adamoDiscoveryBoundStar2024}. 
The global morphology of these galaxies appears to be diverse. In particular, 
a significant fraction of them show flattened shapes and nearly exponential 
light profiles \citep[e.g.][]{ferreiraJWSTHubbleSequence2023,
leeMorphologyGalaxiesJWST2023,kuhnJWSTRevealsSurprisingly2023,
kartaltepeCEERSKeyPaper2023,ormerodEPOCHSVISize2024,
sunStructureMorphologyGalaxies2024,tohillRobustStudyHighredshift2024}.
However, it is unclear if these galaxies are rotation-supported, dynamically 
cold disks. Indeed, some recent analyses demonstrated that many of these 
flattened galaxies are consistent with being dynamically hot systems 
flattened by velocity anisotropy velocity
\citep[e.g.][]{pandyaGalaxiesGoingBananas2024,vega-ferreroNatureDisksHigh2024}. 
This issue is expected to be resolved in the near future as samples  
and analyses of observational data accumulate.

\section{Star formation and the growth of supermassive black holes}
\label{sec:bh_growth}

\subsection{Accretion of gas onto a supermassive black hole}
\label{ssec:bh_capture}

\begin{figure} \centering
    \includegraphics[width=\columnwidth]{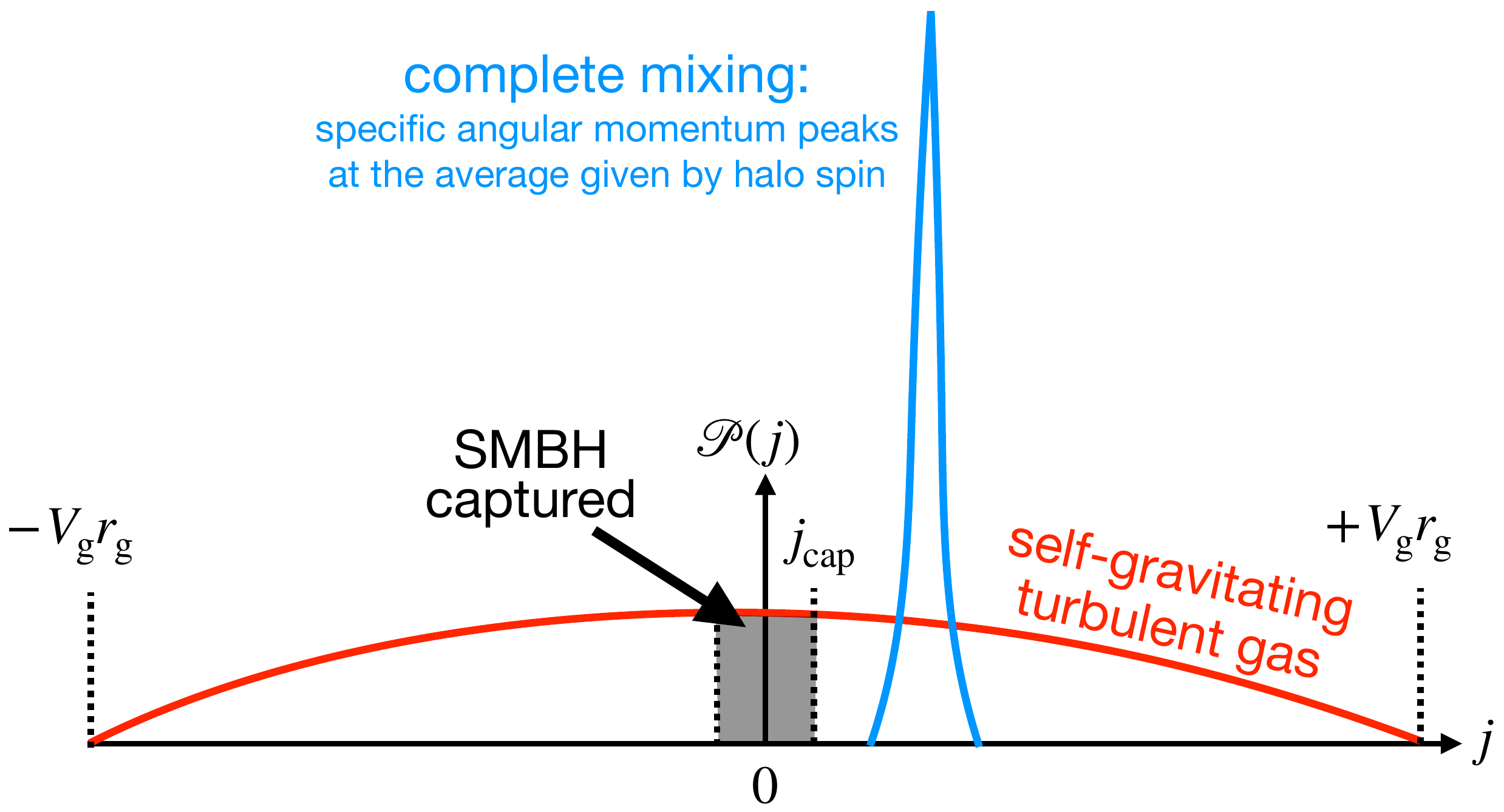}
    \caption{
        A schematic diagram showing the distribution of specific angular momentum 
        ($j$) for gas clouds within a halo. The turbulent motion of gas clouds, 
        driven by fast accretion, yields a broad and uniform distribution 
        ({\bf red curve}) of $j$. The fraction of gas accreted by the 
        SMBH ({\bf gray shaded area}) is determined by the maximum capturing 
        angular momentum, $j_{\rm cap}$. Subsequently, as the driving force of
        turbulent motion diminishes, gas mixing becomes significant, 
        leading to the emergence of an angular momentum barrier and 
        preventing gas accretion ({\bf blue curve}). See \S\ref{ssec:bh_capture} 
        for a detailed discussion on SMBH accretion. This diagram shows the 
        mechanism underlying the formation of dynamically hot systems, 
        the accretion scenario of SMBH within turbulent gas clouds, 
        and the transition to dynamically cold systems.
    }
    \label{fig:angular_momentum}
\end{figure}

When the turbulent medium is fully developed, high-density clouds 
can move almost ballistically, as discussed above. We can then use the loss-cone 
argument \citep{shapiroBlackHolesWhite1986} to estimate the mass that can be captured by a supermassive 
black hole (SMBH) \citep[e.g.][]{hobbsFeedingSupermassiveBlack2011}. The fraction of 
the gas mass in a shell of mass $\Delta M_{\rm g}$ and radius $r_{\rm g}$ that can reach the central 
SMBH of mass $M_{\rm bh}$ at an impact parameter smaller than $r_{\rm acc}$ is 
\begin{equation}
{\Delta M_{\rm acc}\over \Delta M_{\rm g}}
\approx {j_{\rm cap} \over V_{\rm g} r_{\rm g}}
\end{equation}
where $V_{\rm tb}$ is the typical large-scale turbulent velocity, and 
\begin{equation}
j_{\rm cap} = 
\sqrt{2G M_{\rm bh} r_{\rm acc}}\,.
\end{equation}
For turbulence driven by gravitational collapse,  
we expect that $V_{\rm tb}\sim V_{\rm g}$, where 
$V_{\rm g}$ is the typical velocity in the gravitational potential of the proto-galaxy in question. 
The above relation can be understood as follows. Setting $V_{\rm tb} = V_{\rm g}$ in the above relation, 
we have
\begin{equation}
{\Delta M_{\rm acc}\over \Delta M_{\rm tb}}
={j_{\rm cap}\over V_{\rm g} r_{\rm g}}\,.
\end{equation}
 This implies that the distribution of the specific angular momentum 
is uniform in mass between $j\sim 0$ to 
$j=j_{\rm max} \sim V_{\rm g} r_{\rm g}$.
A turbulent medium can produce such a distribution by randomizing the velocity. 
Consider a gas cloud of a given speed $V$ in a spherical shell of a given radius $r$. 
The specific angular momentum of the cloud is 
$j =V r  \sin\theta$, where $\theta$ is the angle between the velocity vector 
and the radial vector at the location of the cloud. If the velocity 
has a random distribution in direction, then the solid angle element 
$2\pi \sin\theta d\theta$ has a uniform distribution.  

We can estimate $r_{\rm acc}$ by requiring that the gravitational 
potential of the SMBH must be able to capture clouds 
moving at a speed $\sim V_{\rm tb}\sim V_{\rm g}$. This gives  
\begin{equation}
r_{\rm acc}\sim {GM_{\rm bh} \over V_{\rm g}^2}
\sim {M_{\rm bh}\over M_{\rm g}} r_{\rm g}\,.
\end{equation}
We thus have 
\begin{equation}
{\Delta M_{\rm acc}\over \Delta M_{\rm g}}
={j_{\rm cap}\over V_{\rm g} r_{\rm g}}
= \alpha_{\rm cap} {M_{\rm bh}\over M_{\rm g}}\,,
\label{eq:def_M_acc}
\end{equation}
where $\alpha_{\rm cap}$ is a constant of order unity.
Thus, if $\Delta M_{\rm acc} = \Delta M_{\rm bh}$, then 
we would have $M_{\rm bh}\propto M_{\rm g}$. 
As we will see later, star formation feedback may change this relationship, 
increasing the value of $\alpha_{\rm cap}$.

Since we are concerned with the feeding of SMBH, it is interesting to 
cast the above relation into a rate relation and write it in terms of 
the Eddington rate of $M_{\rm bh}$,
\begin{equation}
{\dot M}_{\rm Edd} = {M_{\rm bh}\over \tau_{\rm Edd}}\,,
~~~~
\tau_{\rm Edd} \approx 4.4 \times 10^8 \epsilon_{\rm r}\,{\rm yrs}\,,
\end{equation}
where $\epsilon_{\rm r}\sim 0.1$ is the radiation efficiency of a black hole. 
Equation\,(\ref{eq:def_M_acc}) can then be written in the following form
\begin{equation}
{\dot M}_{\rm acc}
= \alpha_{\rm cap} {\tau_{\rm Edd} \over \tau_{\rm g}} 
{\dot M}_{\rm Edd}\,,
\end{equation}
where $\tau_{\rm g}\sim t_{\rm ff}$ is the typical time scale for a gas sub-cloud to move 
across the galaxy once. For an SGC, we can use equations (\ref{eq_rhof}) and (\ref{eq_rhosgc}) 
to obtain
\begin{equation}
t_{\rm ff,SGC} \approx 
\left[{f_{\rm gas}^2 \over 200}\left({c_{\rm f}\over c}\right)^3 \right]^{1/2}
t_{\rm H}\,,
\end{equation}
where $t_{\rm H}$ is the Hubble time. Assuming $f_{\rm gas}=0.15$ and 
$c/c_{\rm f}=2$, we get $t_{\rm ff, SGC}\sim t_{\rm H}/300
\sim 5\times 10^7 {\rm yrs}$. This is comparable to $\tau_{\rm Edd}$ for
$\alpha_{\rm cap}\sim 1$, suggesting that the accretion rate in an SGC can 
reach the Eddington limit. As we will describe below, the value of 
$\alpha_{\rm cap}$ may be increased if supernova feedback 
can replenish the SGC with SMBH-feeding gas, making super-Eddington 
accretion possible.    

Note that the above results are only valid for a fully turbulent 
gas. As the driving force decreases and the mixing of gas becomes important, 
the angular momentum barrier will develop and  
prevent gas accretion. This may occur when feedback can heat most of 
the gas, as well as in late times when the accretion is slow. 
For convenience, Fig.~\ref{fig:angular_momentum} gives a schematic illustration of 
the angular momentum distribution and the SMBH accretion scenario 
described here. 

\subsection{Star formation in self-gravitating clouds (SGC)}     
\label{ssec:star-formation-in-sgc}

The star formation efficiency in individual gas sub-clouds can be defined as 
\begin{equation}
\varepsilon_{\rm sc}
= {{\dot M}_{\rm *,sc} t_{\rm ff, sc}\over M_{\rm sc}}\,,
\end{equation}
where ${\dot M}_{\rm *,sc}$ is the star formation rate, and $t_{\rm ff, sc}$ the 
free-fall time of the sc. Based on observations of star formation in giant 
molecular clouds (GMC), the star formation efficiency in 
general is very low, typically $0.01$
\citep[e.g.][]{krumholzSlowStarFormation2007,
feldmannTimeVariabilityStar2011,
chevanceLifecycleMolecularClouds2020}. 
Suppose that the time scale for the formation of 
a sc is $\tau_{\rm sc}$. 
The efficiency at which gas is converted into stars in an SGC is then  
\begin{equation}
\varepsilon_{\rm SGC}
=\varepsilon_{\rm sc} {t_{\rm ff, SGC}\over \tau_{\rm sc}}\,.
\end{equation}
In a self-gravitating medium, the formation time of a sub-cloud is typically 
its free-fall time, so that $\tau_{\rm sc}\sim t_{\rm ff, sc}$. In 
a turbulent SGC, $t_{\rm ff, SGC}/t_{\rm ff, sc}\sim V_{\rm SGC}/c_s \sim 10$.  
So the typical value of $\varepsilon_{\rm SGC}$ is about $0.1$, comparable to the 
star formation efficiency required by the stellar mass-halo mass relation.   
This suggests that star-forming sub-clouds have to be dispersed and recreated 
about 10 times or more in an SGC.  

A constraint on the star formation scale can also be obtained by using the fact 
that stars in elliptical galaxies are in general enhanced in $\alpha$ elements 
produced mainly by SNa II. The progenitors of these supernovae have masses
$>8 M_\odot$, which have main-sequence lifetime about $10^7{\rm yrs}$. 
Type Ia supernovae, which are the main sources of Fe, have progenitor masses 
about $3 M_\odot$, with lifetime $>10^8{\rm yrs}$.
Thus the time scale of star formation must be about $10^8{\rm yrs}$. This is comparable 
to the free-fall time scale of an SGC. 

For the high-density tail of the gas distribution in an SGC, analytical 
models \citep[e.g.][]{dekelEfficientFormationMassive2023} 
and numerical simulations \citep[e.g.][]{grudicModelFormationStellar2021} 
suggest that the local free-fall time scale can be as short as $1\Myr$ 
and the local star formation efficiency can reach the order of unity.
Such high-density sub-clouds naturally emerge in Q1 as a consequence of the high 
SGC density, strong compression associated with supersonic turbulence, 
and low gas metallicity expected at high $z$ (see \S\ref{sssec:q1-phase}). 
This can lead to the formation of compact stellar systems such as 
young massive star clusters and globular clusters, 
and we will investigate them in 
the third paper of this series \citep{chenTwophasePaper3-2024}.

\subsection{Feedback and quenching}

Both supernovae and accreting SMBHs can deposit energy in the SGC. 
Using a rate of one supernova per $125 M_\odot$ of stars, and a kinetic energy of 
$10^{51}\,{\rm ergs}$ per explosion, we can write the cumulative feedback 
energy gain per gas mass as 
\begin{equation}
{\cal E}_{\rm sn} \approx  (630\,{\rm km\,s^{-1}})^2 
\epsilon_{\rm sn} f_*\,,
\end{equation}
where $f_*=M_*/M_{\rm g}$ and $\epsilon_{\rm sn}$ is the efficiency at 
which the supernova feedback energy can affect the gas component.    
For the feedback by an accreting SMBH, the corresponding expression can be written as  
\begin{equation}
{\cal E}_{\rm bh} \approx  (3\times 10^3\,{\rm km\,s^{-1}})^2 
\epsilon_{\rm bh} f_* \left({f_{\rm bh}\over 10^{-4}}\right)\,, 
\end{equation}
where $f_{\rm bh}= M_{\rm bh}/M_*$, and $\epsilon_{\rm bh}$ is the 
efficiency at which the feedback energy of the SMBH can affect the gas 
component. The main uncertainty is in the values of 
$\epsilon_{\rm sn}$ and $\epsilon_{\rm bh}$, as the feedback energy can 
be lost through radiation and/or leaked without affecting the gas component.
\citet{dekelOriginDwarfGalaxies1986} modeled the evolution of supernova remnants 
in a uniform medium taking into account radiative cooling, and found 
that $\epsilon_{\rm sn}$ depends on the number density of supernova 
explosions. In a high-density, star-forming gas where supernova 
remnants can overlap before radiative cooling becomes important, 
$\epsilon_{\rm sn}\sim 1$. Assuming $\epsilon_{\rm sn} f_*\sim 0.1$, as may 
be appropriate in an SGC, supernova feedback is expected to be effective  
in halos with virial velocity $V_{\rm v} < 200 {\rm km\,s^{-1}}$. 
For $\epsilon_{\rm bh}\sim 0.1$ and $f_* \sim 0.2$, and for 
$f_{\rm bh}$ as low as $10^{-4}$, the corresponding limit is 
$V_{\rm v} \sim 400 {\rm km\,s^{-1}}$, indicating that 
SMBH might be able to drive outflows from relatively massive halos.  

\subsubsection{Feedback from accreting supermassive black holes}

The total energy produced by an accreting SMBH (AGN) is:
\begin{equation}
E_{\rm bh} =\epsilon_{\rm bh} M_{\rm bh} c^2\,,
\end{equation}
where $\epsilon_{\rm bh}\sim 0.1$ is an efficiency factor. We assume that 
a fraction of $f_{\rm E}$ of the feedback energy is coupled to the gas:
\begin{equation}
E_{\rm cp} = f_{\rm E} E_{\rm bh} \,,
\end{equation}
and that the gas coupled to (affected by) the feedback is $M_{\rm cp}$. 
If the feedback energy of the AGN is stored in the gas as thermal energy, we can write 
\begin{equation}
M_{\rm cp}= {f_{\rm E} \epsilon_{\rm bh} M_{\rm bh} c^2 \over w^2}\,,
\end{equation}
where $w$ is given by the gas temperature:
\begin{equation}
w^2 = { kT \over\mu m_p}\,.
\end{equation}
The fraction of gas that is affected by the feedback is then
\begin{equation}
f_{\rm cp}
={M_{\rm cp } \over M_{\rm g}}
={f_{\rm E} \epsilon_{\rm bh} M_{\rm bh} c^2 \over M_{\rm g} w^2}\,.
\end{equation}
Assuming $w^2 \sim V_{\rm g}^2$, with $V_{\rm g}$ being the velocity expected in the 
gravitational potential of the galaxy, the fraction that is not affected by the 
AGN feedback can be written as 
\begin{equation}
F_{\rm agn} (M_{\rm bh}, M_{\rm g}, V_{\rm g}\vert \alpha_{\rm agn}) 
=1- {\alpha_{\rm agn} M_{\rm bh} c^2 \over M_{\rm g} V_{\rm g}^2}\,,
\label{eq:def-f-agn}
\end{equation}
where $\alpha_{\rm agn}$ is a constant parameter. 
To see the consequence of the feedback, let us assume an 
isothermal gas and that all gas not locked 
in stars are affected by the feedback. In this case, we have 
\begin{equation}
M_{\rm cp} = M_{\rm g} - M_*\,,
\end{equation}
and 
\begin{equation}
M_{\rm bh} ={w^2 (M_{\rm g} -M_*) \over f_{\rm cp} \epsilon_{\rm bh} c^2}
=
{w^2 M_{\rm g} \over f_{\rm cp} \epsilon_{\rm bh} c^2}
\left(1- {M_*\over M_{\rm g}}\right)\,,
\end{equation}
where $M_*/M_{\rm g}$ is an overall star formation efficiency. 
If we assume that the AGN feedback quenches the SMBH accretion 
when the gas is heated up to a temperature to escape the 
gravitational potential of the galaxy, we can set 
$w$ to be the velocity dispersion of the gravitational 
potential: $w^2\sim \sigma_{\rm g}^2$. For a self-gravitating gas 
formed in a dark matter halo, 
$M_{\rm g}\sim \sigma_{\rm g}^2 r_{\rm g} \sim \sigma_{\rm g}^3$. We then have 
\begin{equation} \label{eq:bh-sigma-gas-scaling}
M_{\rm bh} \propto \sigma_{\rm g}^b\,,
~~~~(b\sim 5)\,.
\end{equation}
This is actually similar to the observed value, $b \approx 5.3$
\citep[e.g.][]{huBlackHoleMassstellar2008,
mcconnellRevisitingScalingRelations2013,
wooQuiescentActiveGalaxies2013,
sagliaSINFONIBlackHole2016,greeneIntermediateMassBlackHoles2020}.  
If the stellar mass scales with $\sigma_{\rm g}$ as
\begin{equation} \label{eq:faber-jackson-relation}
M_* \propto \sigma_{\rm g}^s\,,
\end{equation}
then
\begin{equation} \label{eq:bh-m-star-scaling}
M_{\rm bh} \propto M_*^{b/s}\,.
\end{equation}
The observed value, $s\sim 3$, then 
implies that $b/s\sim 5/3$, which again is consistent with observational 
results \citep[e.g.][]{greeneIntermediateMassBlackHoles2020,
grahamAppreciatingMergersUnderstanding2023,
zhuangEvolutionaryPathsActive2023}.
More recently, \citet{hongDynamicalHotnessStar2023} adopted a similar approach 
and applied it to MaNGA galaxies. They successfully connected the observed 
Faber-Jackson relation (eq.~\ref{eq:faber-jackson-relation}) and black hole 
scaling relations (eqs.~\ref{eq:bh-sigma-gas-scaling} and 
\ref{eq:bh-m-star-scaling}).

\subsubsection{Ineffective cooling at the high-mass end}

At the massive end, radiative cooling becomes inefficient. 
Feedback may then quench the star formation and SMBH growth in a galaxy,  
even if its halo is still in the fast-assembly phase. The cooling curve suggests that, 
for halos above $M_{\rm cool}\sim  10^{12}M_\odot$ and for the 
density expected for halo gas at the virial temperature, the ratio between 
the cooling time and free-fall time is roughly proportional to halo mass, 
$M_{\rm v}$, quite independent of redshift (see Fig.\,\ref{fig:cooling_diagram}). 
Thus, we need to include an additional halo mass-dependent factor, 
$F_{\rm cool}$, to describe the fraction of cooled gas, as is given by 
Eq.~\eqref{eq:def-f-cool}.


\subsubsection{Modifications at the low-mass end, and the enhancement 
of supermassive black hole accretion}

According to the supernova energy output and rate, the specific energy gained by a 
sub-cloud due to supernovae associated with star formation can be written as 
\begin{equation}
{\cal E}_{\rm sc}
= (40 {\rm km\,s^{-1}})^2 \left({\epsilon_{\rm sc}\over 0.01}\right)\left({\epsilon_{\rm sn}\over 0.5}\right)\,,
\end{equation}
where $\epsilon_{\rm sc}\sim 0.01$ is the star formation 
efficiency in a sub-cloud (sc) and the value is based on that in 
a GMC; $\epsilon_{\rm sn}$ is the fraction of supernova 
kinetic energy that is injected into the gas.   
Thus, supernovae in a sub-cloud are expected to disperse the gas 
from a sub-cloud, rather than to directly drive gas out of the galaxy potential. 
However, such dispersed gas can make the feedback from the central starburst, 
where the effective $\epsilon_*$ is much higher than $\epsilon_{\rm sc}$,  
more effective in loading the feedback energy onto the gas. 
In low-mass halos where the gravitational potential wells are shallow, the 
feedback may drive gas out of the central galaxy, reducing the amount of gas. 
One possibility to model the combined effect of supernova and AGN feedback  
is to replace $F_{\rm agn}$ defined above by  
\begin{equation}
F'_{\rm agn} = F_{\rm agn} F_{\rm sn} \,,
\label{eq:def_feedbacks}
\end{equation}
where 
\begin{equation}
F_{\rm sn}(V_{\rm g}\vert \alpha_{\rm sn}, \beta_{\rm sn}, V_{\rm w}) = 
{\alpha_{\rm sn}  +(V_{\rm g}/V_{\rm w})^{\beta_{\rm sn}}
\over 
1 +(V_{\rm g}/V_{\rm w})^{\beta_{\rm sn}}}\,,
\label{eq:def-f-sn}
\end{equation}
with $\alpha_{\rm sn}\le 1$, $\beta_{\rm sn}>0$, and $V_{\rm w}$ a characteristic 
velocity.    
The assumption made in Eq.~\ref{eq:def_feedbacks} is that supernova feedback and AGN feedback 
operate conditionally, so that the former only operates on the remaining gas 
fraction allowed by the latter, and vice versa. This is similar to the preferred 
scheme proposed by \citet{boothInteractionFeedbackActive2013}, in which the effect of the weaker 
feedback is conditioned on the result of the more dominating one.
As one can see, the combined effect of supernova and AGN feedback is to eject part 
of the gas from (low-mass) halos, while heating part or all of the remaining gas to the 
virial temperature.  

The feedback process can also generate turbulence in the gas. In central 
regions of low-mass halos where radiative cooling is very effective, some of 
the gas affected by the feedback may cool, re-producing clouds with low 
specific orbital angular momentum and enhancing the mass accretion rate of the central SMBH. 
Unfortunately, the details of this `positive' feedback  
are yet to be understood. To model the effect, we include the following factor in the 
accretion rate of SMBH:
\begin{equation}
F_{\rm en} (M_{\rm v}\vert \alpha_{\rm en}, \beta_{\rm en}, M_{\rm en})
= {\alpha_{\rm en} + (M_{\rm v}/M_{\rm en})^{\beta_{\rm en}}
\over 1+ (M_{\rm v}/M_{\rm en})^{\beta_{\rm en}}}\,,
\label{eq:def-f-en}
\end{equation}
where $\beta_{\rm en}$ describes the transition from 
$\alpha_{\rm en}$ at the low-mass end to one at the high-mass 
end, with $\alpha_{\rm en}$, $\beta_{\rm en}$ and $M_{\rm en}$
being model parameters. We assume that this positive feedback is important 
only in the central dense part of an SGC, and will not affect much the total star 
formation rate in the entire SGC.    

\subsection{Co-evolution of different mass components during the fast assembly}
\label{ssec:coevolution-equations}

\begin{figure*} \centering
    \includegraphics[width=0.9\textwidth]{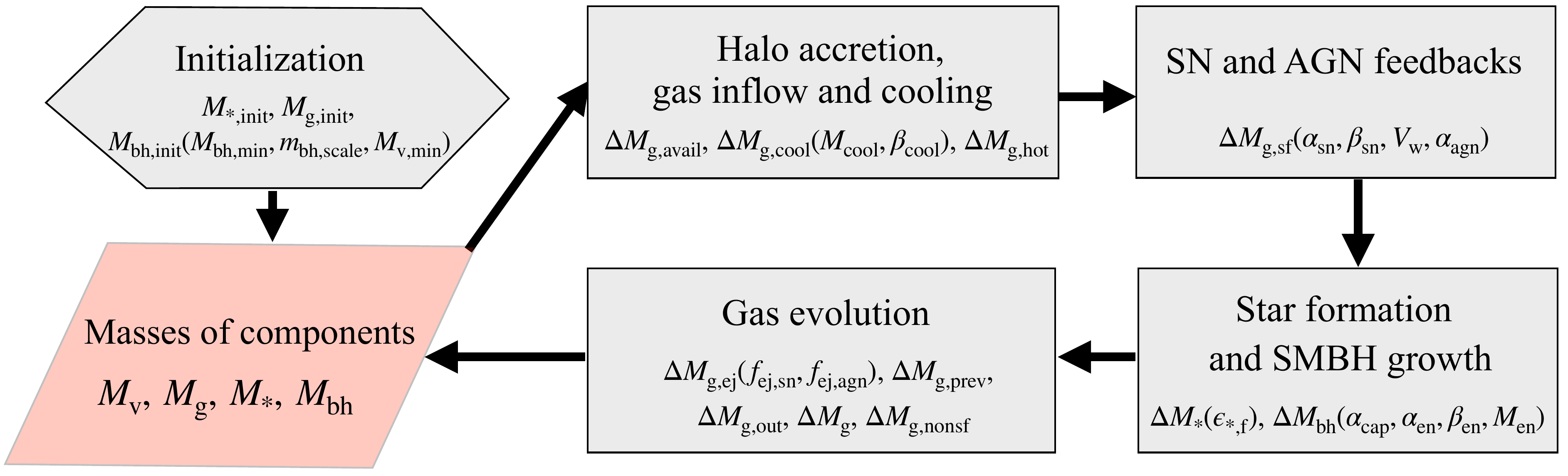}
    \caption{
        A flow diagram showing the model for the evolution of different mass 
        components (shown in {\bf red block}, 
        including halo mass $M_{\rm v}$, gas mass $M_{\rm g}$ remaining in SGC,
        stellar mass $M_*$ and SMBH mass $M_{\rm bh}$) 
        during the fast assembly phase. The algorithm initializes the mass 
        components at the initial snapshot and iteratively updates them 
        until the ending point of the fast assembly. Each {\bf gray block} 
        shows a process, including the relevant physical quantities and 
        model parameters. See \S\ref{ssec:coevolution-equations} for a 
        complete description of the evolutionary equations and 
        Table~\ref{tab:parameters} for a list of the model parameters 
        adopted in this paper.
    }
    \label{fig:flow-chart}
\end{figure*}

In practical applications, the evolution of halos can be modeled reliably using 
merging trees obtained from numerical simulations. 
To model the evolution of baryonic components in a halo, we start by introducing a small 
baryonic seed in the earliest progenitor of the halo and follow the evolution
step by step using a set of equations relating different mass components. 
Based on the discussion presented in the preceding sections, we outline the evolutionary equations 
as follows. Note that the prescriptions are for halos in the fast assembly phase 
(i.e. $z \geqslant z_{\rm f}$). The evolution in the slow assembly phase will be presented later.

\begin{enumerate}[parsep=8pt]
\item {\bf Initialization:} 
In the first snapshot of the main branch of a dark matter halo, we set the initial
stellar mass $M_{\rm *,init}$ and gas mass $M_{\rm g,init}$ in the halo as follows:
\begin{align}\label{eq:init-star-gas}
    M_{\rm *,init} &= 0                            \,;\\
    M_{\rm g, init}(M_{\rm v}) &= f_{\rm B} M_{\rm v}      \,.
\end{align}
When the halo mass $M_{\rm v}$ exceeds a threshold, $M_{\rm v,\,min}$, we 
seed a black hole with a mass $M_{\rm bh, init}$ proportional to $M_{\rm v}$:
\begin{align}\label{eq:init-bh}
    M_{\rm bh, init} & 
        ( M_{\rm v} \vert M_{\rm bh,\,min}, m_{\rm bh,\,scale} ) \nonumber \\
        & = \max(M_{\rm bh,\,min},\, m_{\rm bh,\,scale} M_{\rm v})   \,,
\end{align}
where $m_{\rm bh,\,scale}$ determines the initial black hole mass, 
and $M_{\rm bh, min}$ is a small value used to prevent the black hole mass 
from being too small.

\item {\bf Total amount of available gas:} 
The total amount of gas available 
in a halo is determined by the newly-accreted halo mass, $\Delta M_{\rm v}$, 
in each snapshot, through
\begin{equation}
    \Delta M_{\rm g, avail} = f_{\rm B} \Delta M_{\rm v}\,.
\end{equation}

\item {\bf Cooling of halo gas:} A fraction of the available halo gas cools 
down and contributes to the SGC. We define the amount of `cooled gas' as
\begin{equation}\label{eq:delta-m-gas-cool}
    \Delta M_{\rm g,cool} = F_{\rm cool} \Delta M_{\rm g, avail}       \,,
\end{equation}
where $F_{\rm cool} (M_{\rm v}\vert M_{\rm cool},\beta_{\rm cool})$ is defined 
by equation~(\ref{eq:def-f-cool}). The remaining gas is referred to as the `hot gas':
\begin{equation}\label{eq:delta-m-gas-hot}
    \Delta M_{\rm g,hot} = (1-F_{\rm cool}) \Delta M_{\rm g, avail}    \,.
\end{equation}
It is important to note that once the cooled gas becomes fully turbulent, 
the global cooling function may no longer be valid. For instance, 
\citet{jiSimulationsRadiativeTurbulent2019} suggested that the cooling of 
turbulent gas is dominated by mixing layers with temperatures located at the 
peak of the cooling function.

\item {\bf Feedback mechanisms:} The combined effect of supernova and AGN feedback 
results in the ejection and/or heating of the cooled gas. As a result, 
only a fraction of the cooled gas is able to effectively form stars and   
feed the SMBH. Therefore, we define the `star-forming gas' as
\begin{equation}\label{eq:delta-m-gas-sf}
    \Delta M_{\rm g, sf} = F_{\rm agn} F_{\rm sn} \Delta M_{\rm g, cool}     \,,
\end{equation}
where $F_{\rm agn} (M_{\rm bh}, M_{\rm g}, V_{\rm g}\vert \alpha_{\rm agn})$ 
and $F_{\rm sn}(V_{\rm g}\vert \alpha_{\rm sn}, \beta_{\rm sn}, V_{\rm w})$ are 
defined by Eqs.~(\ref{eq:def-f-agn}) and (\ref{eq:def-f-sn}), respectively. 
In our implementation, we define $V_{\rm g}$ as the maximum circular velocity, 
$V_{\rm max}$, of the host halo, as the turbulent motion of the gas is driven by the 
gravity of the galaxy.

\item {\bf Star formation and SMBH accretion:} The star formation rate (SFR) is 
expected to be related to the amount of cold gas available for star formation. 
In general, we can express this as
\begin{equation}\label{eq:delta-m-star}
    \Delta M_* (\Delta M_{\rm g, sf} \vert \epsilon_{\rm *, f}) = \epsilon_{\rm *, f} \Delta M_{\rm g, sf}    \,,
\end{equation}
where $\epsilon_{\rm *, f}<1$ represents the overall star formation efficiency
in the fast assembly phase. Note that $\epsilon_{\rm *, f}$ is included to reflect 
the fact that a small part of the star-forming gas may be locked in stellar remnants
and in a cold ISM. Stars formed during the fast phase are expected to reside 
in a hot bulge or a hot disk, as indicated by the Q1 and Q4 quadrants, 
respectively, in the quadrant diagram. In what follows, we refer to these two hot 
components collectively as the `bulge'. Since the SMBH growth occurs also during 
this turbulent phase, we can express it using equation~(\ref{eq:def_M_acc}) as
\begin{equation}\label{eq:delta-m-bh}
    \Delta M_{\rm bh} = \alpha_{\rm cap} { M_{\rm bh} \over M_{\rm g} } 
        F_{\rm en} F_{\rm tb} \Delta M_{\rm g, sf}    \,,
\end{equation}
where $F_{\rm en} (M_{\rm v}\vert \alpha_{\rm en}, \beta_{\rm en}, M_{\rm en})$ 
represents the enhancement factor defined by equation~(\ref{eq:def-f-en}), 
and $F_{\rm tb} \equiv \Delta M_{\rm tb}/\Delta M_{\rm g}$ is the mass 
fraction of the turbulent gas, which is set to $1$ due to its degeneracy 
with $\alpha_{\rm cap}$.
If the combined effects of $F_{\rm cool}$, $F_{\rm sn}$ and $F_{\rm agn}$ 
make $\Delta M_{\rm g, sf}$ negligible, we have $\Delta M_{\rm g, sf} \approx 0$. 
Consequently, both $\Delta M_* \approx 0$ and $\Delta M_{\rm bh} \approx 0$, indicating 
the onset of a quenching regime. 
On the other hand, if certain processes lead to a very small value of 
$F_{\rm tb}$, indicating that most of the gas moves in a non-turbulent 
(smooth) fashion, then we have $\Delta M_{\rm bh} \approx 0$ while 
$\Delta M_* \ne 0$. This scenario corresponds to the smooth accretion 
in the slow phase, where the gas is likely to settle into a 
rotation-supported disk. In this case, a dynamically cold system is formed without 
affecting $M_{\rm bh}$.

\item {\bf Evolution of the gas:} The cooled gas ($\Delta M_{\rm g, cool}$) is affected 
by feedback processes, so that part of it is ejected from the 
turbulent region and mixed with the hot gas ($\Delta M_{\rm g, hot}$). 
The part of the gas that has been affected by the feedback but remains 
in the SGC is also prevented from forming stars and feeding the SMBH. 
We refer to it as the `prevented gas' and denote its mass by $\Delta M_{\rm g, prev}$. 
Based on these considerations, we define the amount of ejected gas as
\begin{equation}\label{eq:delta-m-gas-ej}
    \Delta M_{\rm g, ej} = \Delta M_{\rm g, cool} 
    \left[ f_{\rm ej, sn}(1-F_{\rm sn}) + F_{\rm sn} 
    (1-F_{\rm agn}) f_{\rm ej, agn} \right] \,,
\end{equation}
where $f_{\rm ej, sn}$ represents the fraction of SN-affected gas that is 
ejected, while $f_{\rm ej, agn}$ represents the fraction of AGN-affected gas that 
is ejected. Using mass conservation, we can write the amount of the prevented gas as
\begin{equation}\label{eq:delta-m-gas-prev}
    \Delta M_{\rm g, prev} =\Delta M_{\rm g, cool}  
    - \Delta M_{\rm g, ej} - \Delta M_{\rm g, sf} \,.
\end{equation}
The total amount of gas residing outside the turbulent region is given by
\begin{equation}\label{eq:delta-m-gas-out}
    \Delta M_{\rm g, out} = \Delta M_{\rm g, hot} + \Delta M_{\rm g, ej} \,.
\end{equation}
The total amount of gas remaining in the turbulent region is given by
\begin{equation}\label{eq:delta-m-gas}
    \Delta M_{\rm g} =  (\Delta M_{\rm g, sf} - \Delta M_* - \Delta M_{\rm bh}) + 
    \Delta M_{\rm g, prev}.
\end{equation}
And the total amount of non-star-forming gas is given by
\begin{align}\label{eq:delta-m-gas-nonsf}
    \Delta M_{\rm g, nonsf} &= \Delta M_{\rm g, avail} - \Delta M_{\rm g, sf} \nonumber \\ 
                            &= \Delta M_{\rm g, hot} + \Delta M_{\rm g, ej} + \Delta M_{\rm g, prev} \,.
\end{align}     
\end{enumerate}

We note that, due to the limitation of our model in tracking details of the 
feedback on the gas, it remains uncertain whether gas can indeed cool and 
flow into the galaxy before being affected by feedback, or it is continuously 
coupled with the hot outflow and thus prevented from entering the galaxy.
Consequently, the quantity $\Delta M_{\rm g, ej}$ has to be treated as the hot 
gas component that is affected by the feedback in some way and eventually 
resides outside the galaxy, rather than the gas that enters the galaxy 
and is subsequently ejected. The fate of this gas component is not specified: 
it may be incorporated into the hot halo gas outside the SGC, as may 
be the case for high-mass halos, or be ejected from the host halo, as is 
the case for low-mass halos, or may remain in the halo as a part of the 
circum-galactic medium (CGM). 
Note also that $\Delta M_{\rm g, ej}$ and $\Delta M_{\rm g, prev}$ are 
completely degenerate, and the separation of the two is determined 
completely by the choices of $f_{\rm ej, sn}$ and $f_{\rm ej, agn}$
in equation (\ref{eq:delta-m-gas-ej}). Additional modeling and constraints from 
observations of the gas component are needed to separate these different 
possibilities.

Using the equations given above, we can numerically integrate the different components 
until the end of the fast assembly phase. After that, we switch the modeling to 
the slow assembly regime, where a clumpy or a cold disk grows, depending on
whether the galaxy is located in Q2 or Q3 of the quadrant diagram.  
In this paper, we ignore the difference between Q2 and Q3 and generally refer to the stars 
formed in the slow assembly phase as disk stars. As the growth of the SMBH is expected 
to be slow in a less turbulent medium, we freeze the mass of the SMBH at the end of the 
fast assembly phase. In practice, we assume that the transition to a rotation-supported system 
occurs when $f_{\rm ej, sn} \approx f_{\rm ej, agn} 
\approx (f_{\rm B}-\lambda)/f_{\rm B} \approx 3/4$. This choice is made so that 
the gas remains in SGC and available for star formation in the slow phase is 
above $f_{\rm B}/4\sim 0.04$, so that $f_\lambda>1$. 
Our model for the star formation in the slow assembly phase is 
presented below.

\subsection{Star formation in the slow assembly phase}
\label{ssec:slow-phase-model}

For halos in the slow assembly phase, we use the empirical model of 
\citet{luEmpiricalModelStar2014} to model the stellar component. 
The model relates the star formation rate as a function of halo mass and redshift, 
using functional forms characterized by model parameters. 
Specifically, we adopt the functional form and parameters of their Model-II, 
which was calibrated using observed galaxy stellar mass functions at $z \lesssim 4$.
Particularly relevant to the scenario proposed here, 
\citet{luEmpiricalModelStar2014} found that the star formation model 
constrained solely by stellar mass functions is not able to reproduce
the faint-end upturn seen in the local composite conditional luminosity 
function of galaxies in clusters. They empirically proposed a transition 
from an early phase of efficient star formation to a later phase of slow star 
formation to mimic the preventative feedback proposed in the literature 
\citep[e.g.][]{moGalaxyFormationPreheated2002,moGalaxyFormationPreprocessed2004,
luFormationDiscGalaxies2015}, and designed their Model-III to 
replace the behavior of Model-II at high redshift. Our two-phase model provides 
the physical base for the transition. In our implementation, we incorporate a 
log-normal random component with a standard deviation of $\sigma_{\rm *, s}$ in the 
SFR predicted by their Model-II, to account for nuanced factors that are not fully 
captured by the halo mass alone, as described in \citet{chenHowEmpiricallyModel2021} 
using hydrodynamic simulations. 

The SFR is integrated over time in the slow phase to predict the disk mass at each 
snapshot. The final stellar mass, denoted as $M_*$, during the slow assembly 
phase is the sum of the bulge mass, $M_{*, \rm bulge}$, obtained at the last snapshot of 
the fast assembly phase, and the disk mass, $M_{*, \rm disk}$, 
obtained in the slow assembly phase. It should be noted that our modeling does not explicitly 
incorporate galaxy-galaxy mergers. This is because the effects of mergers are implicitly 
captured in the fast assembly phase, and the merger rate is assumed to be low in the slow 
assembly phase.

\bigskip
Fig.~\ref{fig:flow-chart} is a flow chart that summarizes different parts of our model. 
Most of the free parameters in our model have physically motivated ranges. However, some 
parameters are still uncertain and need to be calibrated using observations. 
Quantities that are not used in the calibration process can be considered as model predictions. 
In this paper, we adopt parameters for the slow assembly phase from 
\citet{luEmpiricalModelStar2014}. We manually adjust the remaining parameters 
to match the $M_*$ - $M_{\rm v}$ relation obtained by 
\citet{yangEVOLUTIONGALAXYDARK2012} at $z \approx 0.1$, and to match the 
$M_{\rm bh}$ - $M_{\rm *,bulge}$ relation for elliptical galaxies 
obtained by \citet{grahamAppreciatingMergersUnderstanding2023} at 
$z \approx 0$. The resulting model is referred to as the default model, and 
the corresponding parameter values are given in Table~\ref{tab:parameters}. 
In Appendix~\ref{app:adjustment-model-parameters}, we present the strategy
to calibrate the parameters, as well as the predictions from non-default models
for references. With all these, we can follow the baryonic components in individual dark matter halos
along their assembly histories. In the next section, we will implement these 
prescriptions in a set of simulated dark matter halos 
to predict statistical properties of the galaxy and SMBH 
populations, and to contrast our model predictions with 
with observational data.

\section{Applications to simulated halos}
\label{sec_applications}

\begin{figure*} \centering
    \includegraphics[width=1\textwidth]{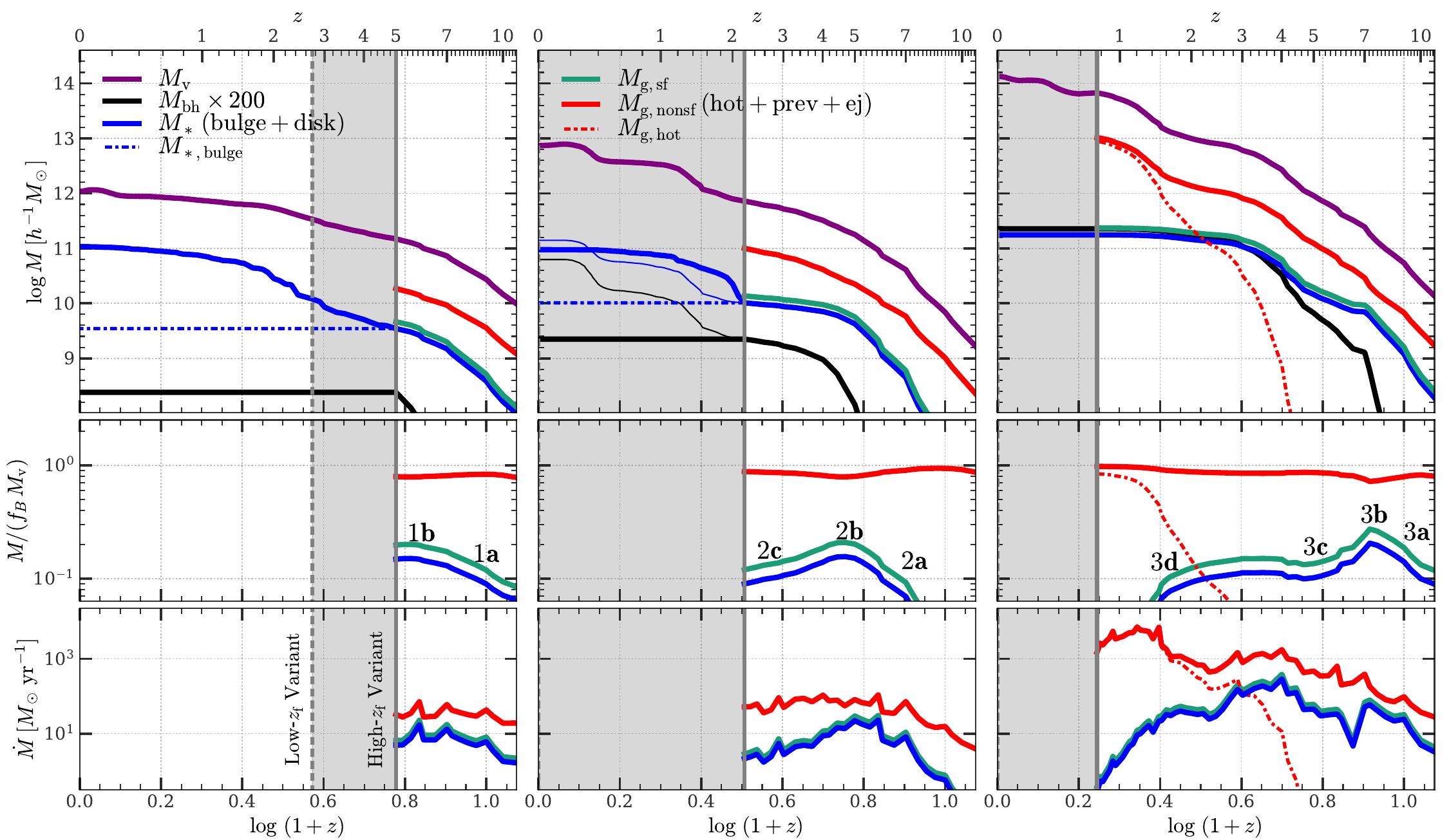}
    \caption{
        Assembly histories of different mass components for three individual
        halos and their central galaxies ({\bf columns}) taken from the 
        sample $S_{\rm h,small}$ 
        (see \S\ref{ssec:halo-sample} for the definition of samples).
        {\bf Top row} shows the masses of dark matter halo, central SMBH, stars and gas. 
        {\bf Middle row} shows the ratio of gas or stellar mass and $f_{\rm B} M_{\rm v}$ 
        in the fast accretion regime.
        {\bf Bottom row} shows the changing rate of gas or stellar mass.
        Masses of different components are shown by curves with different colors 
        and line shapes, as indicated in the legends in the top row.
        Each dashed curve represents a sub-component of that shown by the solid curve 
        with the same color. Specifically, {\bf red solid curve} shows the mass of 
        non-star-forming gas, which includes hot, prevented, and ejected gas, 
        while {\bf red dashed curve} shows the mass of hot gas, which
        is originated from inefficient radiative cooling.
        {\bf Blue solid curve} shows the mass of all stars, while {\bf blue dashed curve}
        shows the mass of bulge stars.
        {\bf Vertical gray solid line} indicates the transition redshift, $z_{\rm f}$, 
        of the High-$z_{\rm f}$ variant, 
        after which the halo enters the slow phase and disk stars begin to form.
        {\bf Vertical gray dashed line} indicates $z_{\rm f}$ of the Low-$z_{\rm f}$
        variant, and the {\bf gray shaded area} marks the uncertainty in the 
        determination of $z_{\rm f}$ bounding by the two variants.
        {\bf Thick curves} are results obtained from the High-$z_{\rm f}$ variant,
        while the {\bf black and blue thin curves} in the top-center panel are 
        obtained from the Low-$z_{\rm f}$ variant.
        In the middle row, {\bf annotations} 
        1a, 2a and 3a mark the SN-regulated regime, 
        1b, 2b and 3b mark the switch point from SN-regulated to AGN-regulated regime,
        2c and 3c mark the AGN-regulated regime, and 3d marks the regime where
        cooling is ineffective. 
        Our model uses different prescriptions for each of these regimes.
        See \S\ref{ssec:coevolution-equations} for the detailed evolutionary 
        equations and Fig.~\ref{fig:flow-chart} for a schematic summary. 
        See \S\ref{ssec:result-growth-of-components} for a detailed discussion 
        of this figure.
    }
    \label{fig:assembly-histories}
\end{figure*}

\begin{figure*} \centering
    \includegraphics[width=0.8\textwidth]{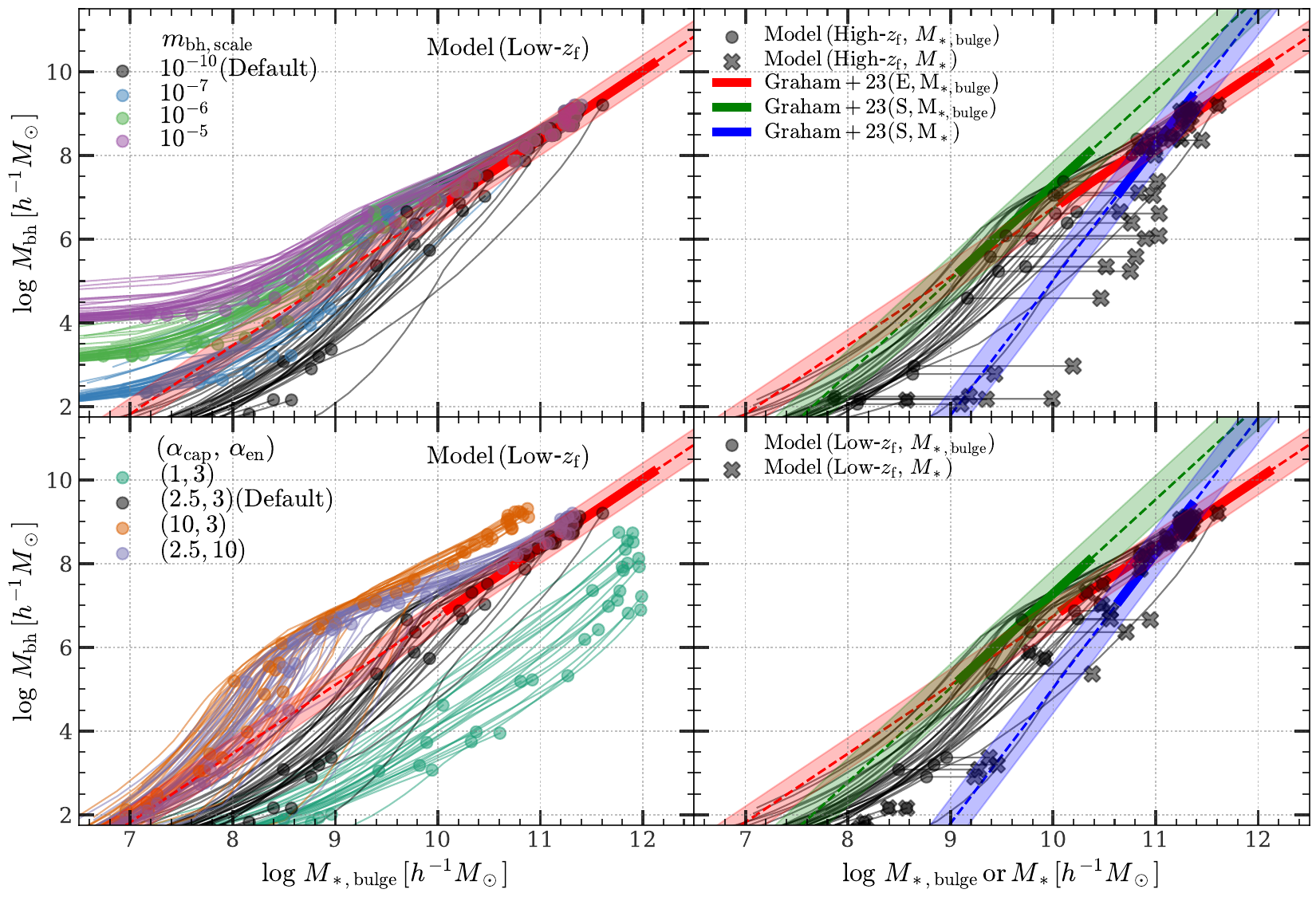}
    \caption{
        Evolutionary paths of individual galaxies in the 
        black hole mass - stellar mass plane.
        {\bf Upper-left panel}: Each bundle of curves with the same color show 
        the $M_{\rm bh}$ - $M_{\rm *,bulge}$ paths
        predicted by a specific model of given black hole seeding 
        parameter $m_{\rm bh,scale}$, as indicated in the legend.
        Other parameters are set to their default values (see 
        Table~\ref{tab:parameters}).
        Transition redshift ($z_{\rm f}$) is defined by the Low-$z_{\rm f}$
        variant. Only fast-phase evolution is shown here. 
        The ending point of the fast phase is marked by a {\bf circle} 
        for each galaxy.
        {\bf Lower-left panel}: Similar to the upper-left panel, but for the 
        combination of black hole mass-capturing capability parameters $\alpha_{\rm cap}$, 
        and $\alpha_{\rm en}$.
        {\bf Upper-right panel} shows the evolution trajectories with
        additional disk growth in the slow phase, as indicated by 
        a {\bf horizontal line} with {\bf a cross} marking the
        the ending point of the slow phase (i.e., $z=0$) for each galaxy.
        Here $z_{\rm f}$ is defined by the High-$z_{\rm f}$ variant.
        {\bf Lower-right panel} is similar to the upper-right panel, but 
        $z_{\rm f}$ is defined by the Low-$z_{\rm f}$ variant.
        In all panels, {\bf colored straight lines} are linear fits to 
        observations obtained by 
        \citet{grahamAppreciatingMergersUnderstanding2023} at $z \approx 0$.
        Specifically,
        {\bf red line} shows the $M_{\rm bh}$ - $M_{\rm *,bulge}$
        relation for elliptical (E) galaxies, while {\bf green line} shows that 
        for spiral (S) galaxies. 
        {\bf Blue line} shows the $M_{\rm bh}$ - $M_*$ (bulge + disk)
        relation for spiral galaxies.
        The solid piece of each line indicates the range reliably covered by 
        observational data, the dashed line indicates the extrapolation, 
        and shading area indicates the residuals of the linear fit. 
        All halos in the sample $S_{\rm h,small}$ are shown (see \S\ref{ssec:halo-sample} 
        for the details of samples).
        See \S\ref{ssec:result-growth-of-components} for a detailed discussion 
        of our results.
        This figure illustrates how the SMBHs in our model converge to the
        observed scaling relations for different types of galaxies and 
        different stellar mass components.
    }
    \label{fig:mbh_ms_paths}
\end{figure*}

\begin{figure*} \centering
    \includegraphics[width=0.8\textwidth]{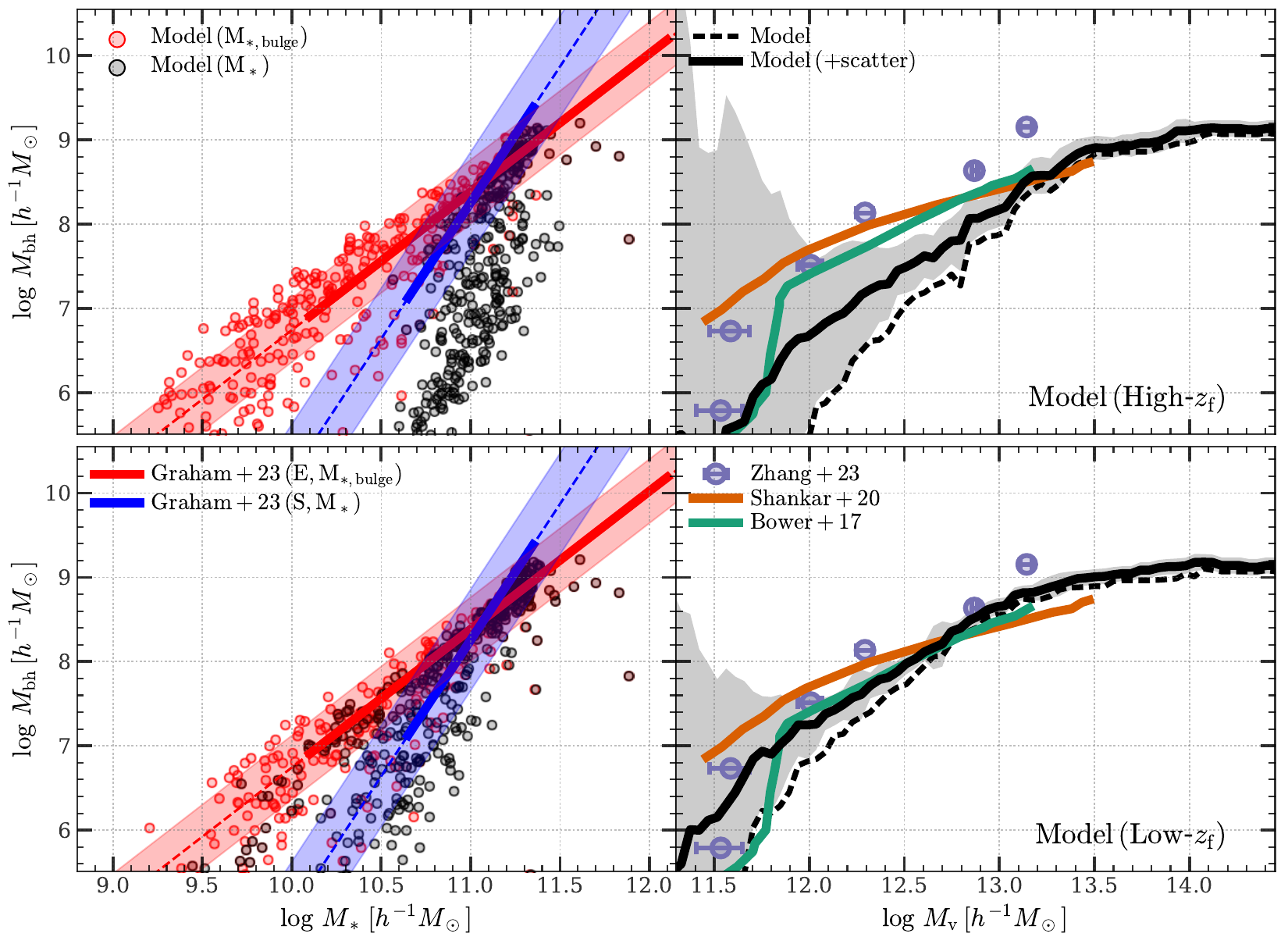}
    \caption{
        SMBH mass to stellar mass ({\bf left column}) and halo virial mass 
        ({\bf right column}) relations at $z \approx 0$. 
        The {\bf upper row} shows the predictions of the High-$z_{\rm f}$ variant, 
        while the {\bf lower row} shows those of the Low-$z_{\rm f}$ variant. 
        The default model parameters are adopted in all panels (see Table~\ref{tab:parameters}). 
        In the left column, the {\bf red circles} show the predicted 
        $M_{\rm bh}$ with $M_{\rm *, bulge}$ of individual galaxies, 
        while the {\bf black circles} show the predicted 
        $M_{\rm bh}$ with $M_{\rm *}$ (bulge + disk). In the right column, 
        the {\bf dashed black line} represents the `true' relation predicted by the model, 
        while the {\bf solid black line} incorporates the model prediction with 
        an up-scatter of $\sigma=3\times10^7\msun$ in linear scale and an 
        up-scatter of $\sigma = 0.5\,{\rm dex}$ in logarithmic scale to mimic 
        the observational systematics near the sharp transition at 
        $M_{\rm v}\approx 10^{11.5}\msun$. For comparison, we show the 
        observational and empirical results from
        \citet{shankarConstrainingBlackHolegalaxy2020},
        \citet{zhangHaloMassobservableProxy2023} and
        \citet{grahamAppreciatingMergersUnderstanding2023},
        and the analytical modeling result 
        from \citet{bowerDarkNemesisGalaxy2017}.
        See \S\ref{ssec:result-bh} for a detailed discussion of our results.
        This figure demonstrates that our model is able to reproduce the 
        local scaling relations of SMBHs, galaxies and halos.
    }
    \label{fig:mbh_ms_and_mh}
\end{figure*}

\begin{figure*} \centering
    \includegraphics[width=0.875\textwidth]{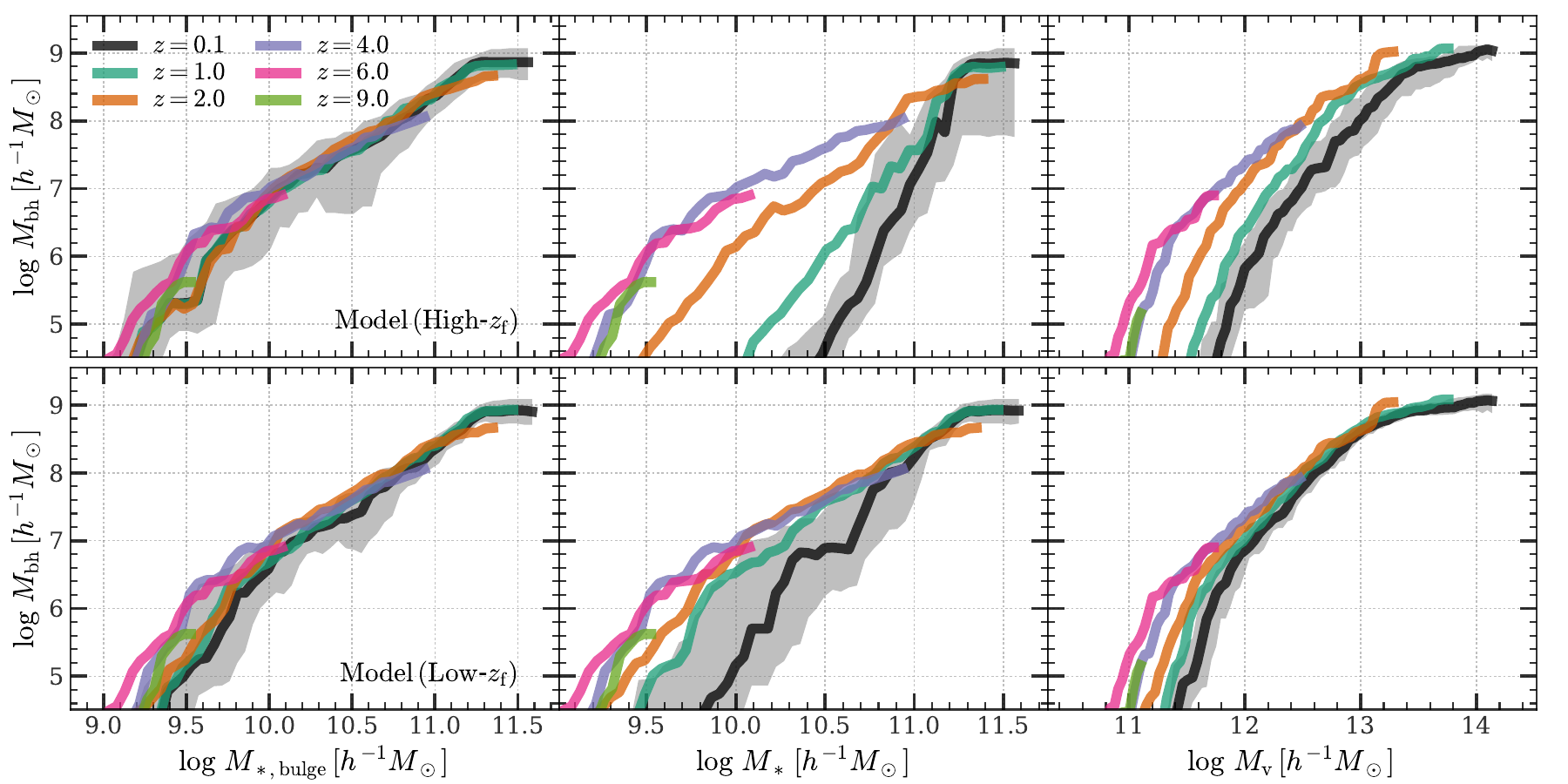}
    \caption{
        Model prediction for the median relations between SMBH mass and bulge 
        mass ($M_{\rm *,bulge}$, {\bf left column}), total stellar mass 
        ($M_*$, including bulge + disk, {\bf middle column}), and halo virial 
        mass ($M_{\rm v} \equiv M_{\rm 200c}$, {\bf right column}) at various 
        redshifts ranging from $z=0.1$ to $z=9$, as indicated in the legend. 
        {\bf Two rows} show the predictions obtained from the High-$z_{\rm f}$ 
        variant and Low-$z_{\rm f}$ variant, respectively, 
        both with default model parameters (see Table~\ref{tab:parameters}). 
        {\bf Shaded areas} surrounding the black curves represent 
        the 1-$\sigma$ (16\% - 84\%) quantiles. See \S\ref{ssec:result-bh} 
        for a detailed discussion of this figure.
    }
    \label{fig:mbh_ms_and_mh_high_z}
\end{figure*}

\begin{figure*} \centering
    \includegraphics[width=0.85\textwidth]{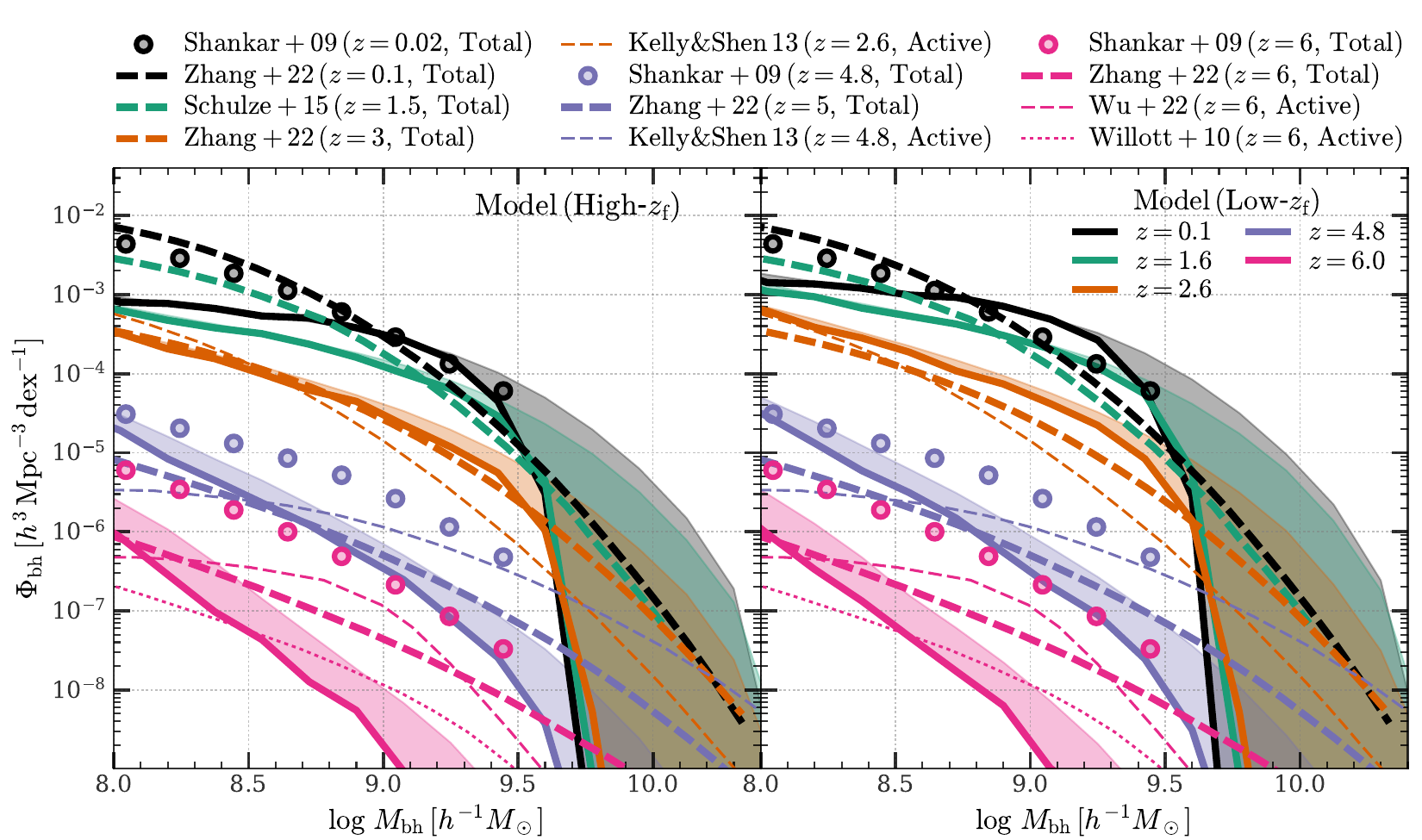}
    \caption{
        SMBH mass functions, 
        $\Phi_{\rm bh} \equiv {\rm d}^2{N_{\rm bh}}/[ {\rm d \log M_{\rm bh}}\, {\rm d}V ]$, 
        at $z \approx 0 - 6$. 
        The {\bf thick solid curves} represent the prediction of our model 
        with default parameters (see Table~\ref{tab:parameters}). The {\bf left 
        and right panels} show the results from the High-$z_{\rm f}$ and 
        Low-$z_{\rm f}$ variants, respectively. The {\bf shaded region} surrounding 
        each curve indicates the variation of $\Phi_{\rm bh}$ when a log-normal 
        random scatter with $\sigma = 0.3$ dex is introduced into 
        $M_{\rm bh}$ to account for the substantial Eddington bias at the 
        high-mass end. Results obtained at comparable redshifts are 
        shown by the same color. The total SMBH mass functions 
        (corrections were made for the AGN duty cycle) are shown by {\bf thicker 
        symbols} (thick solid and dashed curves, as well as circle markers), 
        while active SMBH mass functions (including only active AGN) are 
        represented by {\bf thinner symbols} (thin dashed and dotted curves). 
        The presented observational results include those from
        \citet{shankarSelfConsistentModelsAGN2009},
        \citet{willottEddingtonlimitedAccretionBlack2010},
        \citet{kellyDemographicsBroadlineQuasars2013},
        \citet{schulzeCosmicGrowthActive2015},
        \citet{wuDemographicsQuasarsBlack2022} and
        \citet{zhangTrinitySelfConsistentlyModeling2022}.
        See \S\ref{ssec:result-bh} for a detailed discussion of our results.
        This figure indicates that our model reproduces the observational 
        trend of SMBH mass function at a wide range of redshift.
    }
    \label{fig:bhmf}
\end{figure*}

\begin{figure} \centering
    \includegraphics[width=\columnwidth]{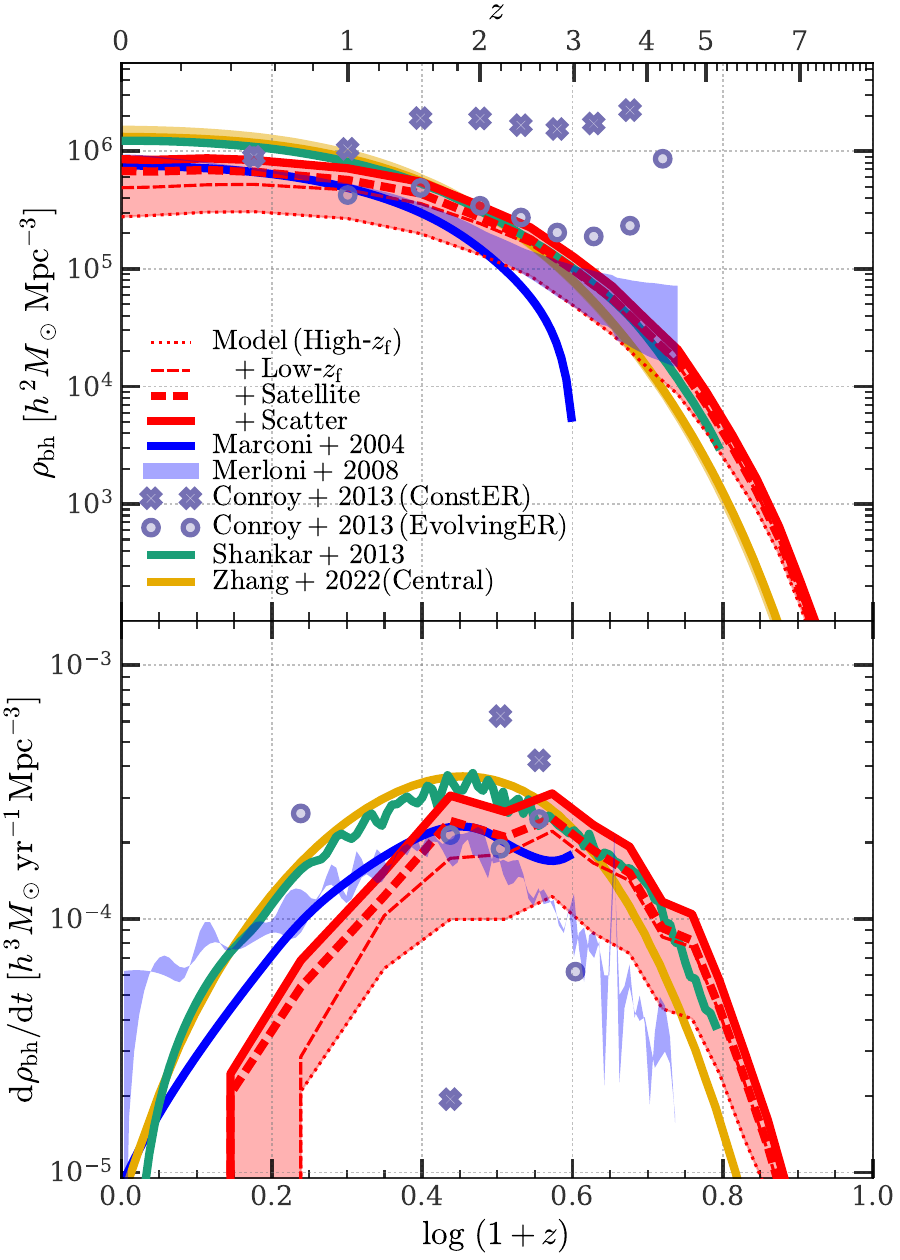}
    \caption{
        Evolution of the cosmic black hole mass density ($\rho_{\rm bh}$, {\bf top panel}) and 
        the black hole mass growth rate density (${\rm d}\rho_{\rm bh}/{\rm d}t$, 
        {\bf bottom panel}). The predictions of the model with default 
        parameters (see Table~\ref{tab:parameters}) are represented by the 
        {\bf red curves} bounding a {\bf shaded area}. 
        The {\bf red dotted curve} shows the case where the transition redshift, $z_{\rm f}$, 
        is defined by the High-$z_{\rm f}$ variant, 
        considering only central galaxies and without incorporating any 
        observational uncertainty. The {\bf red thin dashed curve} represents 
        the result of the High-$z_{\rm f}$ variant, with a further 
        inclusion of satellites resulting in the {\bf red thick dashed curve}. 
        The additional incorporation of a log-normal scatter to 
        $M_{\rm bh}$ with $\sigma=0.3\,$dex is shown by the {\bf red solid curve}. 
        A number of empirical results calibrated by observations are included, 
        such as those from \citet{marconiLocalSupermassiveBlack2004}, 
        \citet{merloniSynthesisModelAGN2008}, 
        \citet{conroySimpleModelQuasar2013},
        \citet{shankarAccretiondrivenEvolutionBlack2013} and
        \citet{zhangTrinitySelfConsistentlyModeling2022}.
        For the study by \citet{conroySimpleModelQuasar2013}, the results obtained 
        by a model with constant and redshift-dependent Eddington ratio 
        distributions are indicated by {\bf cross} and {\bf circle markers}, 
        respectively.
    }
    \label{fig:bhmd}
\end{figure}

\subsection{Halo samples} 
\label{ssec:halo-sample}

The primary samples we use are taken from a dark-matter-only simulation 
conducted as part of the Illustris-TNG project 
\citep{pillepichFirstResultsIllustrisTNG2018, 
nelsonIllustrisTNGSimulationsPublic2019}. To achieve a balance between 
statistical robustness and resolution in halo assembly histories, 
we use the TNG100-1-Dark run (hereafter TNGDark). Specifics of the simulation 
can be found in Appendix~\ref{app:method-halo-sampling}.

To optimize computational efficiency in adjusting our model, we 
work with sub-samples of halos from TNGDark. Details of
the sampling technique and the weighting strategy for computing summary 
statistics, which include both halo sampling rates and contributions from satellite 
subhalos, are given in Appendix~\ref{app:method-halo-sampling}. In summary, 
we define two halo samples at $z=0$ with host halo masses ($M_{\rm v}$, 
defined as $M_{\rm 200c}$) ranging from $10^{10} \msun$ to $10^{14.5} \msun$. 
The first sample, denoted as $S_{\rm h, small}$ and referred to as the small 
sample, consists of 45 halos and is used solely to illustrate the evolution of 
individual galaxies. The second sample, denoted as $S_{\rm h, large}$ and 
referred to as the large sample, includes 720 halos and is used for 
deriving summary statistics. The main branch assembly histories of these halos 
are extracted from merger trees and serve as the input for our model. 
Some examples of the assembly histories drawn from $S_{\rm h, small}$ 
have already been shown in Fig.~\ref{fig:transition_points}.

In cases where summary statistics depend on extremely small or massive halos, 
the simulated sample is insufficient. In such instances, we use analytically fitted 
halo mass functions to sample halos at any desired redshift, and analytically 
fitted halo assembly histories to sample the main branch assembly histories of 
the halos in the sample. We adopt the \softwarenamestyle[Hmf] library 
\citep{murrayHMFHaloMass2014} and the \softwarenamestyle[Diffmah] 
library \citep{hearinDifferentiableModelAssembly2021} to implement halo mass functions 
and halo assembly histories, respectively. Further details about these Monte Carlo samples 
are given in Appendix~\ref{app:method-halo-sampling}.

Throughout the application, we adopt a flat $\Lambda$CDM cosmology with parameters 
obtained from the Planck2015 results \citep{planckcollaborationPlanck2015Results2016}, which 
is also adopted by TNGDark:
the density parameters $\Omega_{\rm M,0}=0.3089$, $\Omega_{\rm B,0}=0.0486$, 
and $\Omega_{\rm \Lambda, 0}=0.6911$, the Hubble constant $H_0 = 100\, h\,{\rm km\,s^{-1}\,Mpc^{-1}}$, 
with $h=0.6774$, the Gaussian initial density field with a power spectrum 
$P(k)\propto k^n$ with $n=0.9667$ and amplitude specified by $\sigma_8=0.8159$. 

\subsection{The growth of different mass components} 
\label{ssec:result-growth-of-components}

Fig.~\ref{fig:assembly-histories} shows the time evolution of different 
mass components in three dark matter halos with present halo mass equal to 
$10^{12.0}$ (left panels), $10^{12.8}$ (middle) and $10^{14.2} \msun$ (right), 
respectively. These halos are taken from $S_{\rm h, small}$, and 
the predictions are made using the High-$z_{\rm f}$ variant.
The transition redshift, $z_{\rm f}$, from the fast to slow 
assembly is marked by the vertical gray solid line for each halo. 
To demonstrate the uncertainty in the determination of $z_{\rm f}$, we use 
a vertical gray dashed line to mark $z_{\rm f}$ based on the 
Low-$z_{\rm f}$ variant.

In the low-mass halo considered here, the bulge was formed at $z\sim 4$ and has a mass
of about $10^{9.6}h^{-1} {\rm M}_\odot$. The associated SMBH has a mass of 
$\sim 10^{6.1}h^{-1}{\rm M}_\odot$, while the disk mass is about 20 times as large as 
that of the bulge. These numbers are quite similar to that of the Milky 
Way galaxy. For the most massive case, the total stellar mass is about 
$10^{11.2}h^{-1}{\rm M}_\odot$ and completely dominated by the bulge that formed at 
$z\sim 3$. The associated SMBH mass is $10^{8.9}h^{-1}{\rm M}_\odot$.  
Both these masses are about five times smaller than those of M87. The case of the 
intermediate mass is quite special in that its transition from the fast to the slow assembly 
is not well defined, as shown by the broad vertical band. Using the High-$z_{\rm f}$ 
definition of $z_{\rm f}$, the predicted bulge mass is about 
$10^{10}h^{-1}{\rm M}_\odot$, and the SMBH mass about $10^{7} h^{-1}{\rm M}_\odot$. 
The predicted disk mass is quite massive, about $10^{11}h^{-1}{\rm M}_\odot$. 
However, if we extend the fast assembly phase to lower redshift, both 
the bulge and SMBH masses will be larger, and the disk mass smaller,   
as is demonstrated by the thin black and thin blue curves in the middle panel 
in the top row predicted using the Low-$z_{\rm f}$ variant. 
This clearly indicates that using a single value of $z_{\rm f}$ to separate the 
fast and slow assembly is not sufficient for some halos. As pointed out earlier, 
we will use both the High-$z_{\rm f}$ and Low-$z_{\rm f}$ variants to bracket the 
uncertainty. 

As one can see, the gas is dominated by the non-star-forming component ($M_{\rm g,nonsf}$, 
the red solid curve), which closely follows the halo mass (the purple solid curve).
Most of the non-star-forming gas is generated by the SN and AGN feedback
and presented either as ejected gas ($M_{\rm g,ej}$) or as prevented gas 
($M_{\rm g,prev}$). The exception is the high-mass halo (right column) near the
end of the fast assembly, where the hot gas ($M_{\rm g, hot}$, red dashed curve) contributes 
significantly to the non-star-forming gas. This is a direct consequence of the inefficient 
radiative cooling of the halo gas at $M_{\rm halo} > 10^{13} \msun$.

By our model design, the stellar mass (blue curve) in the fast assembly phase 
follows that of star-forming gas, $M_{\rm g, sf}$ (green curve).
In the early growth of galaxies when $M_{\rm v} \lesssim 10^{10} \msun$ (marked 
by 1a, 2a and 3a in the middle row for the three galaxies, respectively), 
the potential well of the halo is not deep enough to resist the SN feedback, 
thus the accumulation of star-forming gas is much slower than that of 
the non-star-forming gas, as seen from the middle and bottom rows. 
Consequently, the overall star formation efficiency during this period is 
suppressed at this stage (marked by 1a, 2a, 3a). 
When the halo mass continues to grow so that its potential is 
deepened, the SFR increases rapidly and reaches a peak value 
of $\sim 10$ - $100 \msunperyr$ at $M_{\rm v} \sim 10^{11} \msun$ (marked 
by 1b, 2b and 3b). At this time, the stellar mass $M_* \sim 10^{9.5} \msun$,
the specific star formation rate ${\rm sSFR} \sim 10 \gyri$, and the star formation timescale is 
about $1/{\rm sSFR} \sim 0.1 \gyr$, comparable to the free-fall timescale of the halo,  
$t_{\rm ff} \sim 0.1 / H(z) \approx 0.1 \gyr$. 
This is a special `feedback-free' regime where both SN feedback and AGN feedback 
are inefficient while radiative cooling is still effective. 
As seen from the middle row, the mass of the star-forming gas at this point 
reaches about 20\% - 30\% of the cosmic baryon fraction in the halo, 
a value significantly higher than that in the low-$z$ universe.  The value could be 
even higher in halos where the on-set of SMBH growth was delayed by supernova (SN) 
feedback or some other factors operating near the center. It is thus possible that 
the high star formation efficiency required to explain recent JWST observations of 
massive galaxies at high-$z$ \citep[e.g.][]{naiduTwoRemarkablyLuminous2022,
rodighieroJWSTUnveilsHeavily2022,
donnanEvolutionGalaxyUV2022,
finkelsteinLongTimeAgo2022,
bouwensEvolutionUVLF2023,
finkelsteinCEERSKeyPaper2023,
harikaneComprehensiveStudyGalaxies2023,
labbePopulationRedCandidate2023,
xiaoMassiveOpticallyDark2023,
lovellExtremeValueStatistics2022,
dekelEfficientFormationMassive2023,
masonBrightestGalaxiesCosmic2023,
yungAreUltrahighredshiftGalaxies2023,
boylan-kolchinStressTestingLambda2023,
chenMassiveDarkMatter2023,
shenImpactUVVariability2023} 
is produced in this way. After the peak, the ratio 
$M_{\rm g,sf}/(f_{\rm B} M_{\rm v})$ declines rapidly, because of the onset of the 
SMBH growth (see the black lines in the top panels) and associated feedback. 
There is a delay in the onset of SMBH growth relative to that of the
star formation, caused by the period required to grow a SMBH from a low seeding 
mass assumed in our model. The ratio $M_{\rm g,sf}/(f_{\rm B} M_{\rm v})$ 
then settles down to a roughly constant value at about $0.15$ (marked by 2c and 3c; 1c
does not exist). For the massive case, $M_{\rm g,sf}/(f_{\rm B} M_{\rm v})$ can 
decline further at some point (marked 3d) when the gas can no longer cool.

Let us examine the growth of SMBH in more detail. Fig.~\ref{fig:mbh_ms_paths} 
shows the evolutionary paths of individual systems in the SMBH mass-stellar mass plane.
The upper-left panel shows the impact of varying the seeding mass, as represented 
by the value of $m_{\rm bh, scale}$, while keeping all other model parameters at their 
default values (see Table~\ref{tab:parameters}) and using $z_{\rm f}$ given by High $z_{\rm f}$. 
As one can see, all evolutionary trajectories converge 
to the same fiducial relation represented by the red lane obtained by 
\citet{grahamAppreciatingMergersUnderstanding2023} from low-$z$ data, 
independent of the seeding mass. However, depending on the initial 
seeding mass, it will take some time for a SMBH to reach the fiducial relation. 
Thus, at any given time, particularly at an early time, some SMBH can deviate 
significantly from the fiducial relation, especially in the low-mass end. 
The convergent behavior of the SMBH growth shown here is fully consistent with
the indication obtained by \citet{zhuangEvolutionaryPathsActive2023},
who used the stellar and SMBH masses and growth rates to infer the 
evolutionary trajectories of individual SMBHs and their host galaxies.

In the lower-left panel, we show the impact of changing the values of 
$\alpha_{\rm cap}$, which specifies the ability of a SMBH to capture 
mass in the SGC, and $\alpha_{\rm en}$, which describes the enhancement 
of the capture ability produced by the SN feedback in low-mass halos. 
Our default model assumes $\alpha_{\rm cap}=2.5$ and $\alpha_{\rm en}=3.0$. 
In this model, the SMBH mass increases rapidly with time when the bulge mass reaches
$\sim 10^9 h^{-1} {\rm M}_\odot$, and merges from below to the red lane (the fiducial relation)
at a bulge mass of $\approx 10^{10.5} h^{-1}{\rm M}_\odot$. 
The green bundle (assuming $\alpha_{\rm cap}=1$) and orange bundle  
(assuming $\alpha_{\rm cap}=10$) of lines in the lower-left panel show the 
effects of changing $\alpha_{\rm cap}$. 
As can be seen, a higher value of $\alpha_{\rm cap}$ leads 
to faster growth of SMBH at high-$z$ (corresponding to the lower
end of the bulge mass). However, due to regulations by the AGN feedback, 
the evolutionary trajectories join a lane that is parallel to the red lane 
but has a higher amplitude. In contrast, a value as low as $\alpha_{\rm cap} = 1$ 
results in an inefficient growth at all redshifts, eventually giving 
a $M_{\rm bh}$ - $M_{\rm *,bulge}$ relation that is about $1 {\rm dex}$ lower than the red lane.
In this case, the evolutionary trajectories do not seem to merge, as they 
have not yet reached the point where regulations by the AGN feedback are important. 
Thus, while the logarithmic slope of the $M_{\rm bh}$-$M_{\rm *,bulge}$ relation is 
produced by the AGN feedback, the fiducial value of $\alpha_{\rm cap} \approx 2.5$
is actually required by the observed amplitude of the relation.  
Increasing the value of $\alpha_{\rm en}$ (purple bundle) makes black holes 
grow faster in their early stage, which is similar to the effect of increasing
$\alpha_{\rm cap}$ (orange bundle). However, at a later time when the enhancement 
becomes inefficient, the evolutionary trajectories join the red lane, but with 
a delay in comparison to the default model. The value of $\alpha_{\rm en}$ 
may thus be constrained if the slope of the $M_{\rm bh}$-$M_{\rm *,bulge}$ 
relation becomes steeper when the bulge mass is lower than some characteristic 
value. For $\alpha_{\rm en}=3$, this characteristic mass is about 
$10^{10}h^{-1}{\rm M}_\odot$, which seems to be needed by observational 
data, as we will see below.  

In the upper-right panel, the green line shows the observational results 
for bulges of spiral (S) galaxies. This line has a logarithmic slope of 
$\sim 2.25$, significantly steeper than that for early-type (E) galaxies. 
Our model provides an explanation of this steepening, by assuming  
that bulges in spiral galaxies had not reached the stage of AGN regulation before 
they entered the slow assembly phase.

The blue line in the upper-right panel shows the observed relation between the 
SMBH mass and the total stellar mass relation for spiral galaxies. This line has a slope of
about $3.23$, much steeper than the red line. This steepening is expected  
in our model, where $M_{\rm bh}$ remains constant in the slow assembly
phase while the stellar mass increases due to star formation in the disk. 
This produces a horizontal shift, as indicated by the crosses connected to
the bulge mass (filled circles) by horizontal lines. Since lower-mass 
halos on average move to the slow assembly phase earlier and their star 
formation efficiency in the slow phase is higher, they can grow larger disks (relative
to the bulge component) than higher-mass halos. This leads to a larger horizontal shift for
lower-mass systems and gives rise to the observed steepening of the relation.
The High-$z_{\rm f}$ variant seems to over-predict the shift, resulting in a relation 
that is too steep, while the predictions of Low-$z_{\rm f}$, 
shown in the lower-right panel, work much better.   

\subsection{The supermassive black hole population} 
\label{ssec:result-bh}

We apply our model to the halo sample $S_{\rm h, large}$ to generate galaxies and 
SMBHs to investigate their statistical properties. Fig.\,\ref{fig:mbh_ms_and_mh} shows the correlation of 
the SMBH mass with the stellar mass of galaxies (left column) and with the halo mass (right column). 
Results using default parameters (see Table~\ref{tab:parameters}) are shown for the 
High-$z_{\rm f}$ variant (upper row) and Low-$z_{\rm f}$ variant (lower row), respectively. 
The correlation with the stellar mass is shown separately for 
the bulge stellar mass $M_{\rm *, bulge}$ (red circles) and the total stellar 
mass $M_*$ (black circles). 
As one can see, our model predictions for the $M_{\rm bh}$ - $M_{\rm *,bulge}$ relation 
match well the observational 
data, shown by the red line and band, in both slope and amplitude. The predicted 
scatter increases in the low-mass end, because many systems have not yet reached 
the stage of AGN regulation where $M_{\rm *} > 10^{10.5}$. 
The two variants of $z_{\rm f}$ yield very different 
$M_{\rm bh}$ - $M_{\rm *}$ relations, with High-$z_{\rm f}$ predicting a much
larger disk mass for disk-dominated galaxies and thus a much steeper 
$M_{\rm bh}$ - $M_{\rm *}$ relation than Low-$z_{\rm f}$. 
As discussed before, the fast assembly phases of some halos 
end much earlier in High-$z_{\rm f}$ than in Low-$z_{\rm f}$ 
(see \S\ref{ssec:result-growth-of-components} and Fig.~\ref{fig:assembly-histories}), 
giving much longer time for the disk to grow in these halos. 
The Low-$z_{\rm f}$ variant appears able to match the observational data nicely, indicating that the 
growth of the bulge and SMBH components can continue at
$z<z_{\rm f}$ defined by High-$z_{\rm f}$. When a moderate random scatter
is included in $M_{\rm bh}$ to mimic observational uncertainties,
the $M_{\rm bh}$ - $M_{\rm v}$ relation predicted by the Low-$z_{\rm f}$ variant 
also matches observational results well, while High-$z_{\rm f}$ seems to 
under-predict $M_{\rm bh}$ for given $M_{\rm v}$ in low-mass halos
(see left panels of Fig.\,\ref{fig:mbh_ms_and_mh}).
The difference again has its origin in the fact that the time for 
growing the bulge and SMBH components is shorter in High-$z_{\rm f}$ than in 
Low-$z_{\rm f}$ for some halos.  

A recent spectroscopic study of \citet{maiolinoJADESDiversePopulation2023}
based on JWST/JADES survey reported a number of over-massive black holes 
at $z > 4$. At $ M_{\rm *}\lesssim 10^{9} \Msun$, the estimated SMBH masses 
are orders of magnitude higher than the extrapolation of the local scaling relation
shown in Fig.~\ref{fig:mbh_ms_and_mh}. This discrepancy could be caused by 
two main factors. Firstly, since the local scaling relation is not well constrained 
at $M_{\rm *}\lesssim 10^{9} \Msun$, the extrapolation may not be reliable. 
As our fiducial model of SMBH seeding is based on the local scaling relation, 
it would under-predict SMBH masses at high $z$. Indeed, the change in 
SMBH seeds has significant impacts on the mass evolution of SMBHs in our 
model, as shown in Fig.~\ref{fig:mbh_ms_paths}.
Secondly, observational estimates of the SMBH mass at high $z$ are still quite uncertain.
The mass estimator used by \citet{maiolinoJADESDiversePopulation2023}
relies on assumptions of virialization, a redshift-independent geometry of 
broad-line regions and the extrapolation of locally calibrated scaling relations. 
As shown by the case study of \citet{abuterDynamicalMeasureBlack2024}, 
a break of these assumptions can lead to large systematic changes in the 
estimated SMBH mass. Given the uncertainties both in current observations 
and in our model, we do not make a detailed comparison of our model 
with the observational data. 

Fig.~\ref{fig:mbh_ms_and_mh_high_z} shows the correlation of $M_{\rm bh}$ with 
bulge mass (left), total stellar mass (middle) and halo mass (right) at different 
redshift. The predicted $M_{\rm bh}$-$M_{\rm *, bulge}$ relations by both 
High-$z_{\rm f}$ and Low-$z_{\rm f}$ are quite independent of redshift, 
except at the low-mass end where the amplitude of the relation appears higher
at higher $z$. This redshift-independence of the $M_{\rm bh}$-$M_{\rm *, bulge}$ relation 
is actually seen in observational data \citep[e.g.][]{schulzeAccountingSelectionEffects2014,
zhuangEvolutionaryPathsActive2023}. The middle-column
panels show that the predicted $M_{\rm bh}$-$M_*$ relation depends significantly 
on redshift, becoming progressively steeper at lower $z$, and the predicted evolution is 
stronger in the High-$z_{\rm f}$ variant than in Low-$z_{\rm f}$.  This steepening 
is completely produced by the growth of the disk mass with time after the 
formation of the bulge and SMBH in the fast assembly phase. 
Finally, the predicted $M_{\rm bh}$ - $M_{\rm v}$ relation also steepens with redshift, 
but the change is much weaker for  Low-$z_{\rm f}$ than for High-$z_{\rm f}$. 
In a recent paper by \citet{shimasakuBlackDarkRapid2019}, a weak redshift-dependence 
was found in the observed SMBH mass versus halo mass relation, again favoring a more 
extended growth of $M_{\rm BH}$ than that assumed in High-$z_{\rm f}$. 
 
In Fig.~\ref{fig:bhmf} we show our model predictions for the SMBH mass functions 
at different redshift, plotted as solid curves of different colors. Data for comparisons 
are also shown using symbols and broken lines with similar color coding. As one can see, 
model predictions of High-$z_{\rm f}$ and Low-$z_{\rm f}$
are similar, except at low $z$ where the mass function at the low-mass end predicted by  
High-$z_{\rm f}$ is significantly lower. Our model predictions follow the general 
trend in the observational data, except at high-$z$ where the predicted mass 
function appears significantly steeper than the observational results. 
Unfortunately, the data are still very uncertain; there are large 
discrepancies between different observational results. This is clearly 
seen from the change in the high-mass end of the predicted SMBH mass function 
caused by introducing a moderate amount of scatter, $\sigma = 0.3\,{\rm dex}$, in the predicted $M_{\rm bh}$ 
to mimic the Eddington bias. A horizontal shift of $\approx 1\,{\rm dex}$ 
is produced at $\Phi_{\rm bh} = 10^{-8} h^3{\rm Mpc}^{-3} {\rm dex}^{-1}$
in the low-$z$ mass function, making it difficult to draw any strong conclusions 
(see \citet{conroySimpleModelQuasar2013} for a similar result). A more robust quantity
is the SMBH mass density, $\rho_{\rm bh}$, which is obtained by integrating the 
mass-weighted mass function. 
This density as a function of redshift is shown in the upper panel of 
Fig.\,\ref{fig:bhmd} for different cases, in comparison with various 
empirical results calibrated with observational data 
\citep{marconiLocalSupermassiveBlack2004, merloniSynthesisModelAGN2008, conroySimpleModelQuasar2013, 
shankarAccretiondrivenEvolutionBlack2013,zhangTrinitySelfConsistentlyModeling2022}. 
Note that these empirical results are obtained by different sets of
assumptions that are not necessarily consistent with each other. For example, 
the starting redshift for black hole growth is assumed to be $\approx 3$ by 
\citet{marconiLocalSupermassiveBlack2004}, $\approx 6$ by 
\citet{shankarAccretiondrivenEvolutionBlack2013},
and $\gtrsim 15$ by \citet{zhangTrinitySelfConsistentlyModeling2022}.
As seen from the figure, these assumptions give different evolution
trajectories at high redshift. 
Another source of systematic arises from the degeneracy of model parameters, 
e.g. between the mean relation and assumed variances, which can be seen from the two divergent 
curves obtained by \citet{conroySimpleModelQuasar2013} (crosses and circles). 
However, as shown in the lower panel of Fig.\,\ref{fig:bhmd}, our model 
appears to under-predict the growth rate of SMBH mass at $z<1$. Although 
the fraction of the SMBH mass density generated at $z<1$ may be small,
the discrepancy indicates that our current model may miss some channels of 
SMBH mass growth at low $z$. Indeed, as discussed in \S\ref{sec:bh_growth}, 
our model only includes SMBH growth in Q1 of the quadrant diagram. In reality, 
later mergers of cold gaseous disks, corresponding to low-mass halos  
making an excursion from Q3 to Q4, may feed SMBH with cold gas and increase 
their masses. In addition, gas-rich disks living near the border of Q2 and 
Q3, may become unstable due to non-axisymmetric perturbations from  
their halos and/or other objects, forming bar-like structures that can also 
feed the central SMBH. It seems clear that these channels need to be included 
in order to explain the SMBH mass growth at $z<1$. 

\subsection{Stellar components} 
\label{ssec:results-stellar}

\begin{figure} \centering
    \includegraphics[width=0.845\columnwidth]{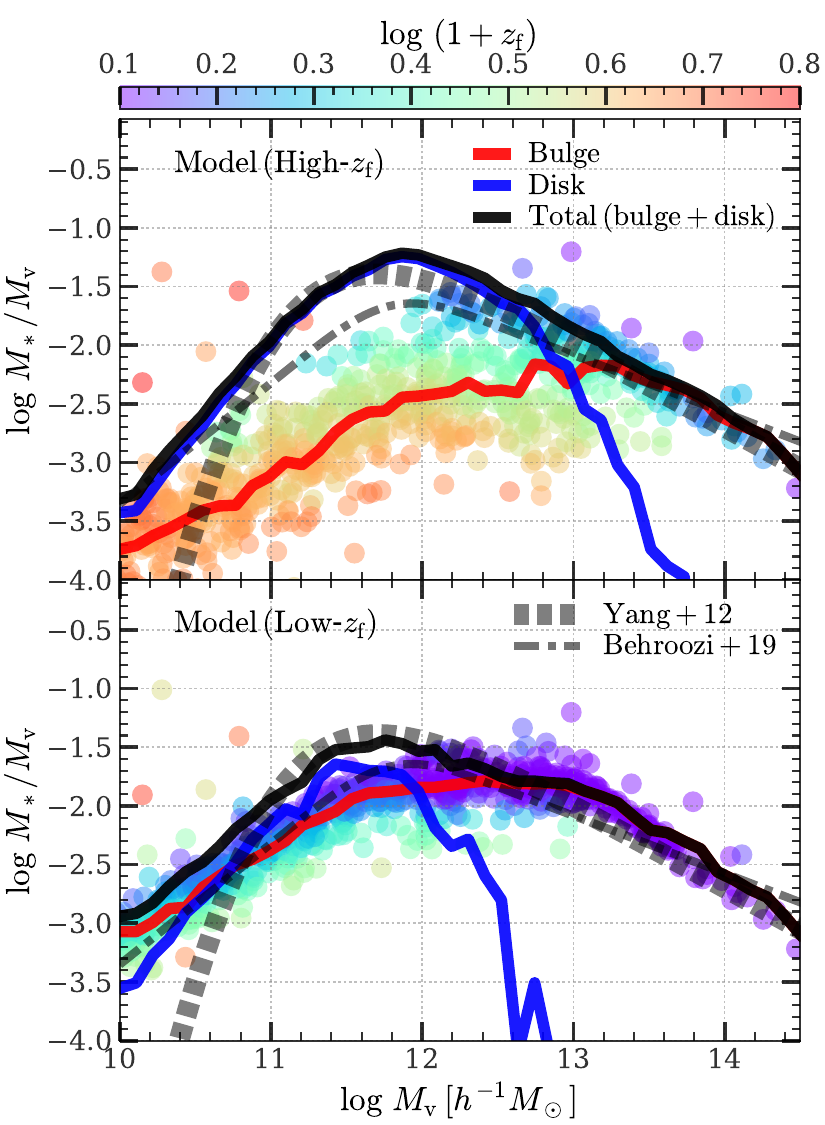}
    \caption{
        Stellar mass to halo mass ratio as a function of halo mass 
        ($M_{\rm v}\equiv M_{\rm 200c}$) at $z \approx 0.1$. The default model 
        parameters (see Table~\ref{tab:parameters}) are adopted, 
        and the halo sample $S_{\rm h, large}$ (see \S\ref{ssec:halo-sample}) 
        is used to derive the statistics. The {\bf upper} and {\bf lower panels} 
        show the High-$z_{\rm f}$ variant and the Low-$z_{\rm f}$ variant, 
        respectively. In each panel, model predictions for bulge, disk, and 
        total (bulge + disk) stellar mass are shown by {\bf red, blue, and black curves}, 
        respectively. {\bf Scatter points} represent the results for the bulge mass of 
        individual galaxies, color-coded by their transition 
        redshift, $z_{\rm f}$, according to the color bar. 
        The {\bf gray thick dashed curve} represents the total stellar mass - 
        halo mass relation obtained by \citet{yangEVOLUTIONGALAXYDARK2012} 
        with conditional stellar mass function modeling. The {\bf gray thin dashed curve} 
        represents the total stellar mass - peak halo mass relation obtained by 
        \citet{behrooziUniverseMachineCorrelationGalaxy2019} with abundance-based 
        empirical modeling. For further discussion, see 
        \S\ref{ssec:results-stellar}. This figure highlights that older halos 
        (with early transition) have centrals with lower bulge mass and 
        larger disk mass, which is crucial for reproducing the observed
        bimodal distribution of galaxies shown in 
        Fig.~\ref{fig:ms_to_mh_colored_f_bulge}.
    }
    \label{fig:ms_to_mh_colored_z_f}
\end{figure}

\begin{figure} \centering
    \includegraphics[width=0.925\columnwidth]{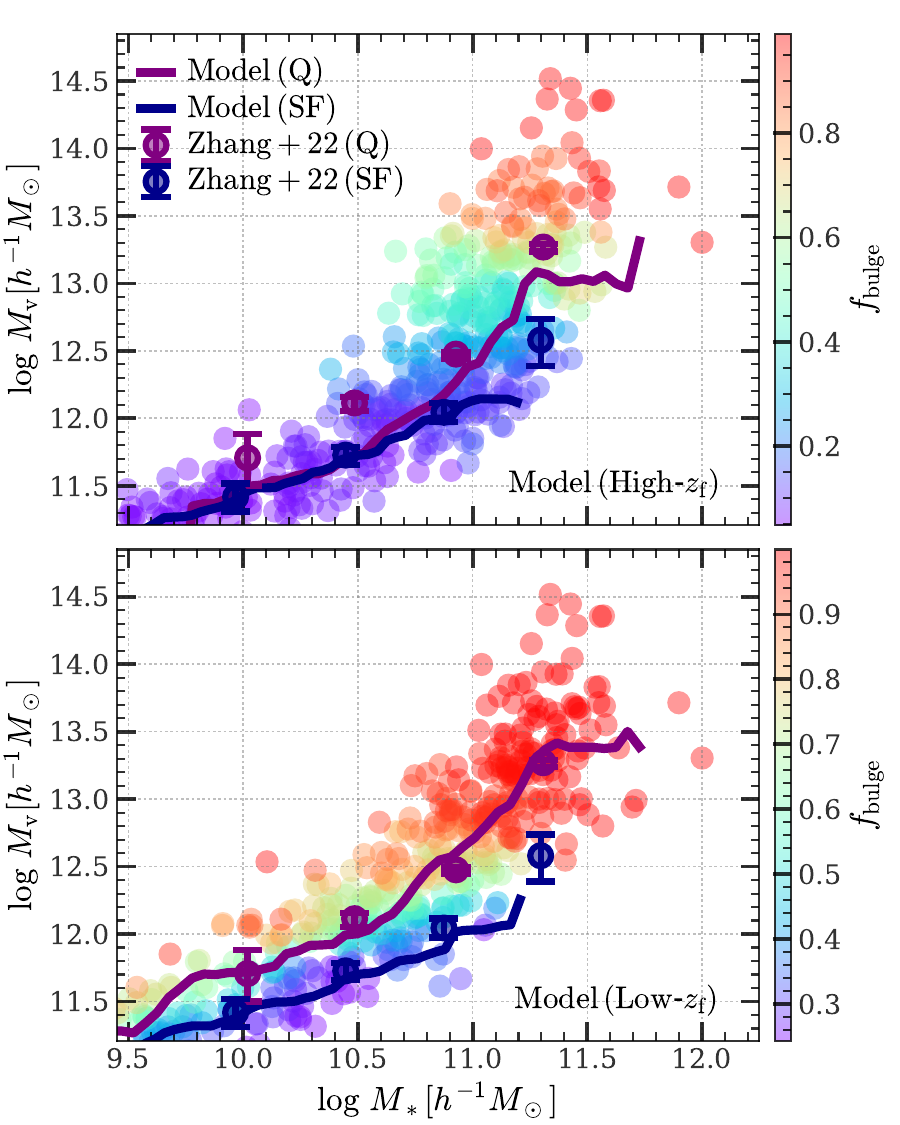}
    \caption{
        Similar to Fig.~\ref{fig:ms_to_mh_colored_z_f}, but here we show the 
        $M_{\rm v}$ - $M_{\rm *}$ relation at $z \approx 0.1$ for direct 
        comparison with observational data. Individual galaxies, 
        represented by {\bf scatter points}, are color-coded based on 
        their bulge fraction, $f_{\rm bulge} \equiv M_{\rm *,bulge}/M_*$, 
        as indicated by the color bar. Galaxy-galaxy lensing measurements 
        of halo mass obtained by \citet{zhangMassiveStarFormingGalaxies2022} 
        for quenched and star-forming galaxies are shown by {\bf purple and blue markers}, 
        respectively. Our predictions for these two types of galaxies are shown 
        by {\bf solid curves}. To mimic observational uncertainties, a log-normal 
        random scatter with $\sigma = 0.15\,$dex is added to both $M_*$ and $M_{\rm v}$. 
        This figure suggests that our model can naturally reproduce the observed
        bimodal distribution of galaxies. For further discussion, 
        see \S\ref{ssec:results-stellar}.
    }
    \label{fig:ms_to_mh_colored_f_bulge}
\end{figure}

\begin{figure} \centering
    \includegraphics[width=0.925\columnwidth]{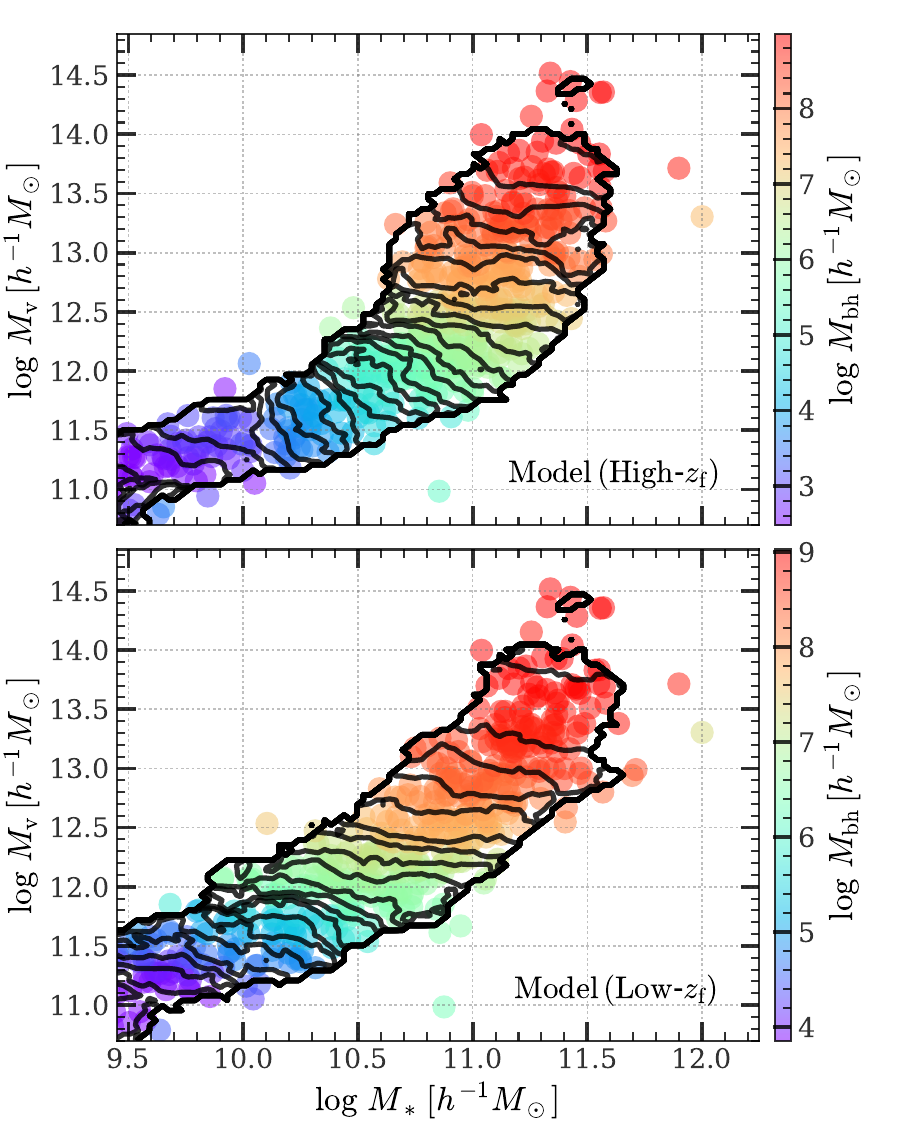}
    \caption{
        Similar to Fig.~\ref{fig:ms_to_mh_colored_f_bulge}, but here 
        {\bf scatter points} are color-coded by SMBH mass $M_{\rm bh}$. 
        Contours with equal SMBH mass are overplotted. 
        This figure indicates that the strong dependency of 
        $M_{\rm bh}$ with $M_{\rm v}$ in high-mass halos is reproduced 
        if the SMBH growth in high-mass halos is regulated 
        by the AGN feedback controlled by gravitational potential.
        See \S\ref{ssec:results-stellar} for a detailed discussion.
    }
    \label{fig:ms_to_mh_colored_m_bh}
\end{figure}

\begin{figure*} \centering
    \includegraphics[width=0.675\textwidth]{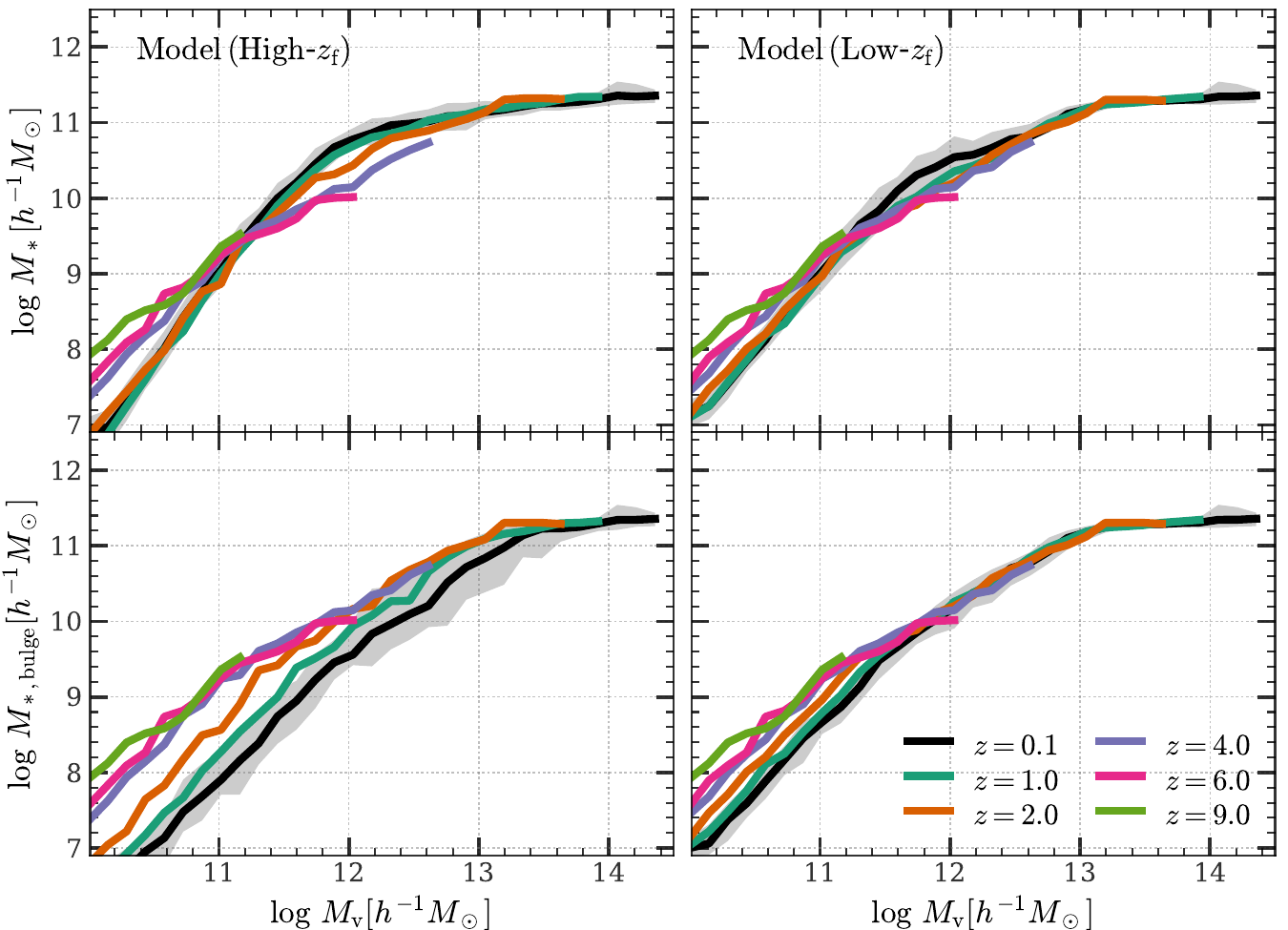}
    \caption{
        The stellar mass to halo mass relation predicted by our model at 
        various redshifts ({\bf different colors}) from $z=0.1$ 
        to $z=9$, as indicated in the legend. The {\bf upper row} shows the 
        total stellar mass ($M_*$), while the {\bf lower row} shows the 
        bulge stellar mass ($M_{\rm *,bulge}$). The {\bf left column} shows 
        the results derived from the High-$z_{\rm f}$ variant of our model, 
        while the {\bf right column} shows those obtained from the 
        Low-$z_{\rm f}$ variant. The {\bf solid lines} represent the median 
        relations, and the {\bf shaded area} surrounding the $z=0.1$ curve 
        shows the 1-$\sigma$ ($16\%$-$84\%$) quantiles at this redshift.
        See \S\ref{ssec:results-stellar} for a detailed discussion.
    }
    \label{fig:ms_to_mh_high_z}
\end{figure*}

\begin{figure*} \centering
    \includegraphics[width=0.825\textwidth]{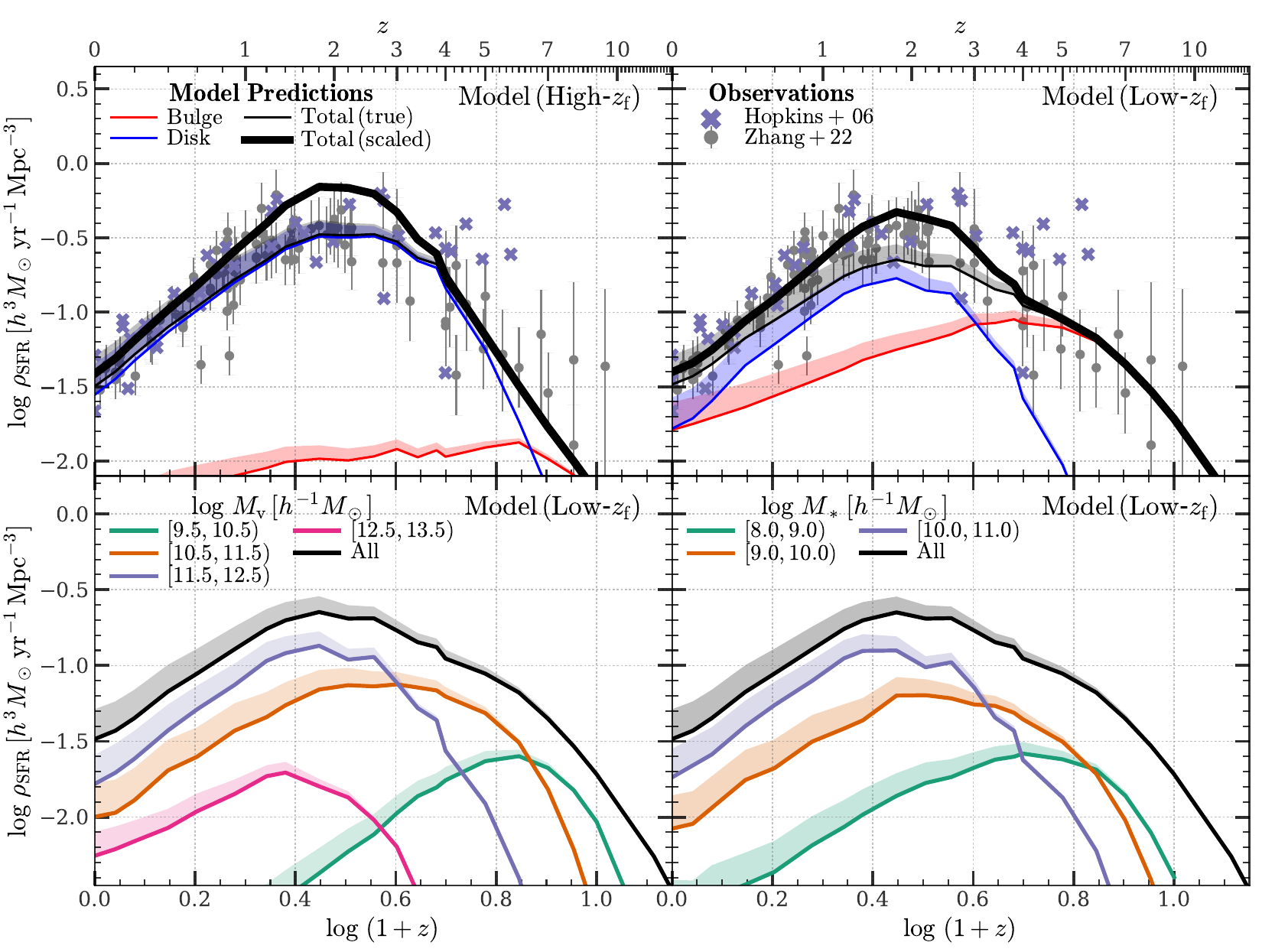}
    \caption{
        Cosmic star formation rate density $\rho_{\rm SFR}$. 
        The {\bf upper-left panel} shows the total $\rho_{\rm SFR}$, 
        and those decomposed as bulge and disk components, all 
        obtained by the High-$z_{\rm f}$ variant. 
        {\bf Thin curves} represent the true values predicted by our model, 
        while the {\bf thick black curve} shows the scaled values for 
        comparison with observational data, corrected according to 
        \S3.5 of \citet{behrooziUniverseMachineCorrelationGalaxy2019}
        to account for the intrinsic inconsistency between the 
        observed evolution of stellar mass functions and 
        cosmic star formation rate density. Observational data points 
        compiled by \citet[{\bf cross markers}]{hopkinsNormalizationCosmicStar2006} 
        and \citet[{\bf dots with errorbars}]{zhangTrinitySelfConsistentlyModeling2022} 
        are shown. The {\bf upper-right panel} is similar,
        but with the Low-$z_{\rm f}$ variant. The {\bf lower panels}, 
        obtained using the Low-$z_{\rm f}$ variant, show a decomposition 
        by halo mass in the {\bf left panel} and a decomposition by stellar mass 
        in the {\bf right panel}. In all panels, the shaded area around 
        each curve indicates the uncertainties contributed by satellite galaxies.
        This figure shows that it is the same population of halos that drive 
        the cosmic star formation throughout the entire history of the 
        Universe, and that the star formation activity of the Universe is not 
        concentrated around a well-defined period around $z \sim 2$ 
        but rather has a more extended distribution in time.
        See \S\ref{ssec:results-stellar} for a detailed discussion.
    }
    \label{fig:sfrd}
\end{figure*}

Fig.~\ref{fig:ms_to_mh_colored_z_f} shows the stellar mass versus halo mass relation 
at $z \sim 0$, predicted by both the High-$z_{\rm f}$ (upper panel) and Low-$z_{\rm f}$ 
(lower panels) variants. Here default model parameters are adopted (see Table~\ref{tab:parameters})
and sample $S_{\rm h, large}$ (see \S\ref{ssec:halo-sample}) is used to derive the statistics.
Results are shown for the total stellar mass ($M_*$, solid black curve), and 
separately for the bulge mass ($M_{\rm *,bulge}$, red curve) and disk mass ($M_{\rm *,disk}$, blue curve). 
The data points are individual model galaxies, color-coded here according 
to $z_{\rm f}$. For comparison, we include the observational results of 
\citet{yangEVOLUTIONGALAXYDARK2012}, obtained 
from fitting galaxy stellar mass functions and conditional stellar mass functions of group 
galaxies. As one can see, for $M_{\rm v}>10^{11}h^{-1}{\rm M}_\odot$
both High-$z_{\rm f}$ and Low-$z_{\rm f}$ variants predict 
a similar $M_*$ - $M_{\rm v}$ relation that closely matches the observational relation. 
For lower $M_{\rm v}$, the predicted $M_*$ is higher than that of \citet{yangEVOLUTIONGALAXYDARK2012}. 
However, the relation for such low-mass halos is still poorly constrained observationally.     
For example, empirical results obtained from other combinations of observational data 
show that the stellar mass in low-mass halos could be much higher, as shown by the thin 
dashed curve obtained more recently by \citet{behrooziUniverseMachineCorrelationGalaxy2019}.

In contrast, the $M_{\rm *, bulge}$ - $M_{\rm v}$
relation at $M_{\rm v}<10^{13}h^{-1}{\rm M}_\odot$ predicted by Low-$z_{\rm f}$ 
is about a factor of three as high as that predicted by High-$z_{\rm f}$. This 
difference is produced by the longer time in Low-$z_{\rm f}$ for halos to stay in the 
fast assembly phase, where stars are assumed to reside in the bulge component. 
The fact that both Low-$z_{\rm f}$ and High-$z_{\rm f}$ predict very 
similar $M_{\rm *}$ - $M_{\rm v}$ relations thus indicate that our star formation 
model matches the empirical model of \citet{luEmpiricalModelStar2014}. 
In their Model-III, a significant fraction of stars in present-day low-mass halos 
are assumed to have formed at redshift above a critical value $z\sim 2$, 
and thus it requires a low star formation rate at lower $z$ to make up 
the total stellar mass. Such a low rate, combined with the shorter time 
available for disk star formation in the case of Low-$z_{\rm f}$, produces 
only small amounts of disk mass in low-mass halos. There is now evidence 
that present-day low-mass galaxies may contain large amounts of old stars
\citep[e.g.][]{zhouStarFormationHistories2020,sacchiStarFormationHistories2021,
mengGalaxyPopulationsGroups2023,weiszReionizationeraGlobularCluster2023}. 
Whether these stars formed in dynamically hot `bulges' or in dynamically cold `disks'
is still unclear. The answer to this question is observationally challenging 
because it is difficult to distinguish random motion from rotation in a low-mass 
halo where the circular velocity is low. 

From the color coding of $z_{\rm f}$ for individual galaxies shown 
in Fig.~\ref{fig:ms_to_mh_colored_z_f}, one can see that, for a given 
$M_{\rm v}$, older halos (those with higher $z_{\rm f}$) have centrals
with lower bulge mass and larger disk mass. This is an interesting prediction 
of the model, and may shed some light on the connection between halos 
and galaxies beyond abundance matching based on masses alone. 
To demonstrate this, we show in Fig.\,\ref{fig:ms_to_mh_colored_f_bulge}
the relation of $M_{\rm v}$ with $M_*$, separately for star-forming and quenched galaxies. 
Observations suggest that quenching is tightly related to dynamical hotness 
\citep[e.g.][]{bluckAreGalacticStar2020,bluckQuenchingGalaxiesBulges2022,hongDynamicalHotnessStar2023}. 
Here we use the bulge fraction, $f_{\rm bulge} \equiv M_{\rm *,bulge}/M_*$, 
as a proxy of dynamical hotness, to separate quenched galaxies from
star-forming ones by a stellar-mass-dependent boundary, $f_{\rm bulge}(M_*)$, 
chosen to match the observed quenched fraction as a function of stellar mass 
obtained by \citet{bauerGalaxyMassAssembly2013}.
Fig.\,\ref{fig:ms_to_mh_colored_f_bulge} shows a clear separation in 
the $M_{\rm v}$ - $M_*$ relation between the two types of galaxies, 
particularly in the Low-$z_{\rm f}$ variant. At a given $M_*$, the halo mass is on average larger for 
galaxies with higher $f_{\rm bulge}$ (thus more quenched). 
Equivalently, for a given $M_{\rm v}$, the total stellar mass 
$M_*$ is on average higher for galaxies of lower $f_{\rm bulge}$ (thus more star-forming). 
This prediction can be tested by measuring the halo mass for galaxies of the same 
stellar mass but with different morphology and different star-formation 
status. Galaxy-galaxy lensing results of \citet{mandelbaumStrongBimodalityHost2016}
and \citet{zhangMassiveStarFormingGalaxies2022} show that star-forming 
galaxies (presumably dominated by the disk component) have lower halo mass for their
stellar mass in comparison to quenched galaxies (dominated by bulges). 
These observational results are qualitatively reproduced by our model, 
particularly by the Low-$z_{\rm f}$ variant, as shown by 
the curves in the figure. We also note that the predicted 
$M_{\rm v}$ - $M_*$ relation has large scatter at the high-mass end, 
produced by the more diversity in their assembly histories.
This may have important implications for models based on abundance matching.

The predicted anti-correlation between star formation rate and $z_{\rm f}$ is somewhat 
counter-intuitive, as one would assume that older halos should preferentially host older-type 
galaxies. This anti-correlation is produced naturally in our model by 
the longer time for disk formation in halos of higher $z_{\rm f}$, provided 
the AGN feedback does not preferentially suppress star formation in old halos.
Our model thus provides a boundary condition to model the observed anti-correlation, 
e.g. using cosmological hydrodynamic simulations. For example, the SIMBA simulation 
\citep{daveSimbaCosmologicalSimulations2019} reproduces the anti-correlation 
\citep{cuiOriginGalaxyColour2021}, as it satisfies the condition. 
In SIMBA, earlier-formed (older) halos host galaxies with later transitions to the slow 
growth of the SMBH. This has the effect of delaying the onset of the `jet' mode 
feedback in older halos, 
thus giving longer time for star formation in dynamically cold disks. 
However, this anti-correlation is not seen in other cosmological 
simulations that implement different feedback models. For example, the trigger 
of the kinetic mode of AGN feedback in the TNG simulation \citep[e.g.][]{weinbergerSupermassiveBlackHoles2018}
was tuned to depend on the mass of SMBH, so that the observed anti-correlation 
between halo formation time and star formation rate is suppressed. 
On the other hand, the EAGLE simulation has only one feedback channel throughout 
the entire halo assembly history \citet{schayeEAGLEProjectSimulating2015}, 
thus mixing the growth paths of galaxies in halos of different formation times. 
As cosmological simulations have not yet reached a consensus on how to model 
the AGN feedback, it is difficult to draw a firm conclusion as to exactly 
which mechanisms can produce the observed anti-correlation. 

Fig.\,\ref{fig:ms_to_mh_colored_m_bh} shows the $M_{\rm v}$-$M_*$ relation color-coded according to the 
value of $M_{\rm bh}$. As expected, $M_{\rm bh}$ increases with both 
$M_*$ and $M_{\rm v}$. However, inspecting the figure in detail reveals 
that $M_{\rm bh}$ is more closely related to $M_{\rm v}$ than to 
$M_*$ in the high-mass end, as shown by the nearly horizontal contours of constant 
$M_{\rm bh}$ at $M_{\rm v}>10^{12}h^{-1}{\rm M}_\odot$. At lower mass, 
$M_{\rm *}<10^{10.5}h^{-1}{\rm M}_\odot$, the contours are nearly vertical, 
indicating that $M_{\rm bh}$ is more closely related to stellar mass than to 
halo mass. These results are produced because the growth of the SMBH 
in high-mass halos is regulated by the AGN feedback efficiency controlled by 
$M_{\rm v}$ (or $V_{\rm max}$). Our test shows that, if the AGN feedback is turned off, 
by setting $\alpha_{\rm agn}=0$, the dominating dependence on 
$M_{\rm v}$ at the high-mass end disappears. When the regulation by AGN feedback 
is insignificant, the growths of both $M_{\rm bh}$ and $M_*$ are driven by the gas 
mass available for them, and the correlation of $M_{\rm bh}$ with $M_*$ becomes more dominating. 
This is seen at the low-mass end in the prediction of the High-$z_{\rm f}$ variant 
but almost absent in that of Low-$z_{\rm f}$. 
Using gravitational lensing measurements of halo mass, 
\citet{zhangHaloMassobservableProxy2023} found that the SMBH mass
depends strongly on $M_{\rm v}$ at $M_{\rm bh}>10^{7}h^{-1}{\rm M}_\odot$, 
while the dependence becomes much weaker at lower $M_{\rm bh}$. This is qualitatively 
in agreement with the prediction of the High-$z_{\rm f}$ variant. 
The discrepancy of the data with the Low-$z_{\rm f}$ variant may 
indicate that the growth of $M_{\rm bh}$ predicted by Low-$z_{\rm f}$, 
which is missing in High-$z_{\rm f}$ but seems to be needed to better match 
other observations, is not able to affect the gas effectively to achieve the regime 
of AGN regulation. The observed transition at $M_{\rm bh}~10^{7}h^{-1}{\rm M}_\odot$
may thus indicate a transition from the regime dominated 
by regulations of AGN feedback to a regime in which AGN feedback is
either no longer important in affecting the gas, or working in a way such that 
its effect is not limited by the potential of halo. Note that the extra growth of 
$M_{\rm bh}$ in Low-$z_{\rm f}$ relative to that in High-$z_{\rm f}$ occurred at relatively 
late time when the condition is already favorable for the formation of a disk-like 
structure and unfavorable for the AGN feedback to affect the gas in a 
self-regulated way. 

In Fig.~\ref{fig:ms_to_mh_high_z} we show the stellar mass - halo mass relation 
for galaxies at different redshift, predicted by our model assuming the 
High-$z_{\rm f}$ variant (left panels)
and Low-$z_{\rm f}$ variant (right panels). Results are shown separately for the 
total stellar mass ($M_*$, upper panels) and the mass of the dynamically hot (bulge) 
component ($M_{\rm *,bulge}$, lower panels). 
Galaxies in massive halos with $10^{13} \msun$ are dominated by the bulge component, 
regardless of the definition of $z_{\rm f}$. The stellar mass - halo mass 
relation is flattened and quite independent of $z$. 
This is the regime of ineffective cooling, which prevents star formation and thus 
`freezes' the stellar mass before these halos enter into the slow regime.
At $M_{\rm v} \lesssim 10^{12.5} \msun$, galaxies are dominated by the disk
component, particularly in the High-$z_{\rm f}$ variant.  The time evolution of 
the stellar mass - halo mass relation then begins to appear, driven by the 
halo-mass-dependent star formation rate of disk formation in the slow phase.
At $M_{\rm v} \lesssim 10^{11}\msun$,
the predicted stellar mass at a given halo mass is higher at higher $z$, 
particularly in the bulge mass. This evolution is driven by the halo-mass 
dependence of the transition time, and consequently, is stronger in the High-$z$ variant. 
These results about the evolution of the $M_*$ - $M_{\rm v}$ relation are 
in qualitative agreement with those obtained using methods of abundance 
matching \citep[e.g.][]{behrooziAVERAGESTARFORMATION2013, 
behrooziUniverseMachineCorrelationGalaxy2019}. 

Fig.~\ref{fig:sfrd} shows the cosmic star formation rate density, $\rho_{\rm SFR}$,
as a function of redshift, in comparison with data compiled by \citet{hopkinsNormalizationCosmicStar2006}
and \citet{zhangTrinitySelfConsistentlyModeling2022}. 
The two upper panels show the total $\rho_{\rm SFR}$ (black thin curve), 
as well as those separated into the bulge (red) and disk (blue) components. 
Results obtained by High-$z_{\rm f}$ and Low-$z_{\rm f}$ are shown in the 
left and right panels, respectively. To deal with the intrinsic inconsistency between the 
observed evolution of the stellar mass function and the cosmic star formation rate density, 
we follow \S3.5 of \citet{behrooziUniverseMachineCorrelationGalaxy2019} and scale the 
predicted `true' star formation rate, ${\rm SFR}_{\rm true}$, to obtain 
a scaled version, ${\rm SFR}_{\rm scaled}$, given by 
\begin{equation}
    \log\left(\frac{{\rm SFR}_{\rm scaled}}{{\rm SFR}_{\rm true}}\right) 
    = \mu(z) + \kappa \exp\left[ - \frac{(z-2)^2}{2} \right]  \,,
\end{equation}
where $\mu(z) = 0.041 - 0.044 z / (1+z)$ and $\kappa = 0.314$. The total scaled
$\rho_{\rm SFR}$ is shown by the thick black curve in the upper panels of Fig.~\ref{fig:sfrd}, 
and this curve should only be used when we compare our results with observational 
measurements on $\rho_{\rm SFR}$.
At $z \lesssim 1.5$, the two variants match the observed density equally well,
with only slight difference that falls within the uncertainty of the observational data.
The divergence emerges at $z \gtrsim 1.5$ and becomes the most significant at $z \approx 2$.
At this epoch, the High-$z_{\rm f}$ variant predicts a much higher 
star formation activity in disks than that in bulges, while the Low-$z_{\rm f}$ variant
has a small but non-negligible bulge contribution.
The observational data seem to favor the Low-$z_{\rm f}$ variant, similar to 
what we have seen in other statistical measures. At $z \gtrsim 3$, the star formation of 
the bulge component dominates the total $\rho_{\rm SFR}$ in the Low-$z_{\rm f}$ variant. 
As observations in such early times suffer from systematic effects, for example, 
in dust correction and sample completeness, it is difficult to obtain 
quantitative constraints on the model.

The lower panels of Fig.~\ref{fig:sfrd} shows the contribution to $\rho_{\rm SFR}$
by halos of different mass (left) and by galaxies of different stellar mass (right),
as function of redshift assuming the Low-$z_{\rm f}$ variant.
At $z \lesssim 3$, most of the cosmic $\rho_{\rm SFR}$ comes from halos with 
$M_{\rm v}$ in the range of $[10^{11.5},\,10^{12.5})\msun$, or galaxies 
with $M_*$ in the range of $[10^{10.0},\,10^{11.0})\msun$. This is exactly the 
population of halos (galaxies) that have peak efficiency in star formation 
(represented by $M_*/M_{\rm v}$) at $z=0$ 
\citep[][see also Fig.~\ref{fig:ms_to_mh_colored_z_f}]{yangEVOLUTIONGALAXYDARK2012}. 
At $z \gtrsim 3$, the contribution to $\rho_{\rm SFR}$ comes mainly from halos 
with $M_{\rm v}$ in the range of $[10^{10.5},\,10^{11.5})\msun$, or galaxies with $M_*$ in the range of 
$[10^{9.0},\,10^{10.0})\msun$. Using halo assembly histories shown in Fig.~\ref{fig:assembly-histories},
one can infer that these halos are exactly the progenitors of the halos 
that contribute the most to the cosmic $\rho_{\rm SFR}$ at $z \lesssim 3$. 
Thus, it is the same population of halos that drives the cosmic star formation
throughout the entire history of the Universe.

One interesting feature in the prediction of the Low-$z_{\rm f}$ variant,  
which is preferred by observational data, is the presence of a slowly evolving 
plateau of the total $\rho_{\rm SFR}$ at redshift $1 \lesssim z \lesssim 5 $. 
This plateau is produced by the combined contribution of two components. 
The first is peaked at $z\approx 5$ with a 
$\rho_{\rm SFR} \approx 10^{-1} h^3 {\rm M}_{\odot}{\rm yr}^{-1}{\rm Mpc}^{-3}$
(red curve in the upper right panel), and 
produced when a significant fraction of 
the $\rho_{\rm SFR}$ contributors enter the `feedback-free' regime  
for the bulge formation (see Fig.~\ref{fig:assembly-histories} and 
\S\ref{ssec:coevolution-equations}).
The second one is peaked at $z \approx 2$ with a 
$\rho_{\rm SFR} \approx 10^{-0.75} h^3 {\rm M}_{\odot}{\rm yr}^{-1}{\rm Mpc}^{-3}$
(blue curve in the upper right panel), 
and produced when these contributors switch to 
the slow assembly phase and begin to form dynamically cold disks.
The predicted peak at $z \approx 2$ has numerous observational supports 
\citep[e.g][]{leborgneCosmicStarformationHistory2009,
dunneStarFormationHistory2009,
kajisawaMOIRCSDeepSurvey2010,
karimStarFormationHistory2011,
cucciatiStarFormationRate2012,
sobralStellarMassFunction2014,
behrooziUniverseMachineCorrelationGalaxy2019}.
The peak around $z \approx 5$ is a prediction of our model and can be tested 
by ongoing and future observations (e.g. DESI, JWST, PFS and Roman). 
If confirmed, the concept of the cosmic noon may need to be re-defined, as 
the star formation activity of the universe is not concentrated around a well-defined 
period around $z \sim 2$, but rather has a more extended distribution in time.

\begin{center}
\begin{table*} 
\caption{List of model components, their functional forms and the default parameters
adopted in this paper for the co-evolution of different mass components. See \S\ref{ssec:coevolution-equations}
for the detailed definitions, Fig.~\ref{fig:flow-chart} for a schematic diagram
and Appendix~\ref{app:adjustment-model-parameters} for the strategy of adjusting
model parameters.
}
\begin{tabularx}{\textwidth}{c | X | X }
    \hline
        {\bf Model Component} 		 
        &
        {\bf Functional}
        &
        {\bf Default Parameter}
        \\
    \hline
    \hline 
        Phase transition
        (eqs.~\ref{eq:def-high-zf}, \ref{eq:def-low-zf})
        &
        $\gamma (z_{\rm f}) = \gamma_{\rm f}$
        &
        $\gamma_{\rm f}=3/8$ (High-$z_{\rm f}$ variant); \newline
        $\gamma_{\rm f}=0$ (Low-$z_{\rm f}$ variant)
        \\
    \hline
        Black hole seeding
        (eq.~\ref{eq:init-bh})
        &
        $M_{\rm bh, init} ( M_{\rm v} \vert 
            M_{\rm bh,\,min}, m_{\rm bh,\,scale} )$ at $M_{\rm v,min}$
        &
        $M_{\rm v,min}=10^9\msun$; \newline 
        $M_{\rm bh,min}=10 \msun$; $m_{\rm bh,scale}=10^{-10}$
        \\
    \hline
        Gas cooling
        (eq.~\ref{eq:def-f-cool})
        &
        $F_{\rm cool} (M_{\rm v}\vert M_{\rm cool},\beta_{\rm cool})$
        &
        $M_{\rm cool} = 10^{13} \msun$; $\beta_{\rm cool}=4$
        \\
    \hline
        Feedbacks
        (eqs.~\ref{eq:def-f-sn}, \ref{eq:def-f-agn}, \ref{eq:def-f-en})
        &
        $F_{\rm sn}(V_{\rm g}\vert \alpha_{\rm sn}, \beta_{\rm sn}, V_{\rm w})$; \newline
        $F_{\rm agn} (M_{\rm bh}, M_{\rm g}, V_{\rm g}\vert \alpha_{\rm agn})$
        &
        $\alpha_{\rm sn} = 0$; $\beta_{\rm sn}=2.5$; $V_{\rm w}=250 \kms$; \newline
        $\alpha_{\rm agn} = 10^{-3}$
        \\
    \hline
        Star formation and SMBH growth
        (eqs.~\ref{eq:delta-m-star}, \ref{eq:delta-m-bh})
        &
        $\Delta M_* (\Delta M_{\rm g, sf} \vert \epsilon_{\rm *, f})$; \newline
        $\Delta M_{\rm bh}(\Delta M_{\rm g, sf} \vert \alpha_{\rm cap}, F_{\rm en})$; \newline 
        $F_{\rm en} (M_{\rm v}\vert \alpha_{\rm en}, \beta_{\rm en}, M_{\rm en})$
        &
        $\epsilon_{\rm *,f}=0.75$; \newline 
        $\alpha_{\rm cap}=2.5$; \newline
        $\alpha_{\rm en} = 3$; $\beta_{\rm en}=2$; $M_{\rm en}=10^{11.5} \msun$
        \\
    \hline
        Gas evolution
        (eq.~\ref{eq:delta-m-gas-ej})
        &
        $\Delta M_{\rm g, ej}(\Delta M_{\rm g, cool} 
            \vert f_{\rm ej, sn}, f_{\rm ej, agn}, F_{\rm sn}, F_{\rm agn} )$
        &
        $f_{\rm ej, sn}=f_{\rm ej, agn}=0.75$
        \\
    \hline
\end{tabularx}
\label{tab:parameters}
\end{table*}
\end{center}

\section{Summary and Discussion}
\label{sec:summary}

In this paper, we have developed a two-phase framework to understand galaxy formation and
the growth of central supermassive black holes (SMBHs) in dark matter halos predicted 
by current cosmology. The main components of the framework and the predictions 
of the model are summarized below.
\begin{enumerate}
\item 
The framework uses the fact that the assembly of a cold dark matter (CDM) halo
in general consists of two phases, a fast phase in which its gravitational potential well
changes and deepens rapidly with time, and a slow phase in which the halo mass increases 
gently without changing much the gravitational potential. We used simulated halo assembly 
histories to design a critical redshift that separates the two phases of a given history 
and to quantify uncertainties in the separation.    
\item 
Our modeling also uses the fact that the universal baryon fraction is high, 
so that cooled gas in a primordial, unprocessed halo becomes self-gravitating before 
it can form a rotation-supported 
disk. The gas associated with the fast assembly is thus expected to form   
a self-gravitating cloud (SGC) that is inherently turbulent because of the fast variation 
of the gravitational potential.  
\item 
The density and size of an SGC in a halo can be estimated from the assembly history of the halo 
and the cold gas fraction. Jeans instability causes the SGC to fragment and form sub-clouds with 
a typical mass of $\sim 10^7\,{\rm M}_\odot$.
\item 
Sub-clouds in SGC are sufficiently compact so that cloud-cloud collision and drag
force on clouds can be neglected. An SGC is thus a dynamically hot system of 
sub-clouds that form stars and move ballistically to feed the central SMBH. 
\item 
Under the assumption of random ballistic motion of sub-clouds, we estimated the 
mass accretion rate of an existing SMBH. We found that such an accretion can 
achieve the Eddington limit, thus providing a viable channel for the growth of SMBH.    
\item 
We assumed that AGN and supernova feedbacks are effective only in the fast phase, 
where sub-clouds have a more isotropic distribution. The feedback efficiency
is also enhanced because individual sub-clouds can be dispersed by supernova (SN) 
feedback associated with the star formation in them.      
\item 
The cumulative effects of AGN and SN feedback in an SGC are twofold: 
regulating star formation and SMBH growth, while simultaneously reducing the 
amount of cold gas within the halo to facilitate the formation of globally 
stable disks. Consequently, the feedback mechanism manifests itself as `ejective' during 
the fast assembly phase and transitions to a `preventive' mode when galactic 
disks begin to form during the slow assembly phase.
\item 
Quenching of star formation through feedback is only effective in dynamically hot 
systems form during the fast assembly phase, but not effective in disk-dominated 
systems that form later in the slow assembly phase.  
\item 
We have applied our model to a set of realistic halo assembly histories to make 
predictions and make comparisons with observational data. 
\item 
We found that an early phase of enhancement in SMBH accretion is needed 
to grow SMBH if the seeding mass is small. We suggested that the SN feedback 
in low-mass progenitor halos, where cooling is very effective,
can quickly re-generate sub-clouds in the turbulent medium to enhance 
the accretion rate of early SMBH.  
\item 
The predicted correlations of SMBH mass ($M_{\rm bh}$)
with the stellar mass of central galaxies ($M_*$) and with the host halo mass 
($M_{\rm v}$) match well observational results. The predicted 
$M_{\rm bh}$-$M_*$ relation for dynamically hot galaxies in the massive end 
follows roughly a power law, $M_{\rm bh} \propto M_*^{5/3}$, expected 
from the regulation by AGN feedback. The predicted relation deviates 
from this relation for galaxies with $M_*<10^{10}{\rm M}_\odot$ because they 
have not reached the regime of the AGN regulation.  
For a similar reason, $M_{\rm bh}$ is predicted to correlate more tightly 
with $M_{\rm v}$ than with $M_*$ in the massive end, 
$M_{\rm bh}>10^7 h^{-1}{\rm M}_\odot$. At lower $M_{\rm bh}$ the stellar  
mass of galaxies becomes the more important factor in producing the correlation,  
as the regulation by the AGN feedback becomes ineffective in the more disk-like 
structure expected for these galaxies. 
\item 
The predicted $M_{\rm bh}$-$M_*$ relation for dynamically hot galaxies is quite 
independent of redshift. The relation steepens significantly with decreasing 
redshift if $M_*$ contains both dynamically hot (bulge) and cold (disk) masses.   
The predicted $M_{\rm bh}$ - $M_{\rm v}$ relation also steepens with decreasing 
redshift, but only mildly. 
\item 
The predicted cosmic SMBH mass density as a function of $z$ matches observational 
results, but the predicted density of mass accretion rate by SMBH at $z<1$     
is lower than that observed. This suggests that other channels of SMBH growth, 
such as those associated with the formation of pseudo bulges, and recent mergers of 
disk galaxies, are needed at low $z$. 
\item 
The stellar mass - halo mass relation predicted by our model is in good agreement  
with those obtained from observational data. Our model predicts that, 
at a given mass, older halos (those with higher formation redshift) tend to host   
galaxies that are more disk-dominated. Similarly, the model predicts 
that, at a given stellar mass, disk-dominated galaxies on average 
have lower halo mass than bulge-dominated ones. This prediction naturally 
emerges as a consequence of the earlier transition of more massive halos, 
and it appears in good agreement with results obtained from galaxy-galaxy 
gravitational lensing. 
\item 
Redshift evolution of the stellar mass - halo mass relation predicted by our model 
is in qualitative agreement with those obtained from abundance matching.
The predicted cosmic star formation rate density as a function of $z$ matches 
observational data at $z \leqslant 5$, and has a slowly evolving plateau
between $ 1 \lesssim z \lesssim $ due to the superposition of a bulge-forming 
peak and a disk-forming peak.

\end{enumerate}

The presentation of our framework given above is based on a set 
of simple assumptions. This makes our model transparent so that 
we can see clearly how different assumptions affect the model 
prediction. We purposely avoid fine-tuning model parameters, 
so that one can see the success and failure of the model faithfully.   
Clearly, many assumptions made in our model need to be checked by 
looking into how the relevant processes operate in realistic settings. The
best approach is perhaps through the use of zoom-in simulations of 
individual halos that can resolve star-forming clouds with masses 
below $10^7{\rm M}_\odot$. The critical things to check are: 
(i) the properties of the turbulent medium associated with the 
fast assembly phase to see to what extent the SGC envisaged in our model 
can be produced;
(ii) the formation, disruption and motion of sub-clouds in the SGC to check 
to what extent the assumption of ballistic motion is valid, and whether 
or not sub-clouds can be produced at sufficient rates to sustain star formation 
and SMBH growth; 
(iii) the rate at which sub-clouds can reach the halo center to be accreted by an 
existing SMBH; 
(iv) the efficiency at which central AGN and SN feedback disperses, heats and 
ejects gas in a dynamically hot SGC of star-forming sub-clouds;  
(v) to what extent the feedback from the SGC can preheat and pre-process 
the halo gas so as to affect the subsequent assembly of gas to form galactic disks. 
Current state-of-the-art computational facilities and numerical methods are powerful 
enough to simulate these processes. The model and results presented in this paper can thus 
be used to motivate, and identify cases for, such simulations.  

One direct consequence of our two-phase model is the inside-out quenching of galaxies, 
as central bulge stars in our model form earlier than the surrounding disk stars.
Indeed, this inside-out trend has been found in observational data. 
For example, using MaNGA data \citet{linSDSSIVMaNGAInsideout2019} found that the inside-out 
quenching is the dominant mode in all environments, and that the signal is stronger in 
more massive galaxies, regardless of environment. They invoked morphological quenching to 
explain their findings. Similar results have also been found by 
\citet{liPMaNGAGradientsRecent2015} using the P-MaNGA data, and by
\cite{nelsonSpatiallyResolvedStar2021} using 3D-HST data.
Such inside-out quenching is expected in our model by the fact that more massive systems 
have later transition times and hence longer time for the bulge component to grow.

Our model also suggests a scenario in which AGN feedback is not expected to 
be significant in affecting the gas (or the feedback is limited to a small fraction 
of the halo gas) in the slow assembly phase. This is consistent with the observational 
results for galaxies at $z \leqslant 0.35$ obtained by 
\citet{zhuangEvolutionaryPathsActive2023}, who concluded that instantaneous 
feedback from accreting SMBH in quenching star formation may not be efficient. 
However, this scenario seems in contradiction with recent cosmological 
simulations \citep{weinbergerSupermassiveBlackHoles2018, daveSimbaCosmologicalSimulations2019}, 
and so the question remains open.

The plateau feature of the cosmic star formation history predicted 
by our model indicates strong star formation activity at $z \sim 5$ in the self-gravitating SGC. 
Recent JWST observations revealed a population of massive galaxies at $z \gtrsim 5$ with significant 
ongoing star formation activities \citep{xiaoMassiveOpticallyDark2023}, which is in 
qualitative agreement with our model prediction. Further information from 
high-redshift observations will be needed to give quantitative constraints on the model.

\section*{Acknowledgements}

YC is funded by the China Postdoctoral Science Foundation (grant No. 2022TQ0329).
HJM thanks T.D. Lee Institute and Shanghai Jiaotong University for hosting his 
sabbatical during which the work was done.
This work is also supported by the National Natural Science Foundation of China 
(NSFC, Nos. 12192224, 11733004 and 11890693) and CAS Project for Young Scientists 
in Basic Research (grant No. YSBR-062).
YC thanks Kai Wang, Hui Hong, Hao Li, Yu Rong, Enci Wang, Wentao Luo and 
Fangzhou Jiang for their valuable insights and discussions. 
The authors would like to express their gratitude to the Tsinghua Astrophysics 
High-Performance Computing platform at Tsinghua University and the 
Supercomputer Center of the University of Science and Technology of China for 
providing the necessary computational and data storage resources that have 
significantly contributed to the research results presented in this paper.
The authors thank \citet{behrooziUniverseMachineCorrelationGalaxy2019}, 
\citet{zhangTrinitySelfConsistentlyModeling2022} and \citet{wuDemographicsQuasarsBlack2022} 
for compiling observational data and making them publicly available. 
The computations and presentations in this paper are supported by various software 
tools, including the HPC toolkits 
\softwarenamestyle[Hipp] \citep{chenHIPPHIghPerformancePackage2023}\footnote{\url{https://github.com/ChenYangyao/hipp}}
and \softwarenamestyle[PyHipp]\footnote{\url{https://github.com/ChenYangyao/pyhipp}},
interactive computation environment 
\softwarenamestyle[IPython] \citep{perezIPythonSystemInteractive2007},
numerical libraries \softwarenamestyle[NumPy] \citep{harrisArrayProgrammingNumPy2020}, 
\softwarenamestyle[Astropy] \citep{
robitailleAstropyCommunityPython2013,
astropycollaborationAstropyProjectBuilding2018,
astropycollaborationAstropyProjectSustaining2022},
\softwarenamestyle[SciPy] \citep{virtanenSciPyFundamentalAlgorithms2020}
and \softwarenamestyle[Pytreegrav]\footnote{\url{https://github.com/mikegrudic/pytreegrav}},
as well as the graphical library 
\softwarenamestyle[Matplotlib] \citep{hunterMatplotlib2DGraphics2007}. 
This research has made extensive use of the arXiv and NASA’s Astrophysics Data System.
Data compilations used in this paper have been made much more accurate and 
efficient by the software \softwarenamestyle[WebPlotDigitizer]. 

\section*{Data Availability}
\label{sec:data-availability}

We provide a code repository, 
\softwarenamestyle[TwoPhaseGalaxyModel]\footnote{\url{https://github.com/ChenYangyao/two-phase-galaxy-model}},
that implements the model described in this paper. The library is written in Python
and can be applied to halo assembly histories from different sources, such as 
N-body simulations or Monte Carlo methods.
All data used in this paper, including halo assembly histories,
modeled galaxies and SMBHs, data points displayed in figures, 
and observational results used for calibration and testing, 
will be distributed along with the repository.



\bibliographystyle{mnras}
\bibliography{references} 




\appendix

\section{The Methods of Sampling Halos and Their Assembly Histories}
\label{app:method-halo-sampling}

\begin{figure}\centering
    \includegraphics[width=0.9\columnwidth]{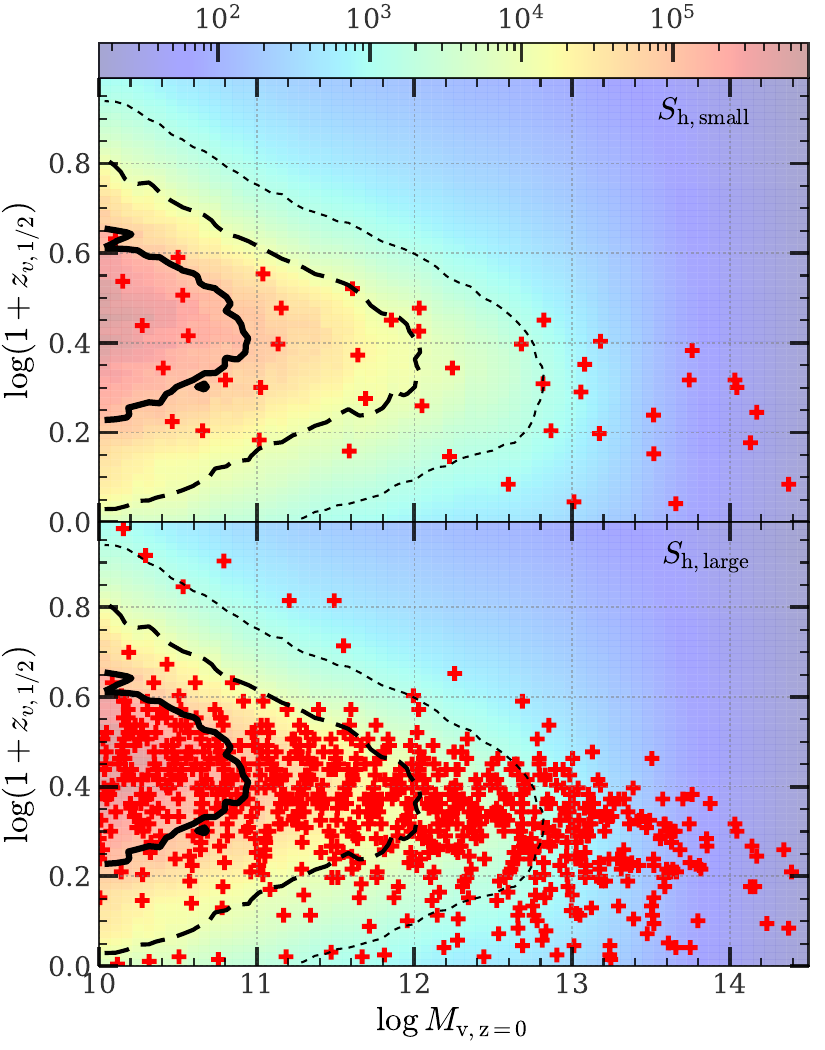}
    \caption{Halo samples extracted from N-body simulation. Each {\bf plus marker}
    indicates an individual halo at $z=0$.
    {\bf Top:} the small sample, $S_{\rm h, small}$, which consists of 45 halos
    with halo mass in the range of $[10^{10.0}, 10^{14.5})\msun$. {\bf Bottom:}
    the large sample $S_{\rm h, large}$, which consists of 720 halos in the
    same mass range.
    Background colors, encoded according to the color bar, indicate the 
    $M_{\rm v,z=0}$ - $z_{\rm v, 1/2}$ distribution 
    of all halos at $z=0$ in the simulation.
    }
    \label{fig:sample-space}
\end{figure}

In \S\ref{ssec:halo-sample}, we briefly introduced the halo samples used in this
paper. In this Appendix, we detail the techniques used to extract halos and their 
assembly histories from either N-body simulations of analytical fittings.

The main samples used in this paper are obtained from the IllustrisTNG100-1-Dark simulation.
The simulation has a simulation box with a side length $75 \mpc$, $1820^3$ 
dark matter particles, each with a mass $6.0\times 10^6 \msun$,
a Plummer equivalent gravitational softening length varying from $1 \kpc$
at high $z$ to $0.5 \kpc$ at low $z$. A total of 100 
snapshots spanning from redshift $z=20.0$ to $0$ have been saved. Halos are 
identified using the friends-of-friends (\softwarenamestyle[FoF]) algorithm
with a scaled linking length of $0.2$ \citep{davisEvolutionLargescaleStructure1985}. 
Subhalos are identified using the \softwarenamestyle[Subfind] 
algorithm \citep{springelPopulatingClusterGalaxies2001,
dolagSubstructuresHydrodynamicalCluster2009}, and subhalo merger trees are 
constructed using the \softwarenamestyle[SubLink]
algorithm \citep{springelCosmologicalSimulationCode2005,
boylan-kolchinResolvingCosmicStructure2009, rodriguez-gomezMergerRateGalaxies2015}. 
The lower limit for FoF halo mass is about $2\times 10^8 \msun$. 
The main progenitor of a subhalo is defined as the one with the most massive 
history among all progenitors \citep{deluciaHierarchicalFormationBrightest2007,
rodriguez-gomezMergerRateGalaxies2015}. The central subhalo of a FoF halo 
is defined as the one with the most massive history among all subhalos within the halo. 
The main branch of an FoF halo corresponds to the main branch of its central subhalo. 
The main information of halo assembly history is contained within the main branch 
\citep[e.g.][]{chenRelatingStructureDark2020}, which we use throughout 
this paper to model the properties of galaxies and their SMBHs.

To enhance the computational efficiency of our model adjustments, we use a 
subsampling technique on the TNGDark halos. This procedure tries to uniformly 
sample the distribution function $p[\mathcal{P}(z_{v, 1/2}) \vert M_{\rm v}]$
by the following steps. First, we select all FoF halos at the desired redshift, denoted as 
$z_{\rm d}$, with a halo mass satisfying 
$M_{\rm v, min} \leqslant M_{\rm v} < M_{\rm v, max}$. 
We partition this range into $n_{\rm bin}$ logarithmically spaced bins 
of equal width. Next, for each halo residing in the $i$-th bin, we trace its 
main branch towards higher redshifts until the halo mass $M_{\rm v}(z)$ drops 
below $M_{\rm v}(z_{\rm d})/2$, thus determining its half-mass formation 
redshift, $z_{\rm v, 1/2}$. We then estimate the cumulative distribution function, 
denoted as $\mathcal{P}_i(z_{\rm v,1/2})$, which characterizes the formation times 
of halos in this particular bin. By enforcing that the extracted halos in this 
bin have equidistant $\mathcal{P}_i(z_{\rm v,1/2})$ values, we select a specific 
number, ${\rm min}(n_{\rm h}, n_i)$, of halos, where $n_i$ represents 
the total number of halos in this bin. Finally, we gather the extracted halos 
from all bins to give the subsample. This approach ensures that our sample 
accurately represents the overall population of halos at redshift $z_{\rm d}$.

In this paper, we use two subsamples of halos from TNGDark at $z_{\rm d} = 0$. 
These subsamples serve different purposes and are defined as follows:
\begin{itemize}
    \item The `small sample' $S_{\rm h, small}$: This subsample consists of halos with a 
    mass range of $(M_{\rm v, min}, M_{\rm v, max})=(10^{10}, 10^{14.5})\msun$. 
    It is further divided into $(n_{\rm bin}, n_{\rm h})=(9, 5)$ bins, resulting in 
    a total of $n_{\rm h,total}=45$ halos. The primary goal of this small sample 
    is to illustrate the assembly histories of individual halos.

    \item The `large sample' $S_{\rm h, large}$: This subsample consists of halos with a mass 
    range of $(M_{\rm v, min}, M_{\rm v, max})=(10^{10}, 10^{14.5})\msun$. It is divided 
    into $(n_{\rm bin}, n_{\rm h})=(18, 50)$ bins, resulting in a total of 
    $n_{\rm h,total}=720$ halos. The large sample is primarily used for statistical 
    analysis. To account for the effect of subsampling in each bin, we assign a 
    weight, $w_i = n_i / {\rm min}(n_{\rm h}, n_i)$, to each halo in the $i$-th bin. 
    This weight is used in computing summary statistics.
\end{itemize}
Figure~\ref{fig:sample-space} shows the distribution of halos in the two samples, 
in comparison with the overall population of halos in the simulation at $z=0$. 
The small sample exhibits a sparse occupation of the halo mass - formation time 
space and misses halos in the extreme tail of the $\mathcal{P}_i(z_{\rm v,1/2})$ distribution. 
In contrast, the large sample provides a finer and more complete coverage, 
following closely with the overall population. It is important to note that both samples 
are biased towards the higher end of the halo mass relative to the underlying distribution 
of all halos. This deliberate choice ensures the inclusion of a sufficient number 
of massive halos, thereby covering their complex assembly histories.

Since observations at high redshift are limited to extremely bright and rare objects, 
it is difficult to obtain robust statistics from above samples when comparing 
with high-z observational results such as black hole mass functions (BHMFs). 
Meanwhile, to derive quantities that rely on small BHs, such as the cosmic BH 
mass densities (CBHMDs) at any given redshifts, the above sample is incomplete 
in the low-BH-mass end ($\lesssim 10^7\msun$).
To bypass these issues, we use analytical fittings to generate halo samples 
whose mass distribution and assembly histories are representative and complete
of BH populations. Specifically, we use the \softwarenamestyle[Hmf] library 
\citep{murrayHMFHaloMass2014} to generate a halo mass function, $\Phi_{\rm c}(M_{\rm v})$, for 
$M_{\rm v, min} \leqslant M_{\rm v} < M_{\rm v, max}$ at the desired redshift 
$z_{\rm d}$. We then sample a total of $n_{\rm h, total}$ halos at 
$z_{\rm d}$, with masses equally spaced between $M_{\rm v, min}$ and $M_{\rm v, max}$, 
and assign a weight $w_{\rm c}$ to each halo by
\begin{equation}
    w_{\rm c} \equiv \frac{\Phi_{\rm c}(M_{\rm v})}{\Phi_{\rm samp}(M_{\rm v})},
\end{equation}
where $\Phi_{\rm c}$ is the halo mass function at $z_{\rm d}$, and $\Phi_{\rm samp}$
is the also mass function evaluated from our sample.
We use the \softwarenamestyle[Diffmah] library \citep{hearinDifferentiableModelAssembly2021} 
to randomly sample a mass assembly history at $z \geqslant z_{\rm d}$ for each halo.
Since our model in the fast phase requires halo $V_{\rm max}$ as input, 
we assume an ideal NFW profile with a concentration parameter 
$c = c_{\rm f} \approx 4$ for each halo throughout the fast phase history, 
We obtain the required halo properties, such as the maximal circular velocity,
$V_{\rm max}$, the halo-centric distance, $r_{\rm vmax}$, 
to achieve this velocity, and the total mass, $M_{\rm vmax}$, enclosed in this radius, 
by 
\citep[e.g.][]{bullockProfilesDarkHaloes2001,klypinResolvingStructureCold2001}:
\begin{align}
    M_{\rm vmax }&= 0.468 \frac{M_{\rm v}}{\mu (c)}; \\
    r_{\rm vmax} &= \frac{2.163 r_{\rm v}}{c}; \\
    V_{\rm max} &= V_{\rm v}\sqrt{\frac{0.216 c}{\mu(c)}}.
\end{align}

With the above semi-analytical procedure, we can define a sample at any desired redshift
by specifying $(z_{\rm d}, n_{\rm h,total}, M_{\rm v,min}, M_{\rm v,max})$.
This sample will be used to derive summary statistics beyond the scope of cosmological
simulations, such as cosmic BH number density, mass density and growth rate density
at $z_{\rm d}$.
Because BHs in satellite galaxies also contribute these quantities 
\citep[see e.g.][]{zhangTrinitySelfConsistentlyModeling2022}, 
for each halo with mass $M_{\rm v}$ at $z_{\rm d}$, we define another 
weight:
\begin{equation}
    w_{\rm s} \equiv 1 + \frac{\Phi_{\rm s}(M_{\rm v})}{\Phi_{\rm c}(M_{\rm v})},
\end{equation}
where $\Phi_{\rm s}$ is the infall mass function of satellite subhalo, 
and $\Phi_{\rm c}$ is the halo mass function, both evaluated at $z_{\rm d}$
from TNGDark and extrapolated to the mass below the resolution limit. The 
total weight, $w = w_{\rm c} w_{\rm s}$, is the final value used in 
computing summary statistics. Note that satellite galaxies do not follow
exactly the same evolution as central galaxies, due to, 
for example, additional environment processes. Thus, our inclusion of satellites
only provides a rough estimate of their contribution, which is small enough
compared to large uncertainties in observations (see, e.g. Fig.~\ref{fig:bhmd}).

\section{Relation of Transition Redshift and Half-mass Formation Redshift}
\label{app:transition-redshift}

\begin{figure*} \centering
    \includegraphics[width=0.8\textwidth]{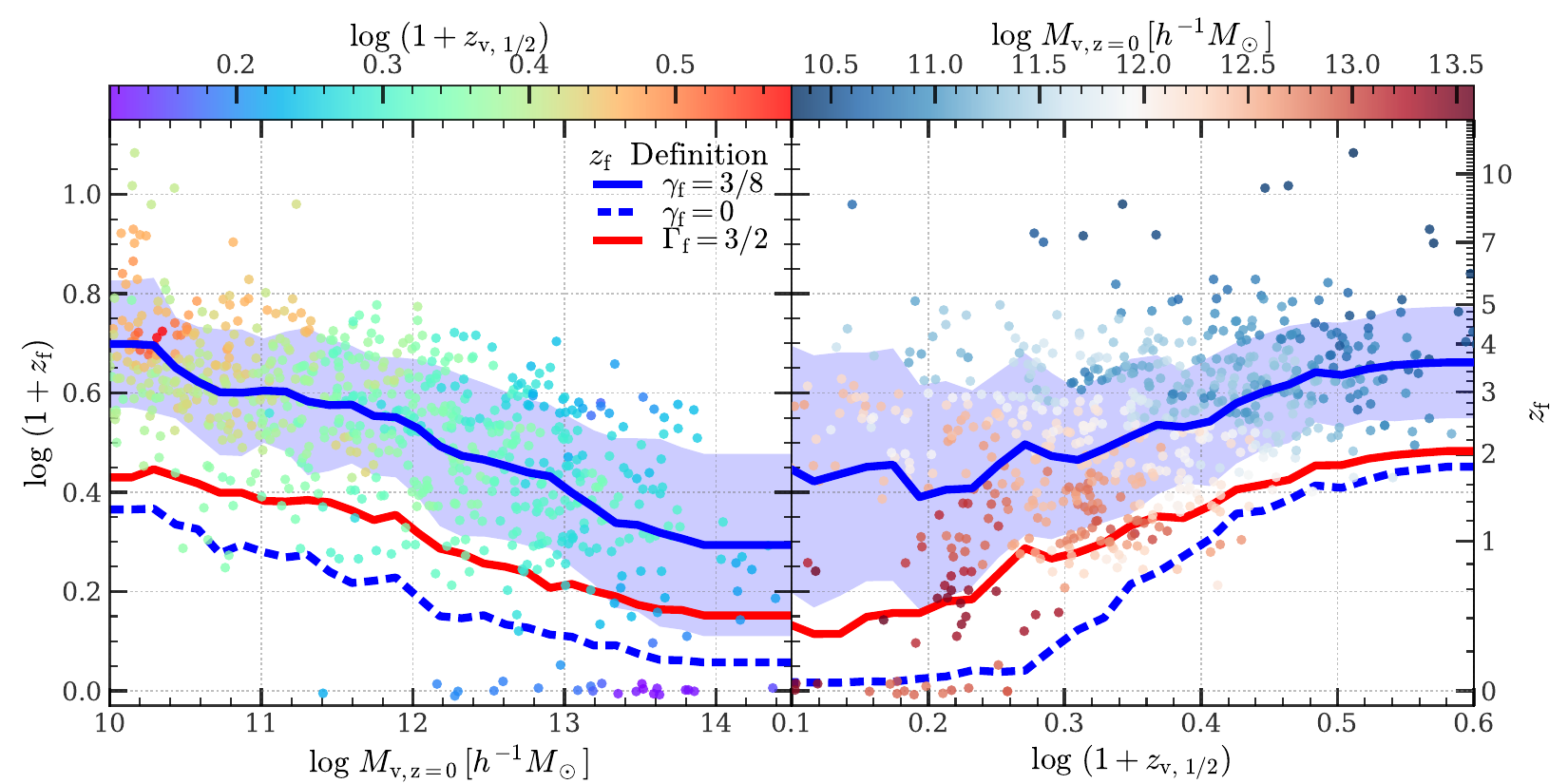}
    \caption{
        The mean transition redshift ($z_{\rm f}$) as a function of halo mass 
        ($M_{\rm v} \equiv M_{\rm 200c}$) at $z=0$ ({\bf left panel}) and 
        half-mass formation redshift ($z_{\rm v,1/2}$, {\bf right panel}). In each panel, three definitions 
        of $z_{\rm f}$ are shown by three curves, respectively, as indicated
        by the legends. For the $\gamma_{\rm f}=3/8$ (High-$z_{\rm f}$) case, 
        {\bf scatter points} represent individual halos, while the {\bf blue shading} 
        indicates the standard deviations. Colors of points are encoded according
        to the colorbar. Halo sample $S_{\rm h,large}$ is used to produce 
        this figure. See \S\ref{sec:halos} for the details of halo transition time. 
    }
    \label{fig:transition_points_vs_mass}
\end{figure*}

The transition redshift, $z_{\rm f}$, is not commonly used in the 
existing literature for the construction of halo-based galaxy models. 
In order to gain a overall feeling of the distribution of 
$z_{\rm f}$, we present in Fig.~\ref{fig:transition_points_vs_mass} 
its correlation with the halo mass, $M_{\rm v}$, at $z=0$, as well as the 
half-mass formation redshift, $z_{\rm v,1/2}$, which is more 
frequently used in halo-based models.

The three definitions of $z_{\rm f}$ yield qualitatively similar results: 
$z_{\rm f}$ consistently decreases with increasing $M_{\rm v}$ and rises 
monotonically with $z_{\rm v,1/2}$. The definition of $z_{\rm f}$ based on 
$\Gamma=3/2$ consistently falls between the other two definitions. 
Consequently, we use the two model variants, High-$z_{\rm f}$ and Low-$z_{\rm f}$, to bracket the systematic uncertainty in the definition of 
$z_{\rm f}$. The relationship between $z_{\rm f}$ and $z_{\rm v,1/2}$ 
conditioned on $M_{\rm v, z=0}$, and vice versa, is shown by the colors of the 
scattered points in the two panels of the figure for the High-$z_{\rm f}$ case. 
At a fixed conditioning variable, the conditional trend is weaker than the 
global trend, but remains statistically significant. 
This suggests that $z_{\rm f}$ provides supplementary information to halo 
assembly that is not entirely captured by $z_{\rm v,1/2}$. 
Given that the `1/2' in the definition of $z_{\rm v,1/2}$ is arbitrary, 
we opt to use $z_{\rm f}$ in this paper to separate the different phases of 
halo assembly.

\section{Fluctuations of Halo Assembly History}
\label{app:fluctuation-mah}

\begin{figure} \centering
    \includegraphics[width=0.95\columnwidth]{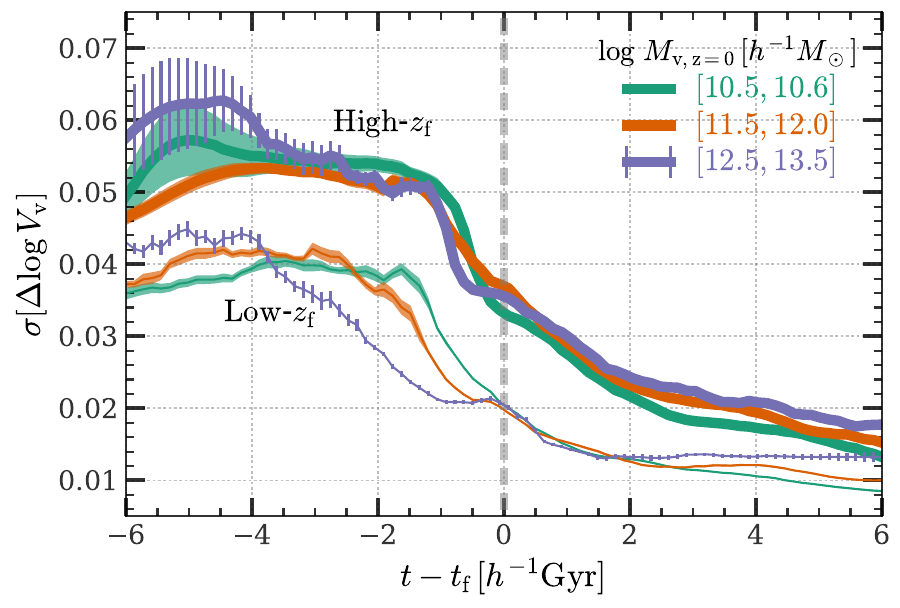}
    \caption{
        Fluctuation of halo assembly history, 
        defined as the standard deviation of 
        $\Delta \log\,V_{\rm v} = \log\,V_{\rm v} - \log\,V_{\rm v}^\text{(fit)}$ 
        at given $t - t_{\rm f}$, where 
        $V_{\rm v}^\text{(fit)}$ is obtained by a parametric fitting described 
        in \S\ref{sec:halos}, and $t_{\rm f} \equiv t(z_{\rm f})$ is the 
        cosmic time at the transition redshift $z_{\rm f}$. 
        Results for halos with different 
        mass are shown by {\bf different colors}. {\bf Thick} and {\bf thin}
        curves are for the $\gamma_{\rm f}=3/8$ (High-$z_{\rm f}$)
        and $\gamma_{\rm f}=0$ (Low-$z_{\rm f}$) variants, respectively.
        Error bars or shadings represent the standard deviations obtained 
        by 20 bootstrap resamplings. This figure shows that the gravitational 
        potential of halo drops significantly near the transition, 
        and appears stabilized after that.
    }
    \label{fig:residual_history}
\end{figure}

\begin{figure} \centering
    \includegraphics[width=0.95\columnwidth]{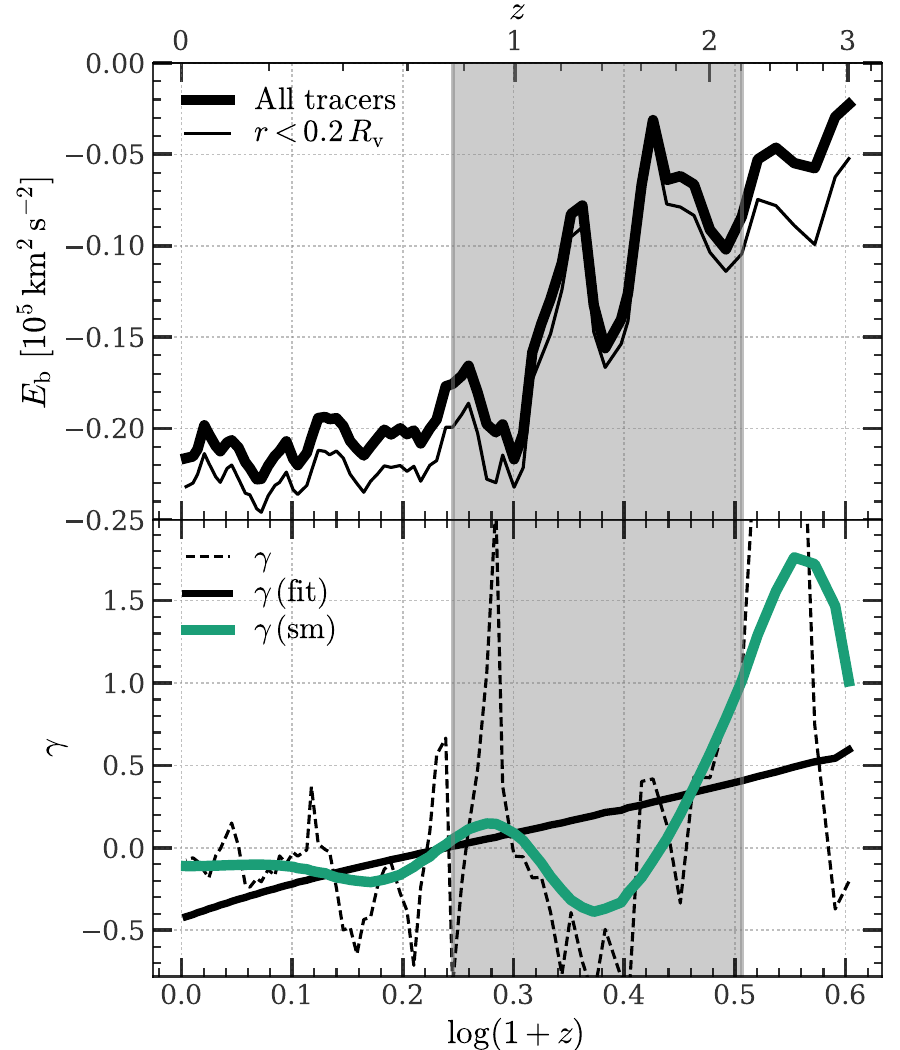}
    \caption{
    {\bf Upper panel} shows the average specific binding energy of dark matter 
    particles as a function of redshift in the main branch of an example halo
    with $M_{{\rm v},z=0} \approx 10^{12}\msun$. 
    {\bf Thick} curve is obtained by using all particles 
    bound to the central subhalo in the range of redshift shown here, while 
    {\bf thin} curve is obtained by only using those within $0.2 R_{\rm v}$.
    {\bf Lower panel} shows the specific growth rate $\gamma$ 
    (defined by Eq.~\ref{eq:def_gamma}) of the same halo. 
    {\bf Dashed}, {\bf solid black} and {\bf green} curves show
    the results obtained from the simulated assembly history, the fitting by a 
    parametric function, and the smoothing by a kernel of size $\tau_{\rm dyn}$ 
    (see Fig.~\ref{fig:transition_points} and \S\ref{sec:halos} for the details).
    {\bf Grey} shaded area indicates the transition redshift bracketed by
    the two variants, Low-$z_{\rm f}$ and High-$z_{\rm f}$.
    This figure suggests that the binding energies of particles respond well 
    to $\gamma$, and provides support to our two-phase separation of galaxy 
    formation by a transition of halo assembly.
    }
    \label{fig:e_bind_vs_z}
\end{figure}

\begin{figure*} \centering
    \includegraphics[width=0.95\textwidth]{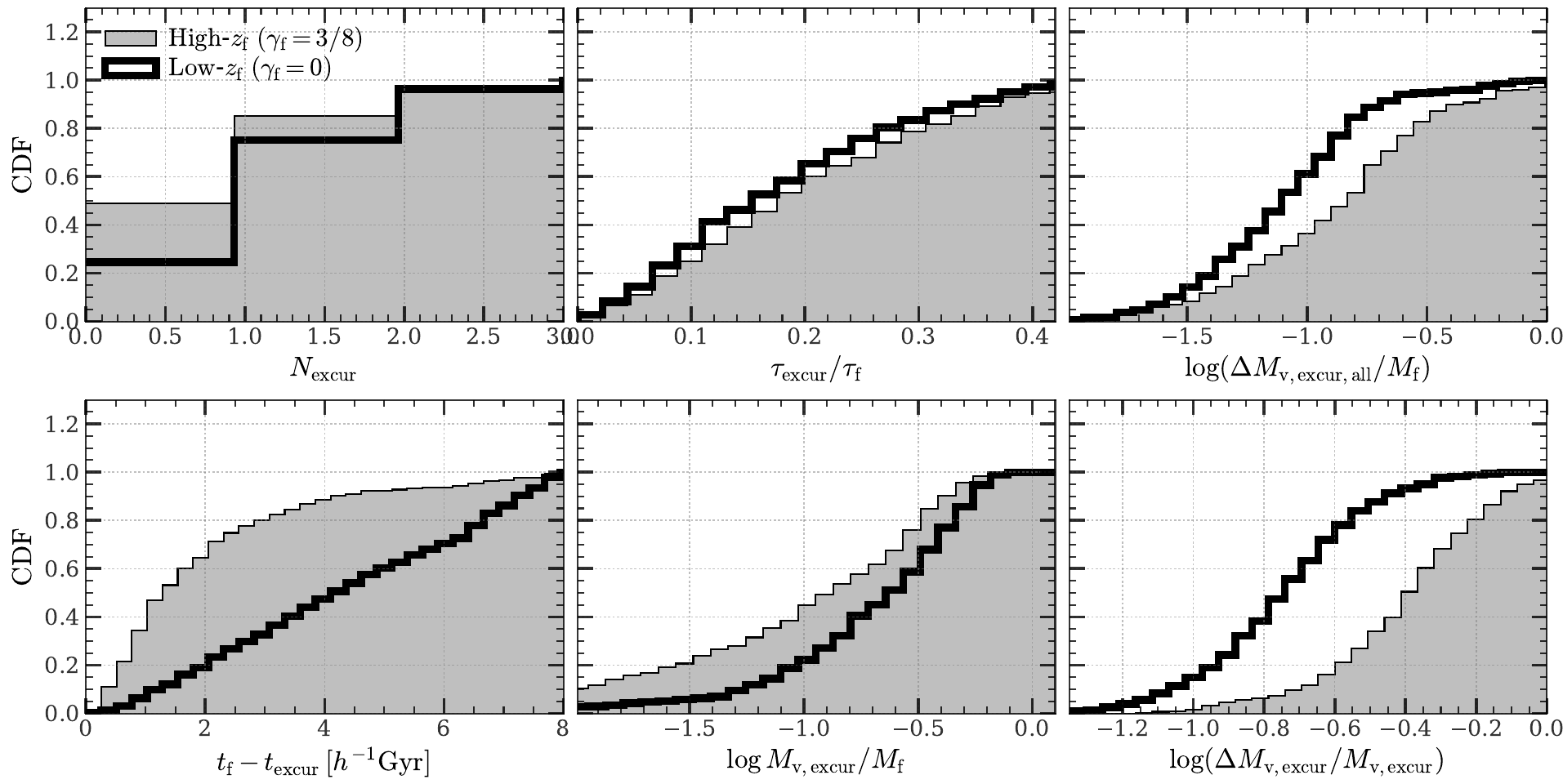}
    \caption{
    The cumulative distribution of quantities describing the excursion of 
    halo assembly to the slow regime ($\gamma^\text{(sm)} < \gamma_{\rm f}$) during
    the fast phase ($z \geqslant z_{\rm f}$).
    The {\bf first row} shows the statistics for main branches,
    while the {\bf second row} shows the statistics for excursions.
    $N_{\rm excur}$ is the number of excursion events of a halo during 
    its fast phase. 
    $\tau_{\rm excur}/\tau_{\rm f}$ is the total time spent by a halo
    in all the excursions divided by the duration of the entire fast phase.
    $\Delta M_{\rm v,excur,all}/M_{\rm f}$ is the total mass accreted
    during all excursions during the fast phase divided by the 
    mass at the end of the fast phase.
    $t_{\rm f} - t_{\rm excur}$ is the difference of cosmic time between 
    the end of the fast phase and an excursion.
    $M_{\rm v,excur}/M_{\rm f}$ is the halo mass at an excursion
    divided by that at the end of the fast phase.
    $\Delta M_{\rm v,excur}/M_{\rm v,excur}$ is the mass accreted 
    during one excursion divided by the mass prior to the excursion.
    Two variants are shown by {\bf thin} and {\bf thick} curves, 
    respectively. This figure gives a summary of excursions in halo assembly,
    which potentially drives the galaxy temporarily to a disk-like morphology.
}
    \label{fig:mah_transitions}
\end{figure*}

To characterize the fluctuation in halo assembly history and the difference between 
the two phases, we compute the logarithmic residual of the simulated virial velocity 
$V_{\rm v}$ relative to $V_{\rm v}^\text{(fit)}$, 
obtained from a parametric fitting of the halo mass assembly history 
(see \S\ref{sec:halos}, Eqs.~\ref{eq:mah-fitting} and \ref{eq:v-vir-fitting}), as 
\begin{equation}
    \Delta \log\,V_{\rm v} = \log\,V_{\rm v} - \log\,V_{\rm v}^\text{(fit)}.
\end{equation}
We then use the standard deviation of $\Delta \log\,V_{\rm v}$ obtained from all 
halo assembly histories at a specific epoch relative to the transition time, 
$t - t(z_{\rm f})$, to quantify the fluctuation in halo assembly. 
The results are shown in Fig.~\ref{fig:residual_history} for the two variants,
$\gamma_{\rm f}=3/8$ and $0$,  and for three bins of halo mass at $z=0$.
For given $\gamma_{\rm f}$, halos of different masses have a similar pattern: 
significant fluctuation prior to the transition, a rapid decline during the 
transition, and subsequent stabilization. 
The fluctuation during the slow phase is about $1/4$ to $1/3$
of that in the fast phase, with the transition period lasting about $1\gyr$. 
The similarity for halos with different masses suggests a unified path of 
deepening in the gravitational potential well in terms of a time variable 
that is measured relative to the transition time. This universality motivates us to 
use a single parametric form (Eq.~\ref{eq:mah-fitting}) to model mass assembly histories 
of all halos, and to separate the two phases by a constant threshold of 
$\gamma_{\rm f}$ (Eqs.~\ref{eq:def-high-zf} and \ref{eq:def-low-zf}).

The rapid change in the gravitational potential well can also drive changes in the 
binding energy of dark matter particles within the halo 
\citep[see e.g. figure 10 of][]{zhaoGrowthStructureDark2003}. 
In Fig.~\ref{fig:e_bind_vs_z}, we present $E_{\rm b}$, the specific (per unit mass) 
binding energy of particles, and $\gamma$, the specific growth rate of the halo, 
as functions of redshift for a Milky Way-mass halo ($M_{{\rm v},z=0}\approx 10^{12}\msun$). 
This particular halo does not experience a `back-splash' event in its history, 
and thus environmental effects do not significantly 
contribute to the change in the particle binding energy. 
The binding energy is computed using all particles bound to the central subhalo. 
The thick curve in the top panel represents the average $E_{\rm b}$ traced by 
all particles bound to the central subhalo over the redshift range shown 
in the figure. The zero point of $E_{\rm b}$ is set such that the maximum value 
among all the tracers is zero. A distinct trend in $E_{\rm b}$ emerges: prior to the 
transition ($z \gtrsim 2$), the particles are nearly bound; within the transition 
period ($0.8 \lesssim z \lesssim 2$, defined by $0 < \gamma^\text{(fit)} < 3/8$), 
the binding energy decreases rapidly and fluctuates significantly; 
well after the transition ($z \lesssim 0.8$), the particles are all 
tightly bound and their binding energy stabilizes at a roughly constant value.
A similar pattern is seen even when considering only particles within $0.2 R_{\rm v}$, 
approximately the upper limit of the size of the SGC (\S\ref{ssec:four-quadrant-gas-evolution}). 
The similarity in the  trajectories of $E_{\rm b}$ and $\gamma$ suggests 
that particles closely respond to changes in the halo potential, supporting our model
assumption on the relation between the two-phase assembly of halos and the dynamical 
hotness of galaxies that form in them.  

The lower panel of Fig.~\ref{fig:e_bind_vs_z} presents three versions of $\gamma$ obtained using: the simulated assembly history; a parametric function fitting of $M_{\rm v}(z)$ 
(denoted as `fit'); and a smoothed version of $M_{\rm v}(z)$ in a Gaussian kernel 
with a width of $\tau_{\rm dyn}(z)$ (denoted as `sm'). 
A notable feature is the presence of fluctuations in $\gamma$ even in the smoothed version. 
During the fast phase, as defined by the Low-$z_{\rm f}$ variant where 
$\gamma^\text{(fit)} \geqslant \gamma_{\rm f} = 0$, $\gamma^\text{(sm)}$ 
makes an excursion below $\gamma_{\rm f}$ at $z \approx 1.7$ and returns at $z \approx 1$ (see also the examples in Fig.~\ref{fig:transition_points}). 
This suggests transformations of galaxies from bulge-like
to disk-like and back to bulge-like, corresponding to 
dynamic oscillations between Q1 and Q2 described 
in \S\ref{ssec:four-quadrant-gas-evolution}.

To quantify such dynamic oscillations, we identify all events of excursion to the slow assembly regime 
during the fast phase of each halo. Here, we define an excursion event as a continuous 
time interval during which $\gamma^\text{(sm)}$ falls below $\gamma_{\rm f}$. 
Fig.~\ref{fig:mah_transitions} shows various statistics of these excursions for 
the halo sample $S_{\rm h,large}$ using both High-$z_{\rm f}$ and Low-$z_{\rm f}$. 
For both variants, the number of excursions ($N_{\rm excur}$) is $\leq 2$ 
for about $80\%$ of the halos, and $\leq 3$ for nearly all halos, indicating that excursions are not rare. 
The total duration of all excursions ($\tau_{\rm excur}$) accounts for $\leq 30\%$ 
of the fast phase for about $80\%$ of halos, and $\leq 40\%$ for nearly all halos, suggesting that these 
periods of temporary excursions away from dynamically hotness are relatively short. 

The starting time of these excursions, $t_{\rm excur}$, is within $3 \gyr$ 
prior to the transition time defined by the High-$z_{\rm f}$ variant for about $80\%$ of halos, 
as one can see from the lower left panel of Fig.~\ref{fig:mah_transitions}). In contrast, the distribution 
of $t_{\rm f} - t_{\rm excur}$ is quite uniform between $0$ and $8 \gyr$ when Low-$z_{\rm f}$ is 
adopted. This difference implies that excursions are frequent when 
$0 < \gamma^\text{(fit)} < 3/8$, suggesting that morphological transformations of 
galaxies are frequent during such periods.
As discussed in \S\ref{sssec:q2-phase}, the transition to Q2 is often associated with 
`wet compactification' that may lead to the emergence of `blue nuggets' and 
enhanced growth of SMBHs. If the AGN feedback is sufficiently strong, 
it may lead to the formation of `red nuggets' even at high redshift. 
The distribution of $M_{\rm v, excur}$, the halo mass during the excursion, is shown in the 
lower middle panel. The result  indicates that excursions are most frequent when $M_{\rm v}$ is about 
$10$ to $20$ percent of $M_{\rm f}$, the halo mass at the end of the fast phase. As shown by the top right panel, 
the typical mass increase during all the excursions, $\Delta M_{\rm v, excur, all}$, is about 6 to 12 percent 
of $M_{\rm f}$. The relatively low values of $M_{\rm v, excur}/M_{\rm f}$ 
and $\Delta M_{\rm v, excur, all}/M_{\rm f}$ indicate that, 
for most systems, the mass of the disk-like component expected to form 
during the excursions is likely much lower than that of the mass component formed 
during the entire fast assembly phase. However, as shown in the lower right panel, for 
a significant fraction of halos, the mass acquisition during an excursion, 
$\Delta M_{\rm v, excur}$, can actually be comparable to the halo mass at the beginning 
of the excursion, $M_{\rm v, excur}$, particularly when the High-$z_{\rm f}$
variant is used. This indicates that the mass acquisition during some excursions 
at high $z$ are significant. Such an excursion is expected to generate a system in Q2, 
producing a disk-like structure that is unstable and clumpy.  

To summarize, the existence of excursions to the slow assembly regime during the fast assembly phase is not 
expected to have a significant impact on our results based on characterizing 
halo assembly histories with the two phases proposed in the main text. The discussion 
above also highlights the connections between the fluctuation in halo assembly and the transformation 
of galaxy morphology. This has significant implications for galaxy formation, particularly at high $z$ 
where such fluctuations are expected to be important. Clearly, detailed investigations are needed to 
quantify such connections and to explore their implications.

\section{Adjustment of Model Parameters}
\label{app:adjustment-model-parameters}

\begin{figure} \centering
    \includegraphics[width=0.9\columnwidth]{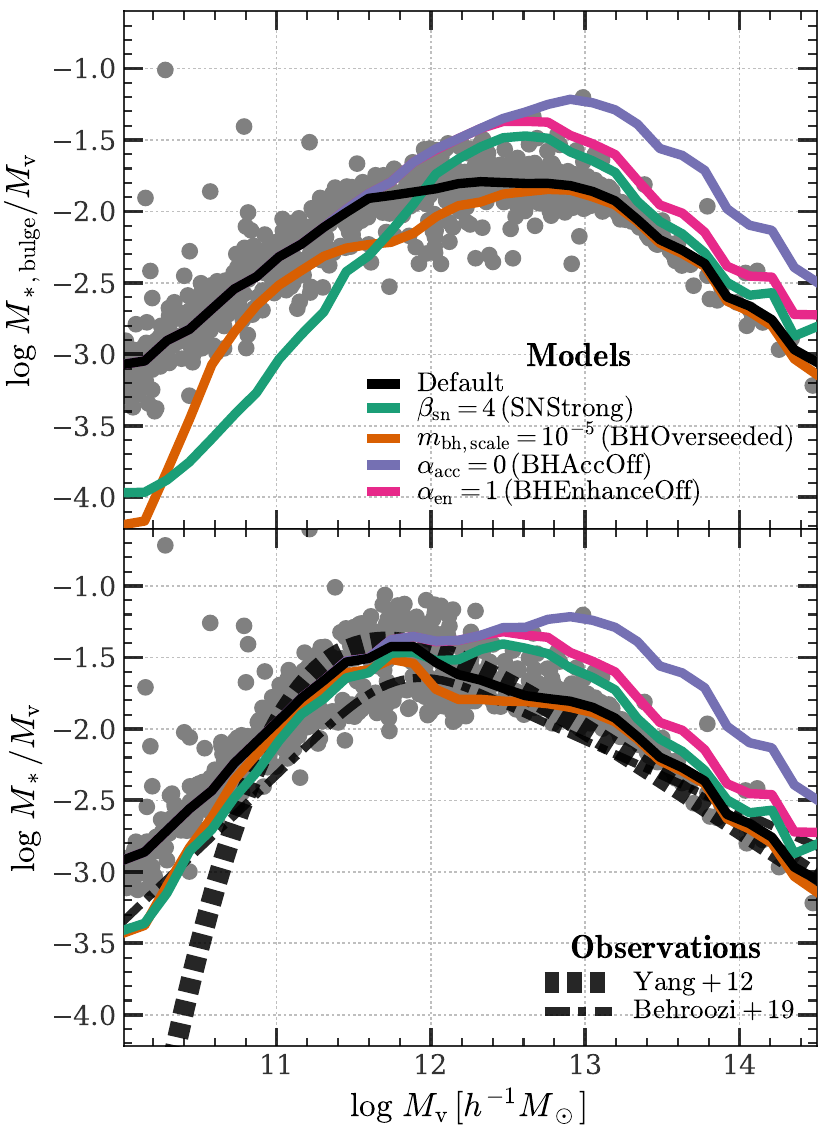}
    \caption{
        Stellar mass to halo mass ($M_{\rm v}\equiv M_{\rm 200c}$) ratio as a function of 
        halo mass at $z=0$ predicted by our models, where $z_{\rm f}$ is defined 
        by the Low-$z_{\rm f}$ variant. 
        {\bf Upper panel} shows bulge stellar mass and
        {\bf bottom panel} shows total stellar mass (bulge + disk).
        Gray scatter points represent individual galaxies obtained from
        default model parameters (see Table~\ref{tab:parameters}), while 
        the {\bf black solid curve} shows the 
        median relation. {\bf Colored curves} are obtained by varing one parameter, 
        as indicated in the legend, from the default model.
        In the bottom panel, {\bf thin black dashed curve} shows the relation obtained by
        \citet{yangEVOLUTIONGALAXYDARK2012} using conditional 
        stellar mass function modeling, while {\bf thick black dashed curve} shows the relation
        obtained by \citet{behrooziUniverseMachineCorrelationGalaxy2019} 
        using an empirical model calibrated to various observations,
        with halo mass defined to be the peak mass.
    }
    \label{fig:ms_to_mh}
\end{figure}

Most of our model parameters have theoretically plausible values. However, 
there remain parameters that are still uncertain. We thus seek observations 
for calibration. In this section, we elaborate our heuristic approach to 
adjust model parameters.

The overall strategy, as stated in \S\ref{ssec:slow-phase-model}, 
involves only using the local stellar mass ($M_*$) to halo mass 
($M_{\rm v}\equiv M_{\rm 200c}$) relation at $z\approx 0.1$ obtained by 
\citet{yangEVOLUTIONGALAXYDARK2012} and the local black hole mass 
($M_{\rm bh}$) to bulge stellar mass ($M_{\rm *,bulge}$) relation for 
elliptical galaxies at $z\approx 0$ obtained by \citet{grahamAppreciatingMergersUnderstanding2023}, to calibrate model parameters. 
We deliberately avoid using high-redshift observations for fine-tuning the 
parameters, since we aim to maximize the predictive power of our model. 
It is also noteworthy that all model parameters in our phase-phase model 
remain constant over the cosmic time, and thus, any redshift dependencies 
of the predicted galaxies and SMBHs come from halo assembly histories. 
This design strategy remains valid as long as our model effectively 
reflects the fundamental physics driving galaxy formation in the self-gravitating 
turbulent medium, which we confirm through all the model-observation comparisons 
presented in this paper.

As described in \S\ref{ssec:result-growth-of-components} and shown in 
Fig.~\ref{fig:mbh_ms_paths}, it is evident that a large value of the 
mass-capturing parameter ($\alpha_{\rm cap}$) is critical for the 
SMBH to evolve into the self-regulating regime that matches the observed 
$M_{\rm bh}$ - $M_{\rm *,bulge}$ relation. However, an 
excessively large value of $\alpha_{\rm cap}$ results in an 
upward shift of the $M_{\rm bh}$ - $M_{\rm *,bulge}$ relation beyond 
the observed range. Consequently, we set $\alpha_{\rm cap} = 2.5$ to ensure 
that the predicted relation eventually converges with that obtained 
by \citet{grahamAppreciatingMergersUnderstanding2023}. As the observed 
$M_{\rm bh}$ - $M_{\rm *,bulge}$ relation for ellipticals is limited to 
massive SMBHs ($M_{\rm bh} \gtrsim 10^7 \msun$), the value of the 
enhancement factor, $\alpha_{\rm en}$, which governs early-time SMBH growth, 
must be calibrated by additional observation. 
We thus use the observed $M_*$ - $M_{\rm v}$ relation for it.

Figure~\ref{fig:ms_to_mh} shows the predicted relation between stellar mass 
(bulge or total) and halo mass according to our model, 
with $\alpha_{\rm en}$ set to 1 to deactivate the enhancement 
(pink curve, labeled as `BHEnhanceOff'). All other parameters are at their 
default values (see Table~\ref{tab:parameters}). Without the fast growth 
resulting from SN-driven turbulence in SGC, SMBHs are unable to grow 
effectively, leading to an excessively fast early growth of 
stellar bulge and an abundance of over-massive galaxies in 
high-mass halos. Conversely, if $\alpha_{\rm en}$ is too large, it would 
result in an expectation of excessively low predicted stellar masses in 
high-mass halos. Therefore, we set $\alpha_{\rm en} = 3$ to match the observation. 
As a self-consistency check, we also show the stellar mass to halo mass relation 
when we completely disable SMBH growth ($\alpha_{\rm acc}$ = 0, shown by the 
purple curve labeled as `BHAccOff'). As expected, the high-mass end of the 
relation is significantly lifted, indicating the necessity of AGN feedback 
from accreted SMBHs to reproduce the observed stellar masses in high-mass halos.

Unlike the BH capturing parameters, the seed mass of SMBH does not impact the 
stellar masses of high-mass galaxies. This is evident in 
Fig.~\ref{fig:mbh_ms_paths}, where SMBHs with arbitrary seeded mass quickly converge 
to the observational line. Therefore, it is expected that the seed mass 
solely affects the stellar mass of low-mass systems, as demonstrated by the 
brown curve labeled as `BHOverseeded' in Fig.~\ref{fig:ms_to_mh} 
through the overseeding of the SMBHs. As the seed mass distribution has 
not been comprehensively understood, we opt to use small stellar-level 
seeds with $M_{\rm bh, min} = 10 \msun$, relying on early-time fast growth to 
boost it to a SMBH. This choice is thought to be both physically 
plausible and numerically stable, hence it is adopted as our default model.

For the parameter $\beta_{\rm sn}$ governing the strength of SN feedback 
in low-mass halos, we expect a value of approximately 2, ensuring that 
the $V^2_{\rm g}$ and $V^2_{\rm w}$ terms in equation~\eqref{eq:def-f-sn} have 
the dimensionality of specific energy. Indeed, this value yields a faint-end 
behavior of the $M_*$-$M_{\rm v}$ relation that well matches Model-III 
of \citet{luEmpiricalModelStar2014}, which concludes that an upturn of the 
faint end is necessary to match the observed galaxy stellar mass function 
in clusters. For comparison, we present in Fig.~\ref{fig:ms_to_mh} a model 
with $\beta_{\rm sn} = 4$ (green curve, labeled as `SNStrong'), which 
obviously eliminates the faint-end upturn and even results in an 
underestimation of $M_*$ for halos $M_{\rm } \leqslant 10^{11} \msun$.

It is important to note that full Bayesian inference based on complete sampling 
of model ensembles and model parameters gives a more reliable parameter calibration 
and model comparison. However, it needs a careful combination of observational 
data to mitigate their systematics and a precise design of the 
likelihood function to account for the error distribution and covariance of 
data points. This is beyond the scope of this paper and will be 
addressed in future work.

\bsp	
\label{lastpage}
\end{document}